\documentclass[twocolumn]{aastex63}

\usepackage{amsmath}
\usepackage{amssymb}
\usepackage{amsthm}
\usepackage{natbib,aasdefs,url,bm}
\usepackage{graphicx}
\usepackage{color}
\usepackage{mathtools,  epsfig, lipsum}
\usepackage{subfigure}
\usepackage{multirow,booktabs,colortbl,tabularx}
\usepackage{enumitem}
\usepackage{booktabs}
\usepackage{hyperref}

\newcommand{\be}{\begin{equation}}
\newcommand{\ee}{\end{equation}}
\def\me{m_{\rm e}}
\def\lth{\ell_{\rm th}}

\def\linj{\ell_{\rm inj}}
\def\sigmaT{\sigma_{\rm T}}
\def\tauT{\tau_{\rm T}}
\def\tauKN{\tau_{\rm KN}}

\def\tauphmat{\tau_{\rm ph-mat}}
\def\tauCeff{\tau_{\rm C, eff}}
\def\tauggth{\tau_{\gamma\gamma, {\rm th}}}

\def\gdag{\gamma_{\dagger}}

\shorttitle{Gamma-ray thermalization and leakage from millisecond magnetar nebulae}
\shortauthors{Vurm \& Metzger }

\begin{document}

\title{Gamma-ray Thermalization and Leakage from Millisecond Magnetar Nebulae:\\ Towards a Self-Consistent Model for Superluminous Supernovae}

\author{Indrek Vurm}
\affil{Tartu Observatory, Tartu University, 61602 T{$\bar o$}ravere, Tartumaa, Estonia}

\author{Brian D. Metzger}
\affil{Department of Physics and Columbia Astrophysics Laboratory, Columbia University, Pupin Hall, New York, NY 10027, USA}
\affil{Center for Computational Astrophysics, Flatiron Institute, 162 5th Ave, New York, NY 10010, USA} 

% These dates will be filled out by the publisher
%\date{Accepted XXX. Received YYY; in original form ZZZ}

% Enter the current year, for the copyright statements etc.
%\pubyear{2020}

% Don't change these lines
%\hypersetup{draft}
%\begin{document}
%\label{firstpage}
%\maketitle

\begin{abstract}
Superluminous supernovae (SLSNe) are massive star explosions that are too luminous to be powered by traditional energy sources, such as the radioactive decay of $^{56}$Ni.  These transients may instead be powered by a central engine, such as a millisecond pulsar or magnetar, whose relativistic wind inflates a nebula of high energy particles and radiation behind the expanding supernova ejecta.  We present three-dimensional Monte Carlo radiative transfer calculations of SLSNe which follow the production and thermalization of high energy radiation from the nebula into optical radiation and, conversely, determine the gamma-ray emission that escapes the ejecta without thermalizing.  Specifically, we track the evolution of photons and matter in a coupled two-zone (``wind/nebula'' and ``ejecta'') model, accounting for the range of radiative processes including (typically multiple generations of) pair creation.  We identify a novel mechanism by which $\gamma\gamma$ pair creation in the upstream pulsar wind regulates the mean energy of particles injected into the nebula over the first several years after the explosion, rendering our results on this timescale insensitive to the (uncertain) intrinsic wind pair multiplicity.  To explain the observed late-time steepening of SLSNe optical light curves as being the result of gamma-ray leakage, we find that the nebular magnetization must be very low $\varepsilon_{\rm B} \lesssim 10^{-6}-10^{-4}$.  For higher $\varepsilon_{\rm B}$ synchrotron emission quickly comes to dominate the thermalized nebula radiation, and being readily photoelectrically absorbed because of its lower photon energies, results in the supernova optical light curve tracking the spin-down power even to late times $\gtrsim 1$ year, inconsistent with observations.  For magnetars to remain as viable contenders for powering SLSNe, we thus conclude that either magnetic dissipation in the wind/nebula is extremely efficient, or that the spin-down luminosity decays significantly faster than the canonical magnetic dipole rate $\propto t^{-2}$ in a way that coincidentally mimicks gamma-ray escape.  Our models predict a late-time flattening in the optical light curves of SLSNe, for which there may be evidence in SN 2015bn. 
\end{abstract}


%%%%%%%%%%%%%%%%%%%%%%%%%%%%%%%%%%%%%%%%%%%%%%%%%%

%%%%%%%%%%%%%%%%% BODY OF PAPER %%%%%%%%%%%%%%%%%%

\section{Introduction} A growing sample of stellar explosions have been discovered with luminosities too high to be powered by traditional energy sources, such as the radioactive decay of $^{56}$Ni.  These include the rare class of stellar core collapse events known as ``superluminous supernovae'' (SLSNe; e.g.~\citealt{Quimby+11,Nicholl+13,Inserra+13,Howell+13,Dong+16,DeCia+18,Lunnan+18,Quimby+18}; see \citealt{GalYam18,Inserra19} for recent reviews) as well as the emerging class of luminous transients with faster evolving light curves indicative of a lower ejecta mass, sometimes referred to as ``fast blue optical transients'' (FBOTs; e.g.~ \citealt{Drout+14,Arcavi+16,Tanaka+16}).   The best studied FBOT is AT2018cow \citep{Prentice+18,Perley+19,Kuin+19}, which peaked on a timescale of only a few days and was accompanied by nonthermal emission across the electromagnetic spectrum, from radio to gamma-rays (\citealt{Margutti+19,Ho+19}).  

Powering the light curves of SLSNe and FBOTs requires prolonged heating of the ejecta by a centrally concentrated energy source.  One such potential source are shocks generated by the collision of the ejecta with a dense circumstellar shell or disk surrounding the progenitor star at the time of explosion (e.g.~\citealt{Smith&McCray07,Chevalier&Irwin11,Moriya+13}).  Circumstellar interaction is most compelling as an explanation for the hydrogen-rich class of SLSNe (which often show emission lines indicative of slow-moving gas; e.g.~\citealt{Smith+07,Fransson+14,Benetti+14,Nicholl+20}), though similar shock-powered interaction could also account for the light curves of at least some hydrogen-poor SLSNe-I (e.g.~\citealt{Sorokina+16,Kozyreva+17,Rest+18}). 

Another commonly discussed model for SLSNe-I/FBOTs invokes energy injection by a young active compact object, such as an accreting black hole or neutron star \citep{Quataert&Kasen12,Woosley&Heger12,Margalit&Metzger16,Moriya+18}, or the rotationally-powered wind of a pulsar or magnetar with a millisecond spin period \citep{Maeda+07,Kasen&Bildsten10,Woosley10,Dessart+12,Metzger+15,Sukhbold&Woosley16,Sukhbold&Thompson17}.  Among the evidence supporting the presence of a central engine are emission line features in the nebular spectra of SLSNe (e.g., \citealt{Nicholl+19}) and peculiar Type Ib SNe (e.g., \citealt{Milisavljevic+18}) similar to those seen in the hyper-energetic supernovae observed in coincidence with long duration gamma-ray bursts.  A central engine in AT2018cow is supported by the rapid time variability and energy spectrum (particularly the Compton hump feature) of the X-rays observed in coincidence with the optical radiation (e.g.~\citealt{Margutti+19}).  

Theoretical models for the visual light curves and spectra of engine-powered supernovae generally assume that the power output of the central engine is thermalized by the surrounding ejecta with 100\% efficiency (\citealt{Kasen&Bildsten10,Woosley10,Dessart+12,Chatzopoulos+13,Metzger+15}).  While this assumption is important to the overall viability of the scenario, and to quantitative inferences drawn about the properties of the central engine (e.g., the dipole magnetic field strength and birth spin period of the neutron star), its justification has thus far received little attention in the literature.

A broad outline for how such a thermalization process might occur is as follows.  A relativistic wind from the neutron star or black hole engine inflates a nebula of relativistic electron/positron pairs and radiation behind the expanding supernova ejecta shell \citep{Kotera+13,Metzger+14,Murase+15}.  Pairs heated near the wind termination shock enter the nebula and quickly radiate their energy via synchrotron and inverse Compton (IC) processes in a broad spectrum spanning the X-ray/gamma-ray band, in analogy with ordinary pulsar wind nebulae like the Crab Nebula (\citealt{Buhler&Blandford14}).  Only the portion of this radiation which can escape the nebula without undergoing $\gamma\gamma$ pair creation, and is in the appropriate energy range to be absorbed and thermalized by the ejecta shell, is available to heat the ejecta and power the supernova emission.  

Thermalization of the nebular radiation will be most efficient at early times after the explosion, when the column density through the ejecta shell and ``compactness'' of the system are at their highest.\footnote{At early times, the nebula can also deposit energy directly by driving a shock into the inner edge of the ejecta shell (e.g.~\citealt{Metzger+14,Kasen+16,Chen+16,Suzuki&Maeda19,Suzuki&Maeda20}).}  However, as the ejecta shell expands and becomes increasingly transparent its radiation field dilutes.  As a result, the efficiency of the thermalization process will drop and the optical supernova light curve will eventually fall below the rate of energy injection from the central engine, as the majority of the engine's power escapes directly from the nebula as high-energy radiation.  

In support of such a picture, the well-studied SLSNe 2015bn (\citealt{Nicholl+16a,Nicholl+16b}) showed a marked steepening of its optical/UV light curve at $t \gtrsim 200$ days below the predicted $\propto t^{-2}$ decay of the magnetar dipole spin-down luminosity, qualitatively consistent with the expectation of radiation leakage \citep{Nicholl+18}.  Searches for this ``missing energy'' by means of X-ray  (\citealt{Margutti+18,Bhirombhakdi+18}) and gamma-ray \citep{Renault-Tinacci+18} observations of SLSNe at late times have thus far only resulted in upper limits (with one possible exception; see  \citealt{Levan+13}).  However, this is not necessarily surprising because the ejecta is expected to remain optically thick to soft X-rays for decades after the explosion \citep{Margalit+18}, while most of the existing gamma-ray upper limits are not deep enough to constrain the expected escaping flux.  

In this paper, we present calculations of engine-powered supernova light curves which for the first time account self-consistently for the thermalization and escape of high energy radiation from the central nebula and the surrounding ejecta shell.  We accomplish this by means of three-dimensional time-dependent Monte Carlo simulations which track the coupled evolution of photons and electron/positron pairs in the nebula and ejecta through the myriad physical processes coupling them.  Such a detailed treatment is necessary because the process of gamma-ray thermalization is complex and non-linear.  

For example, although many processes in the supernova ejecta shell and its radiation field can absorb gamma-ray photons, the initial absorption of a photon is no guarantee of its ultimate thermalization.  Several photon destruction processes result in the creation of high-energy electron/positron pairs.  These pairs in turn radiate their received energy as secondary photons, which themselves can either be absorbed or escape from the ejecta.  Furthermore, over the first several years after the explosion $\gamma\gamma$ pair creation also occurs in the upstream region of the pulsar wind (interior to the termination shock), augmenting the wind mass-loading and regulating the average energy of the pairs the wind releases into the nebula.  Thus, even the inner boundary condition defined by the engine is intimately coupled to the properties of the larger-scale nebula/ejecta during the most accessible observational window.
%Although we primarily focus on SLSNe, a similar model may be applicable to FBOTs and other engine-powered transients, such as tidal disruption events and neutron star mergers (in cases they produce long-lived magnetar remnants). 

This paper is organized as follows.  We begin in Section \ref{sec:preliminaries} with several preliminary considerations which motivate the details of our model.  These include a brief overview of engine-powered supernovae (Section \ref{sec:SLSNe}); properties of the radiation field in the nebula (Section \ref{sec:nebula}); an enumeration of key photon absorption processes (Section \ref{sec:absorption}); the process of thermalization of high-energy photons by the ejecta (Section \ref{sec:thermalization}); and regulation of the wind mass-loading by in-situ $\gamma\gamma$ pair production (Section \ref{sec:gamma}).  In Section \ref{sec:model} we describe and present the results of our numerical radiative transfer calculations.  Further interpretation and implications of our results are discussed in Section \ref{sec:discussion}.  We summarize our findings and conclude in Section \ref{sec:conclusions}.  The Appendices provide several auxiliary calculations which support those in the main text.  

\section{Preliminary Considerations}
\label{sec:preliminaries}

This section introduces the key concepts and physical processes that enter our calculations and provides several analytic estimates that are useful later in interpreting our numerical results.  The reader uninterested in these details on the first pass is encouraged to advance directly to the description and results of the simulations (Section \ref{sec:model}), returning to this section as needed.  Table \ref{tab:timescales} summarizes several of the key timescales in the problem.  

\subsection{Overview of engine-powered supernovae}
\label{sec:SLSNe}

The result of the supernova explosion is the creation of a ballistically\footnote{The high pressure of the nebula may drive a shock into the ejecta shell and lead to its acceleration at early times (e.g.~\citealt{Suzuki&Maeda16}).  However, this acceleration will subside once radiation begins to escape from the nebula at times $t \gtrsim t_{\rm pk}$ (eq.~\ref{eq:tpk}) of greatest interest in this paper.  The ejecta kinetic energy defined here is thus the sum of the initial energy of the explosion and that released by the engine at $t \lesssim t_{\rm pk}$ (e.g., \citealt{Metzger+15,Soker16}).} expanding shell of ejecta of mass $M_{\rm ej} = 10M_{10}M_{\odot}$, mean velocity $v_{\rm ej} = 10^{9}v_{9}$ cm s$^{-1}$, mean radius $R_{\rm ej} = v_{\rm ej}t$, and thickness $\approx R_{\rm ej}$.  The mean electron density in the shell, and its radial Thomson optical depth at time $t = 1 t_{\rm yr}$ yr following the explosion are given, respectively, by
\be
n_{\rm e} \simeq \frac{3M_{\rm ej}Y_e}{4\pi (v_{\rm ej}t)^{3}m_p} \approx 4.6\times 10^{7}M_{10}v_{9}^{-3}t_{\rm yr}^{-3}{\rm cm^{-3}},
\ee 
\be
\tau_{\rm T} \simeq n_{\rm e} \sigma_{\rm T} R_{\rm ej} = \frac{3M_{\rm ej}\kappa_{\rm es}}{4\pi (v_{\rm ej}t)^{2}} \approx 0.95 \, M_{10}v_{9}^{-2}t_{\rm yr}^{-2},
\label{eq:tauT}
\ee
where $\sigma_{\rm T}$ is the Thomson cross section, $\kappa_{\rm es} = Y_e\sigma_{\rm T}/m_p$, and we have assumed an electron fraction $Y_e = 0.5$ appropriate for hydrogen-poor ejecta.  The ejecta becomes optically thin to Thomson scattering ($\tau_{\rm T} = 1$) after a time
\be
t_{\rm T} \simeq 0.97 M_{10}^{1/2}v_{9}^{-1}\,\,{\rm yr}.
\label{eq:tT} 
\ee

A central compact object is assumed to deposit energy into a hot nebula behind the ejecta at a rate
\be
L_{\rm e}(t) = \frac{E_{\rm e}}{t_{\rm e}}\frac{(\alpha-1)}{(1 + t/t_{\rm e})^{\alpha}} \underset{t\gg t_e}\approx \frac{E_{\rm e}}{t_{\rm e}}\frac{(\alpha-1)}{(t/t_{\rm e})^{\alpha}},
\label{eq:Le}
\ee
where $E_{\rm e}$ is the total energy of the engine and $t_{\rm e}$ the duration of its peak activity.  The power-law decay index $\alpha = 2$ is the case of an isolated neutron star or magnetar with a fixed dipole magnetic field and spin inclination \citep{Spitkovsky06} and without significant gravitational wave losses, $\alpha \simeq 2.38$ for a fall-back accreting neutron star \citep{Metzger+18}, $\alpha = 5/3$ for fall-back accretion following a stellar tidal disruption event \citep[e.g.,][]{Phinney89}, and $\alpha \le 5/3$ in supernova fall-back accretion models \citep[e.g.,][]{Coughlin+18} or in the case of accretion fed by a viscously spreading disk \citep[e.g.,][]{Cannizzo+90}.  

Because it provides an acceptable fit to early-time SLSNe light curves (e.g., \citealt{Nicholl+17}) and the combined X-ray/optical luminosity of AT2018cow (e.g., \citealt{Margutti+19}), we hereafter focus on the case of an isolated magnetar ($\alpha = 2$).  Following the convention used by \citet{Nicholl+17}, the engine energy $E_{\rm e} \simeq 2.6\times 10^{52}(P_{0}/1\,{\rm ms})^{-2}{\rm erg}$ is the magnetar's initial rotational energy (for assumed mass 1.4$M_{\odot}$ and birth spin period $P_0$), while the engine lifetime,
\be t_{\rm e} \simeq 1.3\times 10^{5}\left(\frac{P_0}{1\,\rm ms}\right)^{2}B_{14}^{-2}\,{\rm s}, \label{eq:te}
\ee is the magnetic dipole spin-down timescale, where $B_{\rm d}  = B_{14} \times 10^{14}$ G.  The spin-down power at times $t \gg t_{\rm e}$ is thus given by (eq.~\ref{eq:Le})
\be L_{\rm e} = E_{\rm e}t_{\rm e}/t^{2} \approx 3.4\times 10^{42}\,B_{\rm 14}^{-2}t_{\rm yr}^{-2}\,{\rm erg\,s^{-1}},
\label{eq:Le2}
\ee
which is notably independent of $P_0$.  Fitting the magnetar\footnote{Much of the inferred parameter space includes dipole magnetic fields $B_{\rm d} \lesssim 10^{14}$ G outside the range of most Galactic magnetars, instead overlapping those of radio pulsars.  Furthermore, while the usual hallmark of Galactic magnetars is flaring activity driven by the dissipation of magnetic energy, here we are interested in the comparatively steady extraction of rotational energy, again more akin to pulsars than magnetars.  Nevertheless, we use the term ``magnetar'' throughout this manuscript to remain consistent with the SLSNe literature.} model to SLSNe light curve data yield values $P_0 \sim 1-5$ ms and $B_{14} \sim 0.3-3$ (e.g.~\citealt{Kasen&Bildsten10,Chatzopoulos+13,Metzger+15,Nicholl+17}).

The key to maximizing the peak luminosity of the optical transient is to approximately match the engine lifetime $t_{\rm e}$ with the photon diffusion timescale on which the supernova optical light curve peaks \citep{Arnett82},
\be t_{\rm pk} \approx 0.1 M_{10}^{1/2}v_9^{-1/2}\left(\frac{\kappa_{\rm opt}}{\kappa_{\rm es}/3}\right)^{1/2}\,{\rm yr}.
\label{eq:tpk}
\ee 
The latter is set by the condition $\tau_{\rm opt} \sim c/v_{\rm ej}$, where $\tau_{\rm opt} = (\kappa_{\rm opt}/\kappa_{\rm es})\tau_{\rm T}$ is the optical depth of the ejecta to thermal optical/UV photons and $\kappa_{\rm opt}$ is the effective continuum opacity of Doppler-broadened atomic lines (e.g.~\citealt{Pinto&Eastman00}).  

The energy deposition rate $L_{\rm e}$ can also be characterized by the dimensionless ``compactness'' parameter \citep[e.g.][]{Guilbert+83}, which we define as
\be
\linj \equiv \frac{\sigmaT L_{\rm e}}{4\pi\me c^3 R_{\rm ej}} \underset{t \gg t_{\rm e}}\approx 2.3\times 10^{-4}B_{14}^{-2}v_{9}^{-1}t_{\rm yr}^{-3}
\label{eq:ellinj}
\ee
An analogous quantity for the radiation field permeating the ejecta can be defined as
\be
    \ell \equiv \frac{u_{\gamma}\sigma_{\rm T}R_{\rm ej}}{m_e c^{2}} ,
    %\ell \equiv \frac{\sigma_{\rm T}}{4\pi m_e c^{3}} \frac{L_{\rm e}}{R_{\rm ej}},
\label{eq:ell}
\ee
where $u_{\gamma}$ is the energy density of the radiation field in question\footnote{Compactness $\ell$ of a radiation field of energy density $u_{\gamma}$ within a region of size $R$ is numerically equal to the Thomson opacity of the same region that would be obtained if all the radiation energy was converted to electron/positron rest mass.}.
The above quantities are related such that if all the injected energy is converted to radiation in an optically thin environment (with a photon residence time $\sim R_{\rm ej}/c$), then one would have $\ell \approx \linj$.  The main utility of the compactness parameter lies in its direct relation to important quantities within the ejecta and nebula such as the pair-production opacity and electron cooling rate, allowing for a readily scalable description of the physical system. 
%(Section \ref{sec:absorption}; Appendix \ref{sec:app:cooling}, \ref{app:regulation}). 

A portion of the injected luminosity $L_{\rm e}$ will be absorbed by the ejecta and reprocessed to thermal optical/UV wavelength radiation of characteristic thermal temperature $T_{\rm eff} \sim 10^{4}-10^{5}$ K, which is responsible for powering the observed supernova emission.  It is thus useful to introduce a ``thermal compactness'' parameter,
\be
\lth \equiv \frac{u_{\rm th}\sigma_{\rm T}R_{\rm ej}}{m_e c^{2}} = \linj \, f_{\rm th} \, (1+\tau_{\rm opt}),
\label{eq:ellopt}
\ee
where $u_{\rm th} \approx L_{\rm e}(1+\tau_{\rm opt})f_{\rm th}/(4\pi c R_{\rm ej}^2)$ is the energy density of the thermal optical/UV supernova radiation of characteristic photon energy $\epsilon_{\rm opt} \approx 3kT_{\rm eff} \sim 1-10$ eV, and the factor $(1 + \tau_{\rm opt}$) accounts for the additional residence time of thermal photons in the ejecta.

The factor $f_{\rm th}$ entering Equation \ref{eq:ellopt} is the efficiency with which the engine luminosity is thermalized by the ejecta.  A self-consistent determination of $f_{\rm th}$ is non-trivial and will take up the greater part of this paper.  However, for purposes of an estimate connecting to the prior literature, we can parameterize $f_{\rm th} = 1-\exp(-\tau_{\rm therm}) \approx \tau_{\rm therm}$ via a gamma-ray ``thermalization'' optical depth $\tau_{\rm therm} \equiv (\kappa_{\rm th}/\kappa_{\rm es})\tau_{\rm T}$ (\citealt{Wang+15,Chen+15}), which does a good job explaining the late-time decay rate of SLSNe light curves for values of the effective thermalization opacity $\kappa_{\rm th} \sim 0.01-0.1$ cm$^{2}$ g$^{-1}$ (e.g., \citealt{Nicholl+17d}).  Parameterized thus, one finds that at times $t \gg t_{\rm pk}$ ($f_{\rm th} \ll 1$),
\be
%\ell_{\rm th} \approx 10 (1+\tau_{\rm opt})M_{10}\, v_{9}^{-3} \, B_{14}^{-2} \left(\frac{\kappa_{\gamma}/\kappa_{\rm es}}{0.1}\right)^{-1}t_{\rm yr}^{-3}.
\lth \approx
% 2 \times 10^{-4}
2.2 \times 10^{-5}
(1+\tau_{\rm opt})M_{10}\, v_{9}^{-3} \, B_{14}^{-2} \left(\frac{\kappa_{\rm th}/\kappa_{\rm es}}{0.1}\right) t_{\rm yr}^{-5}.
\label{eq:lth}
\ee
One can also define a ``nonthermal'' compactness parameter $\ell_{\rm nth}$ of the nebula, which is analogous to $\lth$ but with $u_{\gamma}$ replaced by the density of nonthermal photons and $R_{\rm ej}$ with the nebula radius $R_{\rm n}$.

As we shall discuss in Section \ref{sec:absorption}, the optical depth across the nebula and ejecta to $\gamma\gamma$ pair production to photons near the threshold energy ($x\theta \approx 1$) can be roughly expressed as $\tauggth \sim \lth/(15\,\theta) \sim (10^{4}-10^{5})\lth$,
where $x \equiv h\nu/\me c^2$, $\theta \equiv kT_{\rm eff}/\me c^2 \sim 10^{-5}-10^{-6}$ and $T_{\rm eff} \sim 10^{4}-10^{5}$ K is the effective temperature of the target radiation field.  The ejecta thus becomes transparent ($\tauggth \lesssim 1$) to photons of energy $m_e c^{2}/\theta \sim 10^{11}-10^{12}$ eV on the timescale
\be
t_{\gamma\gamma} \approx 1.0\,M_{10}^{1/5}v_{9}^{-3/5}B_{14}^{-2/5} \left(\frac{\kappa_{\rm th}/\kappa_{\rm es}}{0.1}\right)^{1/5}\left(\frac{T_{\rm eff}}{10^{4}{\rm K}}\right)^{-1/5}\,{\rm yr},
\label{eq:tgg}
\ee
where at these late times we take $\tau_{\rm opt} < 1$.  Photons of $\sim$TeV energy are thus free to escape to a distant observer starting years following the explosion.

\begin{figure}
\includegraphics[width=0.5\textwidth]{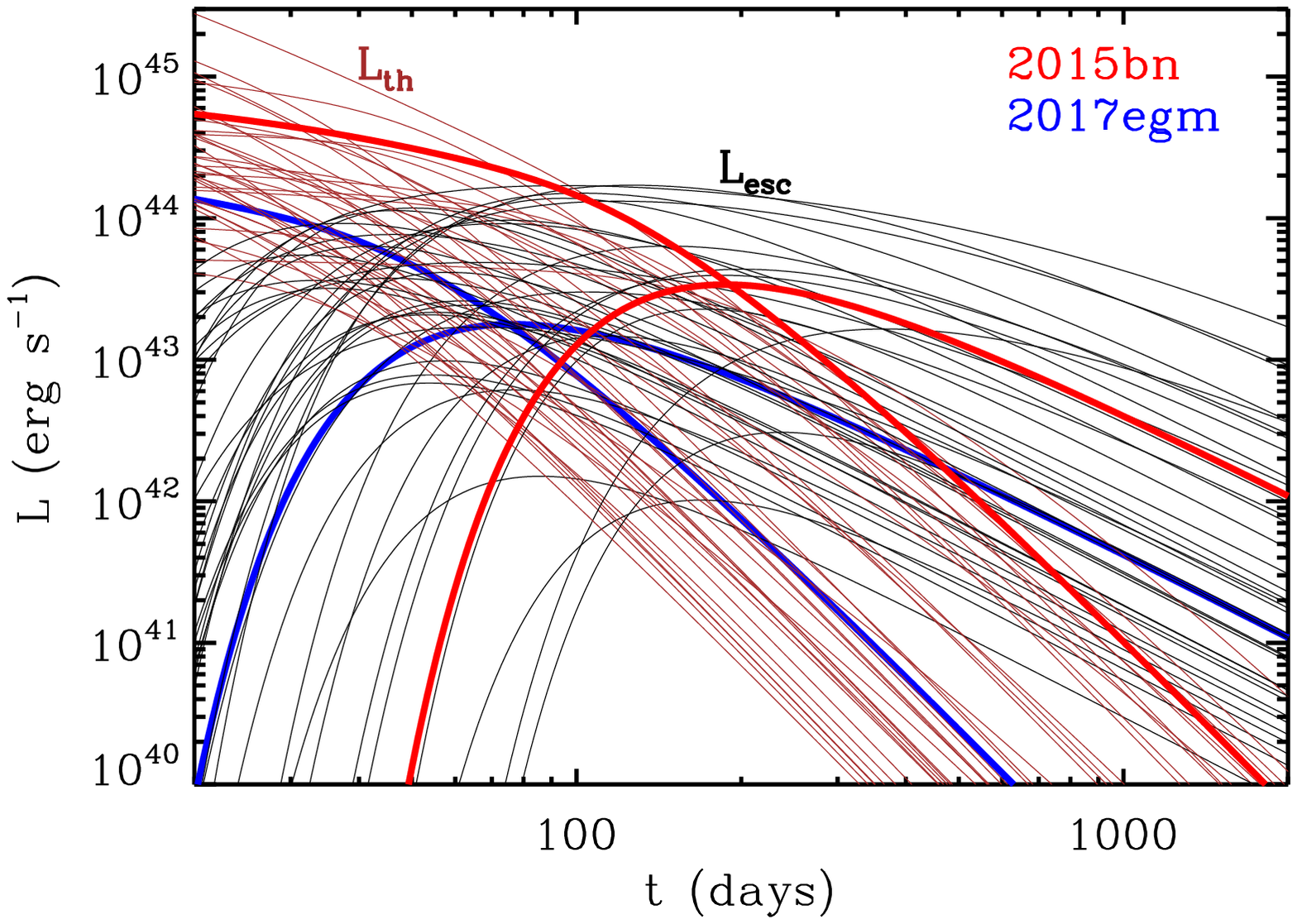}
\includegraphics[width=0.5\textwidth]{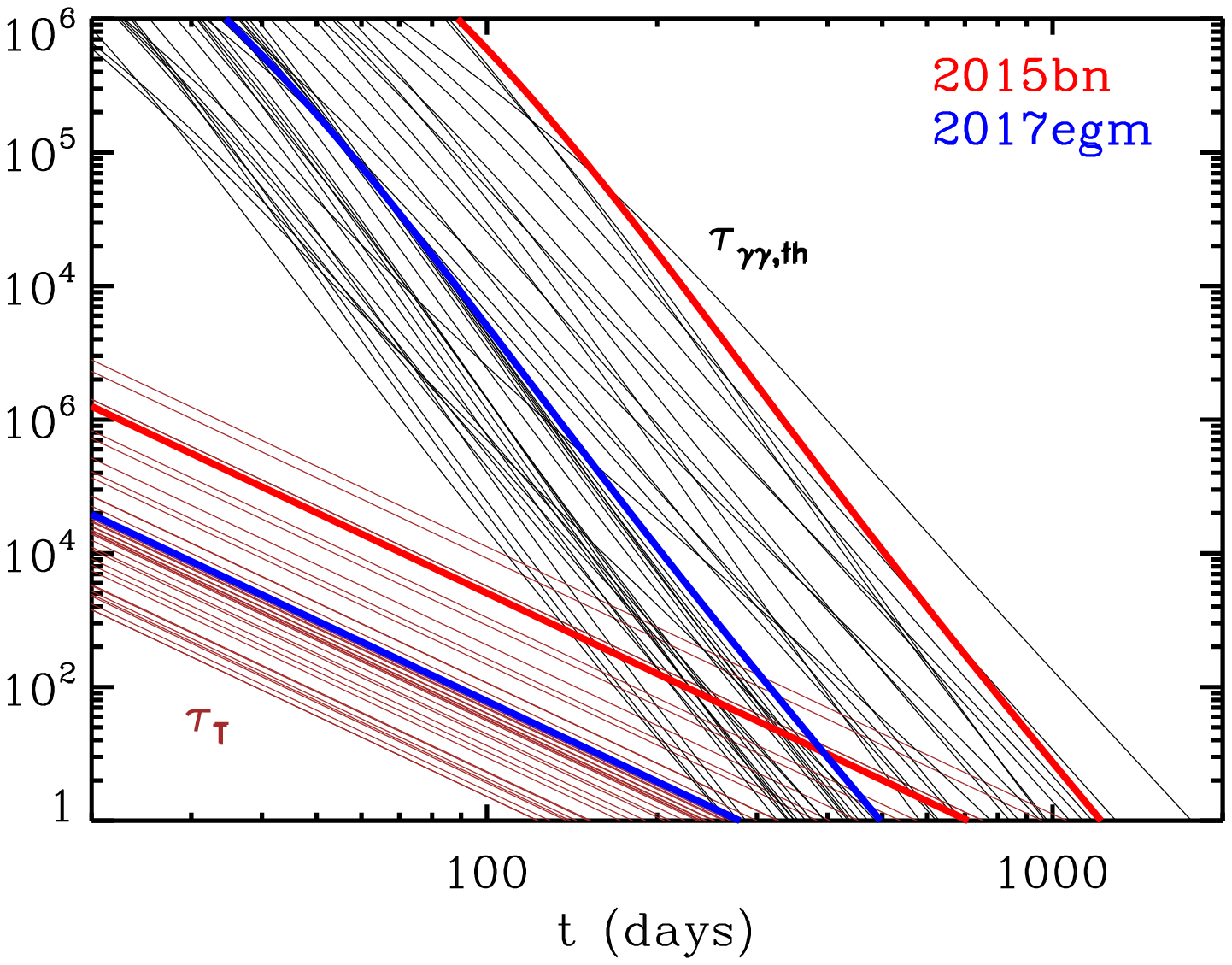}
%\vspace{-0.4cm}
\caption{{\it Top Panel:} Thermal luminosity $L_{\rm th} \equiv L_{\rm e}(1-e^{-\tau_{\rm therm}})$ (brown lines) and inferred ``escaping'' gamma-ray luminosity $L_{\rm esc} \equiv L_{\rm e}e^{-\tau_{\rm therm}}$ (black lines), calculated using ejecta and magnetars parameters inferred from Bayesian fits of optical light curve data to the magnetar model for a sample of 38 SLSNe from \citet{Nicholl+17}.  Here $\tau_{\rm therm}$ is the optical depth of ``thermalization'', which we have calculated assuming a fixed thermalization opacity $\kappa_{\rm th} = 0.01$ cm$^{2}$ g$^{-1}$ motivated by a phenomenological light curve model fit to SN2015bn.  The specific cases of SN2015bn and SN2017egm, whose properties motivate the numerical models presented in this paper (Table \ref{tab:models}), are shown as red and blue lines, respectively. {\it Bottom Panel:} For the same models in the top panel, the Thomson optical depth $\tau_{\rm T}$ (eq.~\ref{eq:tauT}; brown lines) and an estimate of the optical depth of $\sim$TeV photons to $\gamma\gamma$ pair production on the optical supernova light, $\tauggth$ (eq.~\ref{eq:taugg}; black lines).}
\label{fig:SLSNe}
\end{figure}

To connect the above discussion to observations, the top panel of Figure \ref{fig:SLSNe} shows the thermalized luminosity (essentially equal to the optical luminosity after peak), and the inferred escaping engine luminosity, from a sample of 38 SLSNe from \citet{Nicholl+17}, who fit observed multi-band optical light curve data to a modified version of the \citet{Kasen&Bildsten10} magnetar model within a Bayesian framework to determine properties of the engine ($E_{\rm e}$, $t_{\rm e}$ or equivalently $B_{\rm d}$, $P_0$) and the supernova ejecta ($M_{\rm ej}$, $v_{\rm ej}$, $\kappa_{\rm th}$).    %For comparison we show upper limits on the gamma-ray emission obtained by a stacking analysis of {\it Fermi}-LAT observations \citep{Renault-Tinacci+18}.  
The bottom panel of Figure \ref{fig:SLSNe} shows the time evolution of the Thomson optical depth, $\tau_{\rm T}$ (eq.~\ref{eq:tauT}), and the approximate $\gamma\gamma$ optical depth of a $\sim$TeV photon on the supernova optical radiation, $\tauggth$ (eq.~\ref{eq:taugg}).  

\begin{table*}
  \begin{center}
    \caption{Summary of Key Timescales}
    \label{tab:timescales}
    \begin{tabular}{c|c|c} 
      Symbol & Description of Event & Typical Value$^{\dagger}$\\
      \hline
\hline
$t_{\rm e}$ (eq.~\ref{eq:te}) & Engine lifetime (e.g., magnetar dipole spin-down time) & $\sim 0.01-0.1$ yr\\
$t_{\rm pk}$ (eq.~\ref{eq:tpk}) & Peak of the supernova optical light curve & $\sim 0.1$ yr\\
$t_{\rm B}$ (eq.~\ref{eq:tB}) & Synchrotron dominates IC cooling in nebula ($\ell_{\rm B} \approx 0.1\ell_{\rm th}$) & $\sim 0.1\,\varepsilon_{\rm B,-2}^{-1/2}$\,{\rm yr} \\
$t_{\rm T}$ (eq.~\ref{eq:tT}) & Ejecta transparent to Thomson scattering ($\tau_{\rm T} \approx 1$)  & $\sim 1$ yr \\
$t_{\gamma\gamma}$ (eq.~\ref{eq:tgg}) & Wind/nebula and ejecta transparent to $\gamma\gamma$ pair creation ($\tau_{\gamma\gamma,\rm th} \approx 1$) & $\sim 3$ yr  \\
$t_{\pm}$ (Table \ref{tab:regulation}) & Pair-loading of pulsar wind no longer regulated by in situ $\gamma\gamma$ interactions & $\sim 3-10$ yr (Table \ref{tab:regulation}) \\
$t_{\rm c}$ (eq.~\ref{eq:tcool}) & Nebula transitions from radiative to adiabatic evolution & $\sim (10-100)\,\varepsilon_{\rm B,-2}^{1/3}$\,{\rm yr}
    \end{tabular}
  \end{center}
$^{\dagger}$For typical values of the ejecta mass $M_{\rm ej} \sim 10M_{\odot}$ and mean velocity $v_{\rm ej} \sim 5000$ km s$^{-1}$.
\end{table*}

\begin{figure}
\includegraphics[width=0.5\textwidth]{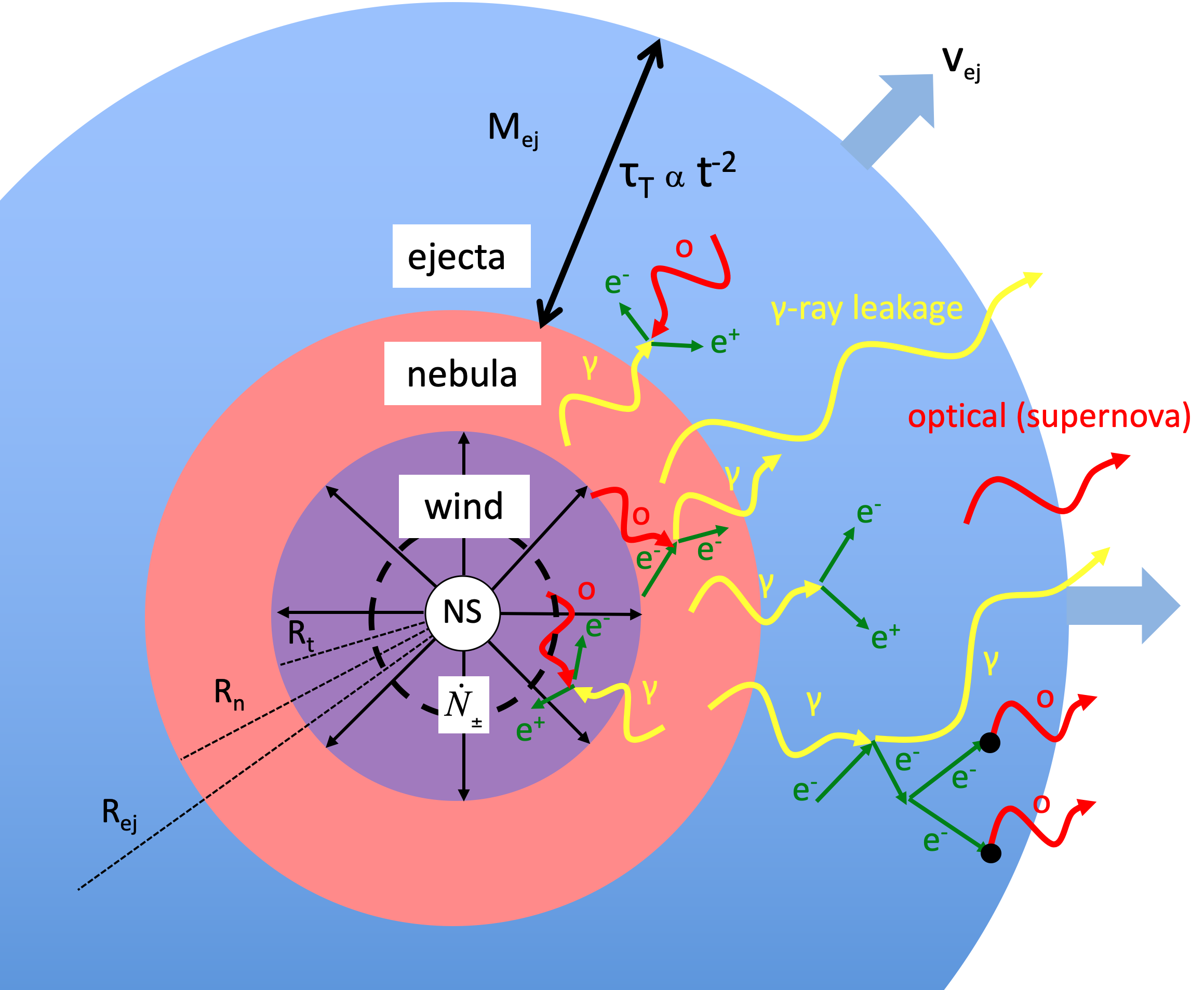}
%\vspace{-0.4cm}
\caption{Schematic illustration of the engine-powered supernova model discussed in this paper.  The pulsar wind of luminosity $L_{\rm e}$ terminates at a shock or region of reconnection (radius $R_{\rm t}$), inflating a nebula of relativistic pairs (radius $R_{\rm n} \simeq R_{\rm t}$) which radiate gamma-rays generated by IC scattering and synchrotron emission.  A fraction of the gamma-ray energy is absorbed and thermalized by the supernova ejecta (of mass $M_{\rm ej}$, mean velocity $v_{\rm ej}$, and radius $R_{\rm ej} \simeq v_{\rm ej} t$) and ultimately reprocessed into optical/UV light, while the remainder escapes directly to the outside observer.  The thermalized fraction of the spin-down luminosity decreases in time as the column through the ejecta shell (Thomson optical dept $\tau_{\rm T} \propto t^{-2}$) and the background thermal and nonthermal radiation fields (e.g. thermal compactness $\ell_{\rm th} \propto t^{-5}$) decrease.  Within the first few years of the explosion, $\gamma\gamma$ interactions between gamma-rays and optical/X-ray radiation load the pre-shock pulsar wind (at radii $\lesssim R_{\rm t}$) with electron/positrons pairs, regulating the flux $\dot{N}_{\pm}$ and mean energy $L_{\rm e}/\dot{N}_{\pm} m_e c^{2}$ of the pairs which enter the nebula and generate the gamma-rays.}
\label{fig:cartoon}
\end{figure}

\subsection{Nebular radiation}
\label{sec:nebula}

This section describes the sources of pairs and radiation in the nebula.  We again focus on the case of a millisecond magnetar, but many of our conclusions would apply to other central engines (e.g. an accretion-powered jet) provided that the mean energy per particle from their relativistic outflows is similarly high to a pulsar wind.  

The characteristic rate of particle injection from a rotationally powered pulsar wind is given by \citep{Goldreich&Julian69}
\begin{eqnarray} 
\dot{N}_{\pm} 
\equiv \mu_{\pm}\left(\frac{I}{e}\right) 
&\sim& 3\times 10^{41}\left(\frac{\mu_{\pm}}{10^{3}}\right)B_{\rm 14}P_{\rm ms}^{-2}\,{\rm s^{-1}}
\nonumber \\
&\underset{t \gg t_e}\approx&  1.3\times 10^{39} \left(\frac{\mu_{\pm}}{10^{3}}\right)B_{14}^{-1}t_{\rm yr}^{-1}\,\, {\rm
s^{-1}},
\label{eq:Ndot}
\end{eqnarray}
where $I \equiv \left. 4\pi c R_{\rm L}^{2}\eta_{\rm GJ}\right\vert_{\rm R_{\rm L}}$, $\eta_{\rm GJ} \approx \Omega B/ 2\pi c$ is the Goldreich-Julian charge density evaluated at the light cylinder radius $R_{\rm L} = P c/2\pi$, $P_{\rm ms} = P/1\,{\rm ms}$, and $\mu_{\pm}$ is the pair multiplicity of the wind (typically, $\mu_{\pm} \lesssim 10^{5}$; e.g.~\citealt{Timokhin&Harding19}; however, see \citealt{Beloborodov19}, who find a different expression for $\dot{N}_{\pm}$ in the case of active magnetars).  

The injected pairs inflate the nebula of radius $R_{\rm n} \approx v_{\rm n}t \lesssim R_{\rm ej}$, where the expansion velocity of the nebula is estimated as $v_{\rm n} \approx v_{\rm ej}/2$ for values of $E_{\rm e}$, $v_{\rm ej}$, $M_{\rm ej}$ characteristic of SLSNe (\citealt{Margalit+19}).  

We can estimate the strength of the magnetic field in the nebula $B_{\rm n}$ by assuming that the magnetic energy $(B_{\rm n}^{2}/8\pi)V_{\rm n}$, where $V_{\rm n} \approx 4\pi R_{\rm n}^{3}/3$ is the nebula volume, is a fraction $\varepsilon_{\rm B} = 0.01\varepsilon_{\rm B,-2}$ of the injected magnetar energy $\sim L_{\rm sd}t$ over the expansion time $\sim t$,
\be
B_{\rm n} \approx \left(\frac{6\varepsilon_{\rm B} L_{\rm e}t}{R_{\rm n}^{3}}\right)^{1/2} \approx 1.3\, \varepsilon_{\rm B,-2}^{1/2}B_{14}^{-1}v_{9}^{-3/2}t_{\rm yr}^{-2}\,{\rm G}.
\label{eq:Bn}
\ee
Magnetization values of $\varepsilon_B \sim 10^{-3}-0.1$ are motivated by modeling the emission of pulsar wind nebulae such as the Crab Nebula (e.g.~\citealt{Kennel&Coroniti84}); however, the physical conditions in the very young pulsar winds considered here may differ markedly from these older sources (Section \ref{sec:discussion}).

In analogy with Equation (\ref{eq:ell}), one can replace $u_{\gamma}$ with $u_{\rm B} = B_{\rm n}^{2}/8\pi$ and $R_{\rm ej}$ with $R_{\rm n}$ to define a ``magnetic compactness'' of the nebula according to
\be
\ell_B \equiv \frac{\sigma_{\rm T}}{m_{\rm e} c^2} \frac{B_{\rm n}^2}{8\pi} R_{\rm n} \approx
8.2 \times 10^{-4} \,\,\varepsilon_{\rm B,-2} B_{14}^{-2} v_{9}^{-2} t_{\rm yr}^{-3}.
\label{eq:lBn}
\ee
The ratio of the magnetic and thermal compactnesses (eq.~\ref{eq:ellopt}) is
\be
\frac{\ell_B}{\lth} \approx \frac{38}{(1 + \tau_{\rm opt})} \varepsilon_{\rm B,-2} M_{10}^{-1} v_{9} \left(\frac{\kappa_{\rm th}/\kappa_{\rm es}}{0.1}\right)^{-1} t_{\rm yr}^2.
\label{eq:lB_lth}
\ee
Once $\ell_B \gtrsim (0.1-1)\lth$, synchrotron emission will overtake IC scattering as the dominant cooling mechanism for relativistic pairs in the nebula, with implications for the photon energy spectrum of the latter.  This critical transition occurs on a timescale
\be
t_B \approx 0.09 \left(\frac{\ell_B}{\ell_{\rm th}}\right)^{1/2}\frac{M_{10}^{1/2}}{\varepsilon_{B,-2}^{1/2}v_{9}^{1/2}}\left(\frac{\kappa_{\rm th}/\kappa_{\rm es}}{0.1}\right)^{1/2}\,{\rm yr},
\label{eq:tB}
\ee
which is less than a month after the explosion for $\varepsilon_{\rm B} \sim 10^{-2}$.  For lower $\varepsilon_{\rm B} \ll 10^{-4}$ (which, as we shall find, may be needed to explain late-time SLSNe light curves; Section \ref{sec:latetime}) the $\ell_B \sim \lth$ transition occurs only after several years.

Pairs enter the nebula at the wind termination shock, where in normal pulsar wind nebulae they are heated.  This heating occurs either at the shock front itself \citep{Kennel&Coroniti84} or by magnetic reconnection in the striped pulsar wind ahead of the shock (\citealt{Sironi&Spitkovsky09,Sironi&Spitkovsky11,Zrake&Arons17}).  Naively, one would expect the freshly injected pairs to acquire a mean random Lorentz factor,
\be
%\Delta \gamma_{\rm sh}
\Delta \gamma = \gamma_{\rm in} \simeq \frac{L_{\rm e}}{\dot{N}_{\pm} m_e c^{2}} \approx
%4\times 10^{9}
3.1\times 10^{8}
\left(\frac{\mu_{\pm}}{10^{4}}\right)^{-1} B_{14}^{-1}t_{\rm yr}^{-1},
\label{eq:gammapm}
\ee
where Eq.~(\ref{eq:Ndot}) is used for the pair injection rate $\dot{N}_{\pm}$.

However, at early times $\Delta \gamma$ will generally be much lower than this estimate because of two effects.  Firstly, the upstream wind prior to the termination shock is loaded by secondary pairs generated by $\gamma\gamma$ processes, increasing the effective value of $\dot{N}_{\pm}$ sharing the pulsar's luminosity.  As we will show in Section \ref{sec:gamma}, at early times after the explosion this additional pair-loading regulates the mean Lorentz factor the pairs entering the nebula to a value $\gamma_0 \sim 10^{2}-10^{6} \ll \gamma_{\rm in}$ (see also Table \ref{tab:regulation}).  This enhanced pair-loading eventually ceases once the compactness of the nebula drops sufficiently, typically on a timescale of several years to a decade.  

%either (1) the upstream wind region becomes transparent to $\gamma\gamma$ interactions ($t \gtrsim t_{\gamma\gamma}$; eq.~\ref{eq:tgg}); or (2) synchrotron emission$-$which does not produce sufficiently energetic photons to pair produce on optical target photons$-$comes to dominate the nebula cooling ($t \gtrsim t_{\rm B}$; eq.~\ref{eq:tB}).  
%Also note that when $\gamma\gamma$ pair production dominates the wind loading, the pairs' energy comes directly from the primary gamma-ray and hence heating at the wind termination shock is less important (the pairs are deposited in the upstream wind already with a relativistic temperature).     

Another effect can in principle also limit the energy of pairs entering the nebula at late times, though it is not typically relevant for the choice of model parameters presented in this paper.  If the particle heating occurs in a region of the wind or termination shock where the electric field obeys $E < B$, then $\Delta \gamma$ is also limited by synchrotron cooling during the particle energization process, to a value (e.g.~\citealt{Cerutti+14})
\be
\gamma_{\rm rad} \approx \left(\frac{3e}{2B_{\rm w} r_{\rm e}^{2}}\right)^{1/2} \approx
9.7 \times 10^7 \left(\frac{R_{\rm t}}{R_{\rm n}}\right)^{1/2} \sigma^{-1/4}v_{9}^{1/2}B_{14}^{1/2}t_{\rm yr},
\label{eq:gammarad}
\ee
where $r_{\rm e} = 2.82\times 10^{-13}$ cm and
\begin{align}
B_{\rm w} = \left[\frac{2L_{\rm e}\sigma}{(\sigma+1) c R_{\rm t}^{2}} \right]^{1/2}
\label{eq:Bw}
\end{align}
is the magnetic field strength near the termination of the wind at radius $R_{\rm t}$, where $\sigma$ is the wind magnetization (the second line of eq.~\ref{eq:gammarad} assumes $\sigma \ll 1$).  
%Typically we have $\gamma_{\rm rad} \gtrsim \gamma_{\rm in},\gamma_0$ for the first few years after the explosion, in which case radiation will not limit the particle energies.

In summary, the mean Lorentz factor of pairs entering the nebula obeys
\be
\Delta \gamma(t) \simeq {\rm min}[\gamma_{\rm in},\gamma_{\rm rad},\gamma_{\rm 0}],
\label{eq:gammapm2}
\ee
where $\Delta \gamma \simeq \gamma_0 \sim 10^{2}-10^{6}$ at early times while the wind experiences significant $\gamma\gamma$ pair-loading, before increasing to $\Delta \gamma \sim \gamma_{\rm in}, \gamma_{\rm rad} \lesssim 10^{8}$ after several years.

%However, because the rate of IC scattering on the $\sim 1$ eV thermal photons by pairs of energy $\gamma_{\pm} \gtrsim 10^{6}$ is suppressed by Klein-Nishina effects, synchrotron cooling can dominate for even smaller values of $\ell_{\rm B}/\ell_{\rm th} \gtrsim 0.1$ at late times once $\Delta \gamma$ has risen to high values.
%(justifying our scaling of $t_{\rm B}$ in Eq.~\ref{eq:tB}).  

Upon entering the nebula with energy $\Delta \gamma$, the pairs cool via synchrotron radiation and IC scattering on lower energy target radiation.  In the Thomson regime, IC cooling will dominate synchrotron cooling for as long as the target photon energy density exceeds that of the magnetic field (e.g., $\ell_{\rm th} \gtrsim \ell_{\rm B}$ for a thermal target field).  At late times $t \gg t_{\rm B}$, synchroton dominates IC scattering as the dominant source of pair cooling.  At some point, however, even synchrotron radiation can no longer efficiently cool the nebula.  The synchrotron cooling time is shorter than the nebula expansion time for pairs above a Lorentz factor,
\be
\gamma_{\rm c} = \frac{6\pi m_e c}{\sigma_{\rm T}B_{\rm n}^{2}t} \approx 15 \varepsilon_{\rm B,-2}^{-1} B_{14}^{2} v_{9}^{3}
t_{\rm yr}^{3}.
\label{eq:gammacool}
\ee
Equating this to $\Delta \gamma$, we see that the nebula will remain fast-cooling until a time
\be
t_{\rm c} \approx 40 \left(\frac{\Delta \gamma}{10^{6}} \right)^{1/3}v_9^{-1}B_{14}^{-2/3}\varepsilon_{B,-2}^{1/3}\,{\rm yr}.
\label{eq:tcool}
\ee
At early times $t \ll t_{\rm c}$ the injected pairs radiate all the energy they receive instead of transferring it to the kinetic energy of the ejecta via adiabatic expansion.  The nebula's luminosity will therefore match that of the central engine for years to decades after the explosion, i.e. on the timescales of relevance to follow-up observations.\footnote{An exception occurs if the nebula magnetization is extremely low ($\varepsilon_{\rm B} \lesssim 10^{-6}$) in which case pairs cool exclusively through IC scattering.  Adiabatic losses then set in on a timescale as short as months: once the IC gamma-rays start to leak out from the nebula the resulting drop in thermalization reduces the target optical radiation background, reducing the IC cooling rate in a runaway process.}

A pair of Lorentz factor $\gamma_{\pm}$ can produce IC radiation up to the same scale as its own energy,
\be
h \nu_{\rm IC,max} \approx \gamma_{\pm} m_e c^{2} \sim 5\times 10^{11}\left(\frac{\gamma_{\pm}}{10^{6}}\right)\,{\rm eV}.
\label{eq:EIC}
\ee
By comparison, the characteristic frequency of synchrotron radiation from the same pair in the nebula magnetic field (eq.~\ref{eq:Bn}) is significantly lower, 
\be
h\nu_{\rm syn} =
{\rm min}[h\nu_{\rm syn,0}, h \nu_{\rm syn,max}],
\ee
where 
\be
h\nu_{\rm syn,0} =
h\nu_{\rm B} \gamma_{\pm}^{2} \approx
%\frac{h}{2\pi}\frac{eB_{\rm n}\gamma_{\pm}^{2}}{m_e c} \approx
%240 \, {\rm GeV}
15\left(\frac{\gamma_{\pm}}{10^{6}}\right)^{2}\varepsilon_{\rm B,-2}^{1/2}B_{14}^{-1}v_{9}^{-3/2}t_{\rm yr}^{-2}\,{\rm keV},
\label{eq:num}
\ee 
$\nu_{\rm B} = eB_{\rm n}/2\pi m_e c$ is the Larmor frequency,
and $h \nu_{\rm syn,max} \approx 160$ MeV is the maximum synchrotron frequency in the radiation reaction limited case (\citealt{Guilbert+83,Cerutti+14}).  For the regulated energies of the injected pairs $\Delta \gamma = \gamma_{\pm} \sim \gamma_0 \lesssim 10^{6}$ expected for several years after the explosion, we have $h\nu_{\rm syn} \lesssim 10$ keV.  Such low energy photons are readily thermalized by the ejecta (Section \ref{sec:thermalization}).  We shall find that the flux of escaping high-energy gamma-rays is thus greatly reduced once synchrotron competes with IC cooling (at $t \gtrsim t_{\rm B}$; eq.~\ref{eq:tB}).

\subsection{Absorption and energy loss of high-energy photons}
\label{sec:absorption}

Energetic photons deposited into the nebula interact with both matter and radiation fields as they diffuse outward through the nebula and ejecta. The relevant processes depend on the photon energy and involve both scattering and absorption, which can lead to the generation of secondary electron-positron pairs and their radiation.  The net effect of these processes is to reprocess the primary photon energy towards lower frequencies.  A fraction of this energy eventually reaches the thermal pool and emerges as optical radiation; quantifying this thermalization efficiency or optical depth self-consistently is one of the goals of our Monte Carlo simulations. This section summarizes the main radiative mechanisms involved.

The main source of opacity for GeV$-$TeV photons is pair production on soft radiation fields and nuclei in the ejecta (Bethe-Heitler process).  The optical depth through the ejecta of the former is given by \citep{Zdziarski_89}
\be
%\tau_{\rm ph-mat}
% \tauphmat = \frac{20}{8\pi}\alpha_{\rm fs}\left[ \ln(2x) - \frac{109}{42}\right]\tau_{\rm T}, % this is with cosmological composition
\tauphmat = \frac{21}{2\pi}\alpha_{\rm fs}\left[ \ln(2x) - \frac{109}{42}\right]\tau_{\rm T},
\ee
where $\alpha_{\rm fs} \simeq 1/137$, $x \equiv h\nu/m_{\rm e} c^2 \gg 1$, and we have assumed oxygen-dominated composition.

Depending on energy, $\gamma\gamma$ pair production can take place on either thermal (e.g., optical/UV) or nonthermal (e.g., X-ray) radiation fields.  For a broad target spectrum of radiation energy density $u_{\gamma}$, a simple approximation for the $\gamma\gamma$ pair production opacity is given by \citep{Svensson_87}:
\be
%\tau_{\rm ph-ph} \approx \frac{x}{6} \, l_{\nu} \left(\frac{2}{x}\right).
\tau_{\gamma\gamma, {\rm nth}} \approx \eta(\alpha) x \, \ell_{\nu} \left(\frac{1}{x}\right),
\label{eq:tauggnth}
\ee
where $\eta(\alpha) \equiv (7/6) (2-\alpha)^{-1}(1-\alpha)^{-5/3}$ and $\alpha$ is defined so that $du_{\gamma}/d\ln{x} \propto x^{\alpha+1}$.  We have also defined the differential compactness of the target radiation field as
\be
\ell_{\nu} = \frac{\sigma_{\rm T}}{m_{\rm e} c^2} \frac{{\rm d}u_{\gamma}}{{\rm d}\ln x}  R_{\rm ej},
\label{eq:ellnu}
\ee
which is a more precise, frequency-dependent version of the compactness introduced in eq.~(\ref{eq:ell}).
%where ${\rm d}u_{\gamma}/{\rm d}\ln x$ is the radiation energy density per logarithmic photon energy interval.
The above approximation for $\tau_{\gamma\gamma}$ relies on the availability of target photons near the threshold energy $1/x$ and is accurate to within $0.3\%$ when $-6<\alpha<0$.
%also, it is  only valid as long as the target spectrum is softer than $F_{\nu}\propto \nu$.

For a thermal target radiation of temperature $\theta \equiv kT_{\rm eff}/\me c^2$ and compactness $\lth$ (eq.~\ref{eq:ellopt})
a simple empirical formula for the pair-production opacity can be written as
\be
\tauggth
= \frac{\lth}{\overline{x}} \frac{\ln(1 + x\theta)}{2x\theta} \exp\left(-\frac{0.9}{x\theta}\right),
\label{eq:taugg}
\ee
which has an error of $<15$\% at $x\theta \gtrsim 0.1$.
Here $\overline{x} \approx 2.7\theta$ is the average thermal photon energy in $\me c^2$ units.

Photons can also lose energy by Compton downscattering off electrons in the ejecta. An effective optical depth for a photon to lose most of its initial energy can be defined as
\begin{align}
\tauCeff
&\equiv \frac{t_{\rm esc}}{t_{\rm C}}
\approx \frac{\dot{x}_{\rm C}}{x} \frac{R_{\rm ej}}{c} \left( 1 + \tauKN \right) \nonumber \\
&\approx
%\tauT \frac{\ln (1 + x)}{1 + 3x} \left[ 1 +  \frac{\ln(1+3x)}{3x} \tauT \right],
\tauT \frac{\ln (1 + x)}{1 + 3x} \left[ 1 +  \frac{3\tauT}{8x} \ln\left(1+\frac{8x}{3} \right) \right],
\label{eq:tauCeff}
\end{align}
where the term $(1 + \tauKN)$ accounts for the enhanced residence (diffusion) time of the photon in the opaque medium.  Here we have approximated the particle energy loss rate $\dot{x}_{\rm C}/x \approx c\sigma_{\rm T} n_{\rm e} \ln(1+x)/(1+3x)$
which is accurate to within $<15\%$ at $x \gg kT/m_{\rm e} c^2$ (where $T \sim 10^{4}-10^{5}$ K is the gas temperature), and
$\tau_{\rm KN}\approx 3\tauT\ln(1 + 8x/3)/8x$ (accurate to within 10$-$15\% and asymptotically approaching the exact $\tau_{\rm KN}$ at $x\ll 1$ and $x \gg 1$). 
%$\tau_{\rm KN}\approx \tauT\ln(1 + 3x)/3x$ (accurate to within 10$-$15\%). 
%The condition $\tauCeff = 1$ determines whether a photon of energy $x$ escapes the medium without losing most of its energy.
% However, if $\tauCeff > 1$, a more detailed calculation is required for determining the escaping energy fraction; simply using $\exp(-\tauCeff)$ would give an erroneous result.

Finally, photons of the lowest energies are absorbed by the photoelectric (bound-free) process, which for oxygen-dominated composition of hydrogen-poor supernovae can be approximated at $E \gtrsim 0.5$~keV by \citep{Verner+96}
\be
\tau_{\rm bf} \approx 2.5\times 10^4 \, E_{\rm keV}^{-3} \, \tauT ,
\ee
where $E_{\rm keV} \equiv h\nu/1\,{\rm keV}$.  Given that $\tau_{\rm T} \gtrsim 1$ for roughly a year after the explosion (eq.~\ref{eq:tT}; Fig.~\ref{fig:SLSNe}), photons with energies $\lesssim$ 10 keV are absorbed for several years after the explosion (\citealt{Margalit+18}).
%and assume X, Y, Z abundance.

\begin{figure}
\includegraphics[width=0.5\textwidth]{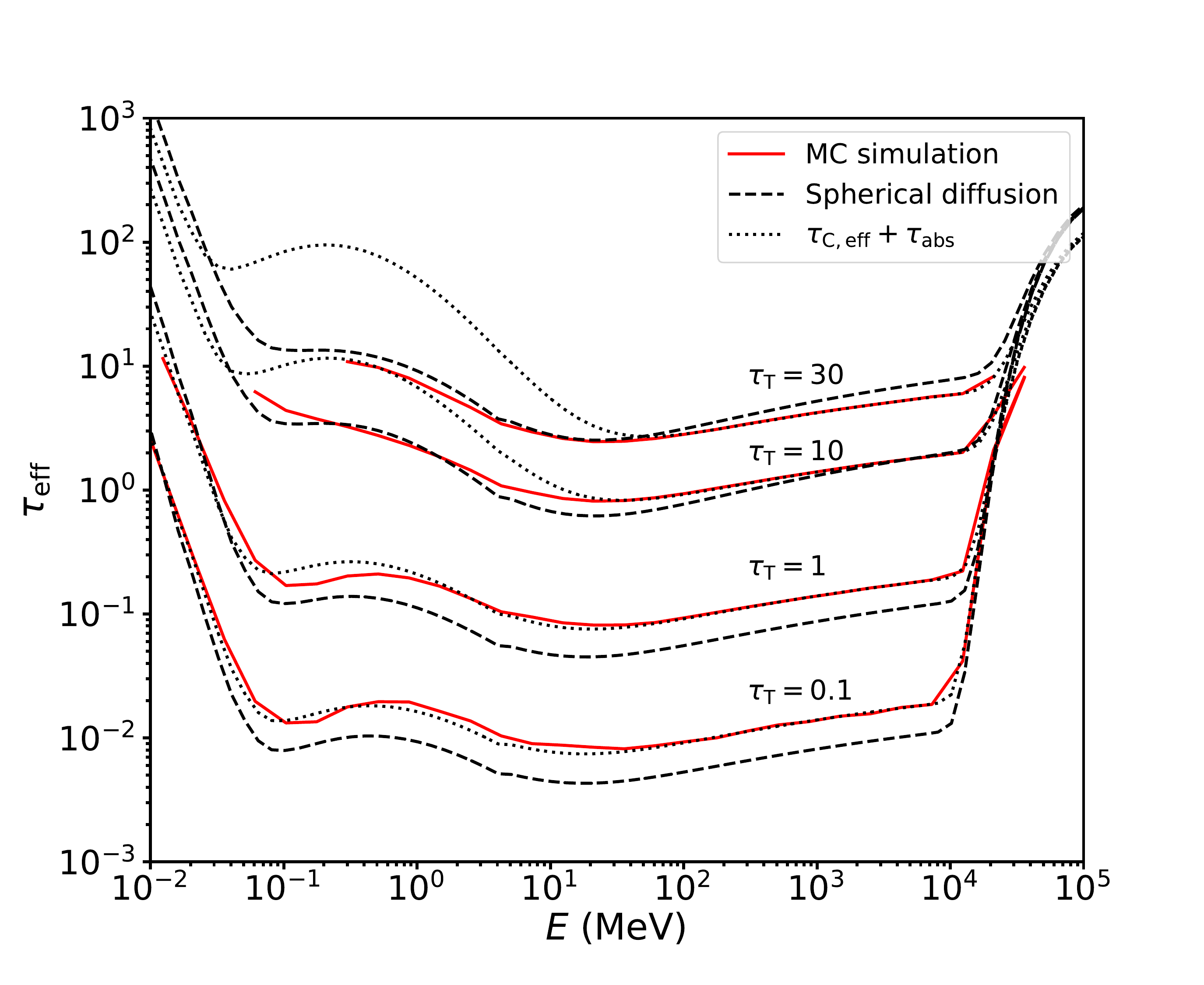}
%\vspace{-0.4cm}
\caption{  Effective optical depth, $\tau_{\rm eff}$, as a function of initial photon energy for diffusion out of a spherical cloud, defined such that $\exp{(-\tau_{\rm eff})}$ is the fraction of energy that escapes without significant downgrading (i.e. at photon energies $E_{\rm out} > E/2$).  Results are shown for different values of the radial Thomson optical depth (from top to bottom): $\tauT = 30, 10, 1, 0.1$.  Red solid lines show the results of full Monte Carlo simulations; dashed lines correspond to an analytical solution to the radiative diffusion equation (Appendix \ref{sec:app:diffusion}); and dotted lines show $\tauCeff + \tau_{\rm abs}$, where $\tau_{\rm abs} = \tauphmat + \tauggth + \tau_{\rm bf}$ is the sum of absorption optical depths and $\tauCeff$ is an estimate of the effective optical depth due to Compton down-scattering (eq.~\ref{eq:tauCeff}).  The thermal compactness in the Monte Carlo calculations is taken to be $\lth = 10^{-2}$, however our results only depend on this assumption at the highest photon energies $E \gtrsim 10^{10}$ eV (which undergo $\gamma\gamma$ interactions on the thermal radiation field).}
\label{fig:taueff}
\end{figure}

Figure \ref{fig:taueff} shows the effective optical depth as a function of the primary photon energy for different Thomson columns, $\tau_{\rm T}$.  From low to high photon energies, the dominant processes responsible for photon energy loss are: bound-free absorption ($E \lesssim 30$~keV); Compton down-scattering ($E \lesssim 100$~MeV); Bethe-Heitler photon-matter pair production ($E \lesssim 10$~GeV);
pair creation on nonthermal radiation (if present); pair creation on thermal supernova radiation ($E \gtrsim 10-100$~GeV).  

Dashed and dotted lines in Figure~\ref{fig:taueff} show different approximate analytical results, which may be compared with exact results obtained from Monte Carlo simulations of photon diffusion (solid lines).  In the Compton-dominated range ($0.1 \lesssim E \lesssim 100$ MeV), the simple analytic expression for $\tauCeff$ (eq.~\ref{eq:tauCeff}) can significantly overestimate the true effective optical depth when $\tauT > 1$.  A more detailed analytical solution to the radiative diffusion equation (described in Appendix \ref{sec:app:diffusion}) does a better job of reproducing the numerical results in the Thomson thick case. 

\begin{figure*}
\includegraphics[width=0.5\textwidth]{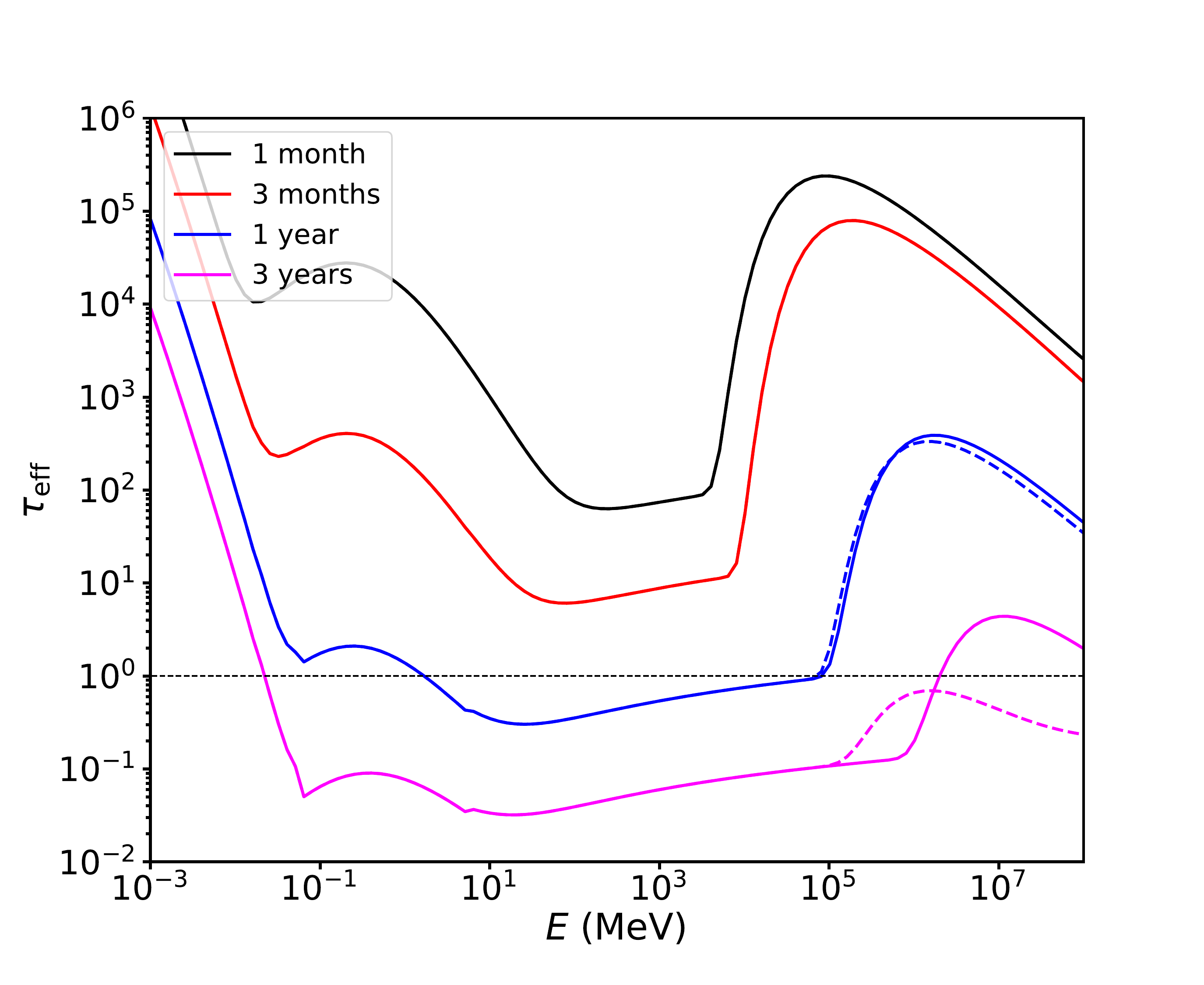}
\includegraphics[width=0.5\textwidth]{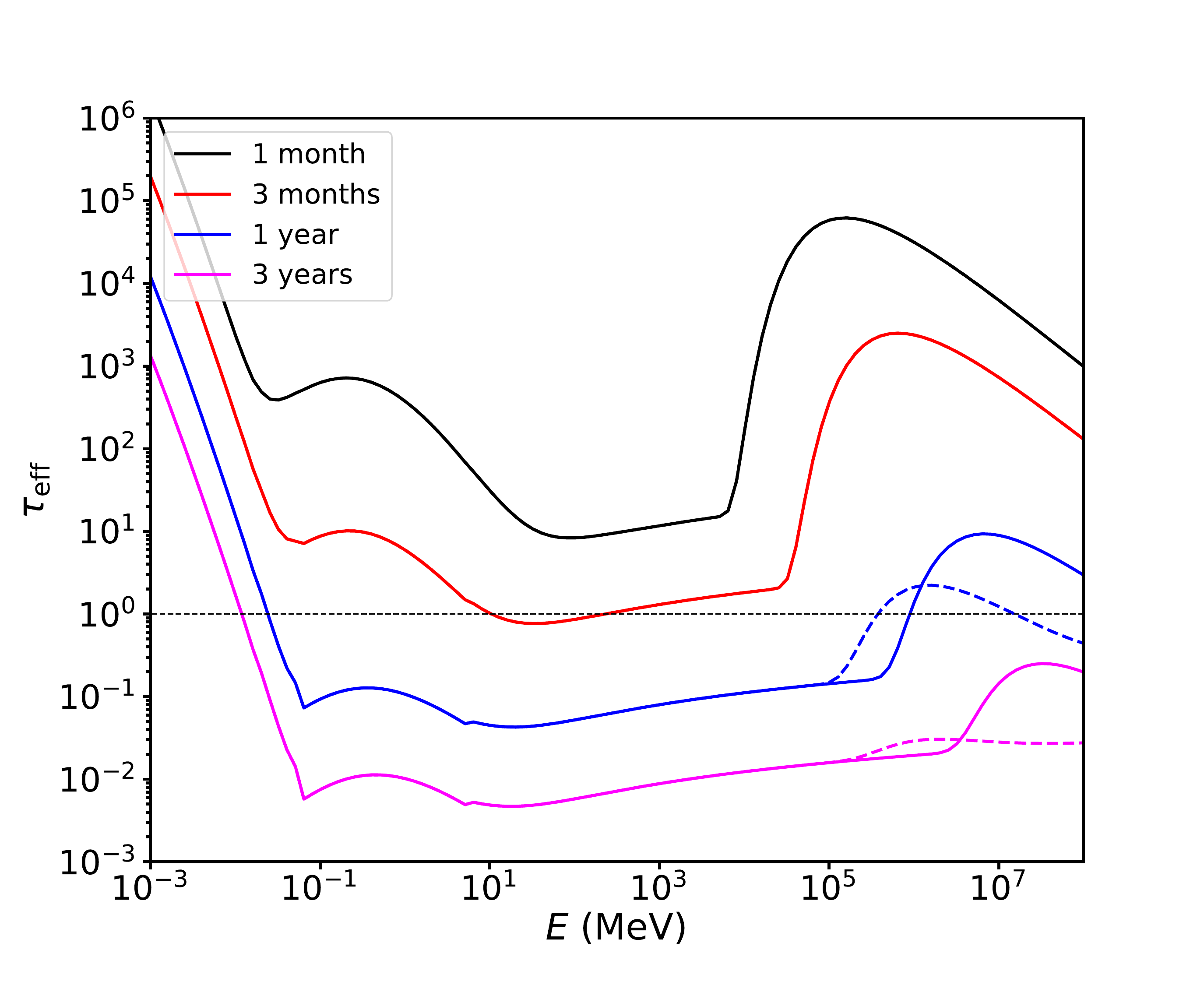}
%\subcaptionbox{SN2015bn}
{\includegraphics[width=0.5\textwidth]{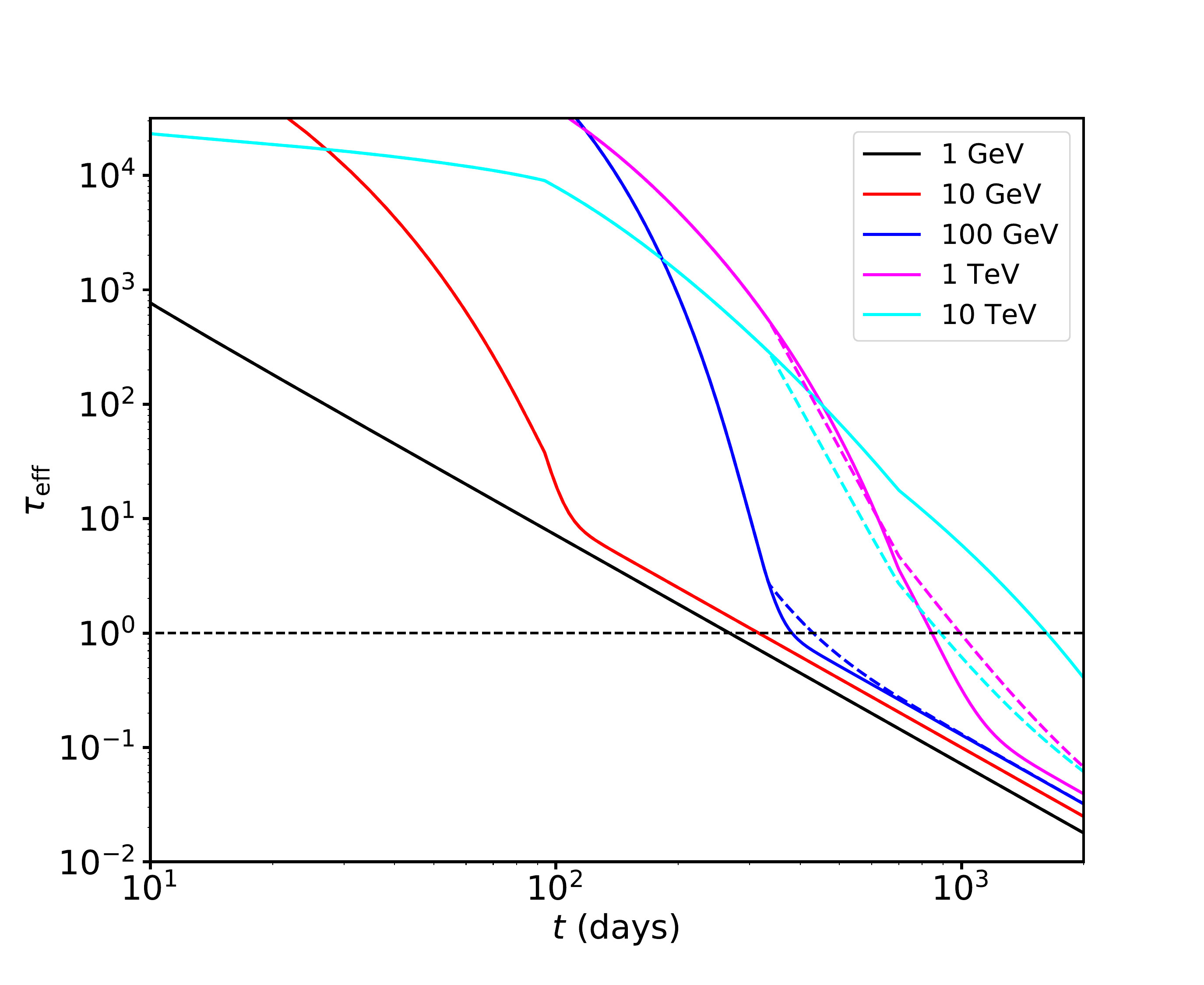}}
%\subcaptionbox{SN2017egm}
{\includegraphics[width=0.5\textwidth]{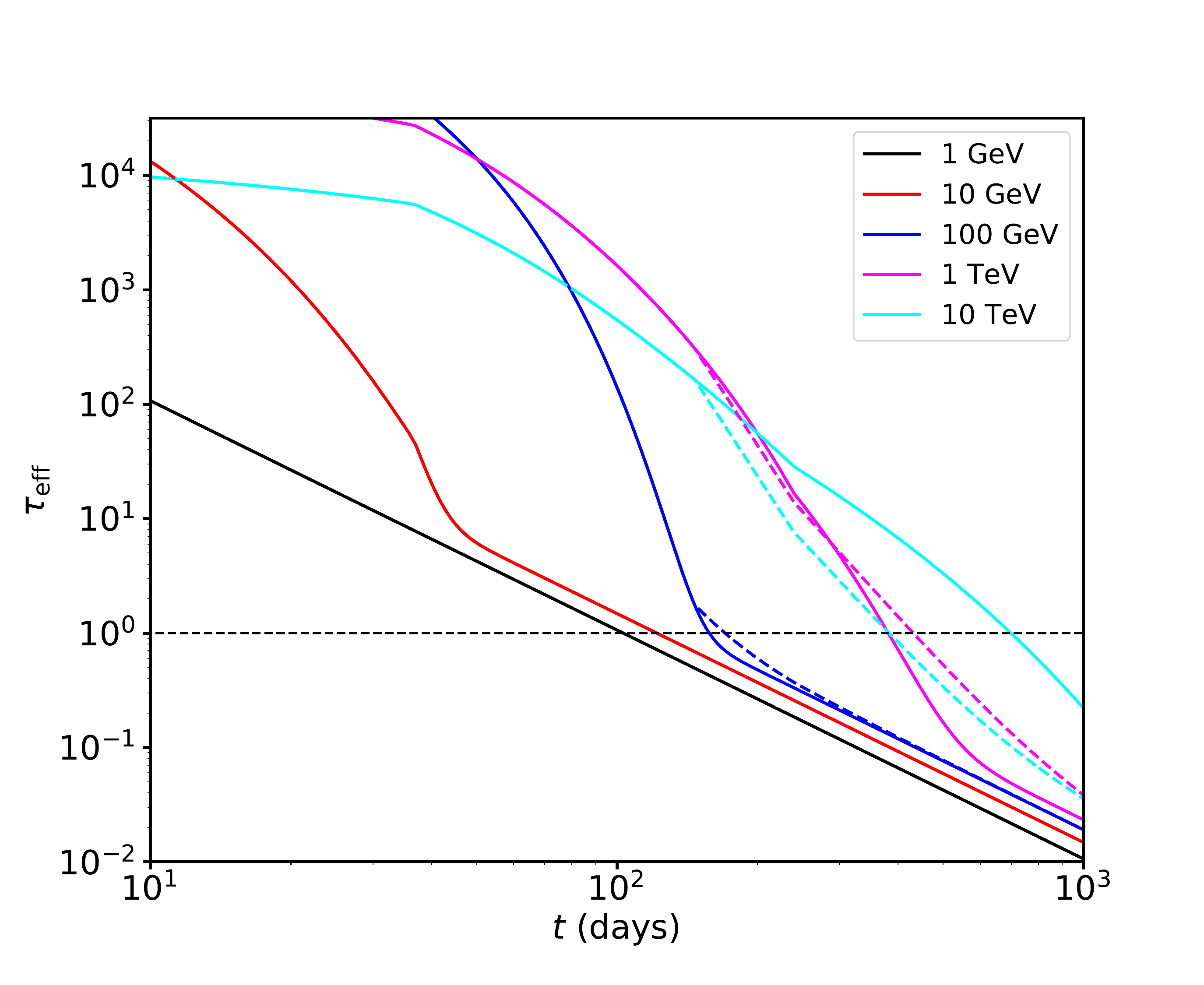}}

\caption{ {\it Top Panels:}  Effective optical depth through the supernova ejecta, $\tau_{\rm eff}$, as a function of photon energy shown at different snapshots in time after the explosion as marked.  The calculations are performed for magnetar and ejecta parameters fit to the optical light curves of SN2015bn ({\it Left}) and SN2017egm ({\it Right}).  We explore the dependence of the results on two approximations for the optical radiation field at late times, which serve as targets for $\gamma\gamma$ pair production.  The cases shown as solid lines assume blackbody emission at the equilibrium temperature $T_{\rm eff} = (L_{\rm opt}/4\pi \sigma R_{\rm ej}^{2})^{1/4}$,
where $L_{\rm opt}$ is the optical/UV luminosity.  The cases shown as dashed lines were instead calculated assuming a floor on the blackbody temperature at $T_{\rm eff} = 4000$ K, as a crude proxy for the supernova nebular spectrum comprised of optical/NIR emission lines.  We neglect nonthermal X-ray target radiation in calculating the $\gamma\gamma$ optical depth, as is generally a good approximation at late times when $\tau_{\gamma\gamma} \lesssim 1$.
{\it Bottom Panel:} Effective optical depth, now as a function of time after explosion for different photon energies as marked.   
}
\label{fig:taueff1}
\end{figure*}

%In order to illustrate these processes, Figures \ref{fig:taueff1}-\ref{fig:taueff3} show
The top panels of Figure \ref{fig:taueff1} shows the effective optical depth $\tau_{\rm eff}$ as a function of photon energy at several snapshots in time after the explosion, for ejecta and magnetar parameters fit to the optical light curve data on SN2015bn and SN2017egm (see Table \ref{tab:models}).  The lowest optical depth occurs in an energy window surrounding $\sim 1$ GeV, which broadens from the MeV to the TeV range as the ejecta expands and the radiation field dilutes over the course of several years.  The bottom panel of Figure \ref{fig:taueff1} shows $\tau_{\rm eff}$ now as a function of time since explosion, for different photon energies.  The ejecta becomes transparent ($\tau_{\rm eff} < 1$) at energies 1$-$10 GeV accessible to {\it Fermi}-LAT by $t \sim 100$ days, while transparency is delayed for years at $\sim$TeV energies because of $\gamma\gamma$ absorption by the supernova optical light.

%\includegraphics[width=0.45\textwidth]{Opacity_SN2015bn_with_nth_pprod.eps}
%%\vspace{-0.4cm}
%\caption{Same as Figure \ref{fig:taueff1} but including a contribution to the opacity from nonthermal radiation. %{\bf BDM: what are the different lines?}}
%\label{fig:taueff2}
%\end{figure}

\subsection{Gamma-ray thermalization}
\label{sec:thermalization}

The efficiency of reprocessing and thermalization of the nebula energy by the supernova ejecta is crucial to explaining the optical light curves of engine-powered supernovae.  However, the mechanism of reprocessing is complex and non-linear.  Even if gamma-rays are attenuated during their escape from the ejecta (Section \ref{sec:absorption}), this does not guarantee that their energy will be thermalized.  In particular, the result of $\gamma\gamma$ and Bethe-Heitler photon-matter processes (which dominate the attenuation of high-energy gamma-rays) is the production of an energetic electron/positron pair.  As shown in Appendix \ref{sec:app:cooling}, this pair will generally experience rapid radiative cooling, producing a second generation of radiation which itself may or may not escape the ejecta (potentially producing yet another generation of pairs, and so on).

Energy is deposited into the thermal radiation field via three main channels: (1) Compton downscattering of photons on cold electrons,
(2) Coulomb collisions whereby nonthermal pairs lose their energy to cold plasma, and (3) photoionization.  

Below a few tens of keV, the dominant opacity source is photoionization.  At these energies the photoelectrons are strongly coupled to thermal electrons; furthermore, the photoionization cross-section generally increases as the radiation energy is reprocessed towards lower frequencies by repeated photoionizations and recombinations.  For these reasons it is reasonable to assume that all energy lost to photoionization is efficiently thermalized, i.e. occurs faster than any other timescale of interest.

Between tens of keV and $\sim 100$~MeV most of the photon energy loss is due to Compton downscattering.  Above $\me c^2 \sim 1$ MeV the photon can lose a significant fraction of its initial energy in a single scattering event; energy conservation then implies that the energy of the upscattered electron can reach a similar fraction of the photon energy.  Following Equations (\ref{eq:t_brem})-(\ref{eq:gdagger}) in Appendix \ref{sec:app:cooling}, the $\sim$~MeV electron subsequently loses its energy either via IC scattering or Coulomb interactions with thermal electrons, both which typically occur faster than the ejecta expansion time over which the pairs would experience adiabatic losses (Fig.~\ref{fig:tcool}).  In the former case the photon is upscattered from the optical/UV to $\sim$~keV energies where it is thermalized by phoionization, while in the latter case the $\sim$MeV electron directly shares its energy with the thermal electron pool.  

Photon-matter pair production begins to dominate the photon energy loss above a few tens of MeV. However, only a fraction of the energy used for pair creation is eventually thermalized. The created secondary leptons have, on average, half of the energy of the primary photon.  They cool mostly by IC and free-free emission, as well as by Coulomb collisions which become increasingly dominant as the lepton energy decreases (eqs.~\ref{eq:coolingratesratio}, \ref{eq:coolingratesratio2}).  The thermalization efficiency of the lepton's energy depends on (1) the fraction\footnote{Roughly given by the ratio of the lepton Lorentz factor below which Coulomb losses dominate, $\gamma_{\star\star}$ (eq.~\ref{eq:gammastarstar}), and the Lorentz factor at which the lepton was created.} of its energy lost to Coulomb collisions; (2) the shape and characteristic frequency of the pair's IC and free-free radiation.  The energy emitted in $\lesssim 10$~keV photons is photoelectrically absorbed and thermalized, while higher energy radiation either escapes, is lost by direct Compton (and thermalized), and/or generates a subsequent generation of electron-positron pairs.

The issue of thermalization is directly addressed by our Monte Carlo simulations in Section \ref{sec:model}, which follow the diffusion of photons out of the spherical ejecta cloud and self-consistently follow the electron-positron pair cascade initiated by the high-energy radiation.  To briefly preface our results here (see Fig.~\ref{fig:Lth} for details), for magnetized nebulae ($\varepsilon_{\rm B} > 0$) Compton and photoelectric thermalization are found to contribute roughly equally to the ejecta heating for the first few months after explosion while the Thomson optical depth is high.  However, photoelectric absorption comes to dominate the ejecta heating at late times, once synchrotron emission contributes a greater fraction of the nebula cooling.  Coulomb heating also contributes appreciably during the transition period between the early Compton- and late photoelectric-dominated epochs.  

\subsection{Pair-loading of the pulsar wind and regulation of the mean energy of pairs entering the nebula}
\label{sec:gamma}

\begin{table*}
  \begin{center}
    \caption{Regimes of $\gamma\gamma$ pair creation regulation of the pulsar wind, in approximate temporal order of dominance (see Appendix \ref{app:regulation} for details)}
    \label{tab:regulation}
    \begin{tabular}{c|c|c|c} 
      High-energy $\gamma$-ray & Target photon & Regulated pair energy, $\gamma_{0}$$^{\dagger}$ & End time of regulation$^{\dagger}$ \\
      \hline
      \hline 
      IC & IC & $\sim 1/(2\sqrt{\theta}) \sim 10^{2}-10^{3}$ &  $t_{\pm,\rm IC} \approx 0.3$ yr (eq.~\ref{eq:tpmnth}) \\
      IC & Synchrotron & $\sim (\me c^2/4\theta h \nu_{\rm B})^{1/4} \sim (10^{4}-10^{5}) \, t_{\rm yr}^{1/2}$  & $t_{\pm,\rm IC-syn} \approx 1$ yr (eq.~\ref{eq:tppICsyn}) \\
      IC & Thermal & $\sim 1/\theta \sim 10^{5}-10^{6}$  & $t_{\pm, \rm th} =$ min[$t_{\gamma\gamma}$, $t_{\rm B}$] $\sim 0.1-3$ yr \\
%      Synchrotron & Synchrotron & $\sim 10/\ell_{\rm inj} \sim 10^{4}t_{\rm yr}^{3}$ & $t_{\pm, \rm syn}\sim 3-10$ yr (eq.~\ref{eq:tppsyn}) \\
      Synchrotron & Synchrotron & $\sim (\me c^2/h \nu_{\rm B})^{1/2} \sim (10^{6}-10^{7}) \, t_{\rm yr}$ & $t_{\pm, \rm syn}\sim 3-10$ yr (eq.~\ref{eq:tppsyn}) \\
    \end{tabular}
  \end{center}
  $\dagger$Numerical estimates assume fiducial parameters: $B_{\rm d} \sim 10^{14}$ G, $v_{\rm ej} \sim 5000$ km s$^{-1}$, $T_{\rm eff} \sim 10^{4}-10^{5}$ K ($\theta \sim 10^{-5}-10^{-6}$).
\end{table*}

%In addition to ``primary'' pairs generated close to the neutron star surface (eq.~\ref{eq:Ndot}), upstream of the termination shock the pulsar wind is loaded with secondary pairs due to $\gamma\gamma$ interactions by high-energy photons from the nebula.  For the first few years after the explosion the compactness of the wind region is sufficiently high that these pairs greatly outnumber the primaries and hence their production regulates the mean energy of the pairs entering the nebula.

Consider a simple model in which the pulsar luminosity is dissipated in a localized radial zone, near or just prior to the termination shock (where the radial momentum flux of the wind matches the nebula pressure) and transferred to the pairs entering the nebula from the upstream wind region (Fig.~\ref{fig:cartoon}; \citealt{Kennel&Coroniti84,Sironi&Spitkovsky11}).  The mean energy gain per particle,
\be \Delta\gamma = \frac{L_{\rm e}}{\dot{N}_{\pm} m_e c^{2}},
\label{eq:deltagamma}
\ee
depends on the ratio of the wind luminosity, $L_{\rm e}$, and the number of pairs carried into the dissipation radius, $\dot{N}_{\pm}$.  The latter is the sum of pairs injected directly from the engine on small scales (eq.~\ref{eq:Ndot}) and those generated {\it in situ} via $\gamma\gamma$ pair creation (eq.~\ref{eq:gammapm}).  The high-energy gamma-rays which create pairs in the wind can originate both directly from the nebula (radiation from freshly heated pairs) as well as from the supernova ejecta (inwards scattering/diffusion following incomplete reprocessing of gamma-rays from the nebula).  This creates a feedback loop between $\Delta\gamma$ and the number of generated pairs in the wind: as the energy gain per particle increases, the resultant higher photon energies emitted by the nebula increase pair production in the wind, which in turn reduces the per-particle energy once those pairs reach the dissipation region.

In a (quasi-)steady state the number ${\cal M}_{\rm e}$ of secondary pairs produced in the wind region per single lepton heated at the termination shock must be unity, i.e.
\be
{\cal M}_{\rm e} \approx 1. \,\,\,\,\,  
\label{eq:regulation}
\ee
If this was not the case (say, each heated lepton were to instead generate more the one secondary), then $\dot{N}_{\pm}$ would increase and $\Delta\gamma$ would correspondingly decrease.  Typically, both the pair-production opacity and the overall ``gain'' factor of the cycle decrease with decreasing $\Delta\gamma$, hence the system would readjust itself to attain a gain factor $\sim 1$.

The source of the high-energy gamma-rays and target radiation on which they produce pairs in the wind can be either the optical/UV thermal background or higher-energy nonthermal X-rays.  The different regimes are described in detail in Appendix \ref{app:regulation} and their key properties summarized in Table \ref{tab:regulation}, in rough time order in which they dominate the wind pair-loading.  We also provide an estimate of the timescale, $t_{\pm}$, after which each pair-loading channel ``breaks'' and $\Delta \gamma$ increases to the next relevant channel.   

At early stages, around and just after optical peak, the value of $\dot{N}_{\pm}$ due to $\gamma\gamma$ pair loading of the wind is so high that leptons entering the nebula do not generate sufficiently energetic photons to interact with the optical/UV radiation field; the generation of secondary pairs at this stage is dominated by interactions of IC gamma rays with
%higher-energy
other nonthermal photons (typically, X-rays and soft gamma rays also of IC origin).  This phase generally includes the epochs at which the ejecta becomes optically thin to photon-matter pair production (Fig.~\ref{fig:taueff}), enabling gamma-ray emission in the {\it Fermi} LAT range to escape.  For sufficiently high values of the nebula magnetization, the interaction of the IC-generated photons with synchrotron targets can also dominate the wind pair loading for times up to about a year.

As the nebula expands and the compactness decreases, the value of $\Delta\gamma$ increases until the cooling pairs radiate photons of sufficient energy to also pair produce on the $\sim 1-10$ eV thermal optical photons.  We explore this case here in some detail here, as it typically dominates the wind pair-loading on timescales of years after the explosion, when the system is becoming transparent at the highest gamma-ray energies.  A more detailed exploration of the results described below is provided in Appendix \ref{app:regulation}.

Figure \ref{fig:pairmult} shows the numerically calculated multiplicity ${\cal M}_{\rm e}$ due to $\gamma\gamma$ pair creation in the pulsar wind by a single pair injected to the nebula.  We use exact IC/synchrotron cooling rates and spectra assuming a target radiation field dominated by thermal radiation of dimensionless temperature $\theta \equiv kT_{\rm eff}/m_e c^{2}$ and show the results as a function of the initial injected Lorentz factor $\gamma_0$ of the pair.  Different lines show the results for different values of the $\gamma\gamma$ optical depth $\tau_{\gamma\gamma,\rm th}$ through the wind region and for different ratios of the nebula's magnetic to thermal compactness, $\ell_{\rm B}/\ell_{\rm th}$ (eq.~\ref{eq:lBn}).  

Unsurprisingly, Fig.~\ref{fig:pairmult} shows that the multiplicity increases with the initial particle energy $\gamma_0$, the mean energy of the target radiation $\theta$, and the gamma-ray optical $\tau_{\gamma\gamma,\rm th}$.  On the other hand, $\mathcal{M}_e$ decreases with increasing $\ell_{\rm B}/\ell_{\rm th}$.  This is because synchrotron photons (which dominate IC in cooling the pair for $\ell_{\rm B}/\ell_{\rm th} \gtrsim 0.1-1$) have lower energies which are insufficient to produce pairs on the optical photons (see eq.~\ref{eq:num} and surrounding discussion) .  

At early epochs $t_{\pm, \rm IC} \lesssim t \lesssim t_{\rm B}$ (eq.~\ref{eq:tB}) when the nebula is still weakly magnetized ($\ell_{\rm B}/\ell_{\rm th} \lesssim 0.1$), Fig.~\ref{fig:pairmult} shows that the regulation condition $\mathcal{M}_e \sim 1$ (eq.~\ref{eq:regulation}) is always achieved for particle energy $\gamma_0\theta \sim few$.  The mean Lorentz factor of particles entering the nebula, as well as the characteristic energy of the photons they radiate (but not necessarily those that escape), will thus remain close to $\Delta \gamma = \gamma_0 \sim 1/\theta$ as long as the wind is opaque to gamma-rays above this energy (i.e. $\tau_{\gamma\gamma,\rm th} > 1$).  Rescaling equation (\ref{eq:tgg}) slightly to represent the optical depth of the nebula\footnote{The thermal compactness of the ejecta $\lth$ must be replaced by the somewhat higher compactness of the wind region, $\ell_{\rm th,n} \approx (R_{\rm ej}/R_{\rm n})\ell_{\rm th}$ arising due to its smaller radius $R_{\rm n} \approx R_{\rm ej}/2$.} instead of the entire ejecta we see that the condition $\tau_{\gamma\gamma,\rm th} > 1$ will remain satisfied for a few years following the explosion. 
Taking $\theta = kT_{\rm eff}/(m_e c^{2}) \sim 10^{-6}$ for $T_{\rm eff} \approx 6000$ K, the regulated particle/photon energy during this epoch is $\gamma_0 m_e c^{2} \sim m_e c^{2}/\theta \sim$ 1 TeV just after the source is becoming transparent to $\gamma\gamma$ pair production at these energies ($t_{\gamma\gamma} \sim$ yrs; eq.~\ref{eq:tgg}).  

On the other hand, Figure \ref{fig:pairmult} also reveals that for high magnetization $\ell_B/\ell_{\rm th} \gtrsim 0.1$ ($t \gtrsim t_{\rm B}$; eq.~\ref{eq:tB}) the self-regulation condition $\mathcal{M}_e \sim 1$ cannot be achieved for any value of $\gamma_0$.  
%Qualitatively, this is because synchrotron photons$-$which increasingly dominate the nebula cooling for larger values of $\ell_B/\ell_{\rm th}-$are not energetic enough to produce pairs on the thermal background (see eq.~\ref{eq:num}).  
It would then appear that the pair regulation process should break down at $t \sim t_{\rm B}$ even if $\tau_{\gamma\gamma,\rm th} > 1$ is still satisfied.  In practice, however, we find that if $\varepsilon_{B}$ is sufficiently large for $t_{\rm B} \ll t_{\gamma\gamma}$ then synchrotron radiation dominates over thermal photons as the targets for pair creation anyways (Table \ref{tab:regulation}).

%In summary, once $\tauggth \ll 1$ becomes satisfied (and potentially starting earlier, once $\ell_B/\ell_{\rm th} \ll 0.1$ if $t_{B} < t_{\gamma\gamma}$), the energy per particle at the shock will start to increase from its self-regulated value $\gamma_0 \lesssim 1/\theta \sim 10^{6}$ to the much larger value set either by either the intrinsic pair loading (eq.~\ref{eq:gammapm}) or as limited by synchrotron cooling (eq.~\ref{eq:gammarad}).  The minimum timescale of this transition,
%\be
%t_{\pm} \approx {\rm max}[t_{\gamma\gamma},t_{\rm B}],
%\label{eq:tpm}
%\ee
%can range from months (for large $\varepsilon_B \gtrsim 10^{-2}$; eq.~\ref{eq:lB_lth}) to a few years (when $\tauggth \lesssim 1$; eq.~\ref{eq:tgg}).

Finally, at late times $t \gg t_{\gamma\gamma}$ once the wind is transparent to IC radiation, a final stage of regulated pair production can occur, this time by synchrotron gamma-ray photons on synchrotron X-ray targets.  This regulation phase can last for a time $t_{\pm, \rm syn} \sim $ 10 yr, after which the mean pair energy $\Delta \gamma$ increases to the completely ``unregulated'' value set by the intrinsic wind pair loading ($\gamma_{\rm in}$; eq.~\ref{eq:gammapm}) or by the synchrotron burnoff limit ($\gamma_{\rm rad}$; eq.~\ref{eq:gammarad}). 

%One caveat to this conclusion, revealed by our numerical models, is that for higher values of the nebula magnetization $\varepsilon_{B} \gtrsim 10^{-4}$, the pair regulation is dominated by nonthermal radiation in the nebula and hence $\gamma\gamma$ regulation is sustained even at times $t \gtrsim t_{\rm B}$ in these cases. 

%One caveat to this conclusion, revealed by our numerical models, is that for higher values of the nebula magnetization $\varepsilon_{B} \gtrsim 10^{-4}$, the pair regulation is dominated by nonthermal radiation in the nebula and hence $\gamma\gamma$ regulation is sustained even at times $t \gtrsim t_{\rm B}$ in these cases.
%However, over the first months after the explosion, pair production on nonthermal target photons can dominate over thermal targets (Section \ref{sec:results}) and larger value of $\theta \approx \theta_{\rm nth} \sim 10^{-2}$ for nonthermal radiation corresponds to a lower regulated pair energy $\sim 1/\theta_{\rm nth} \sim 100$ MeV at early times.

\begin{figure}
\includegraphics[width=0.5\textwidth]{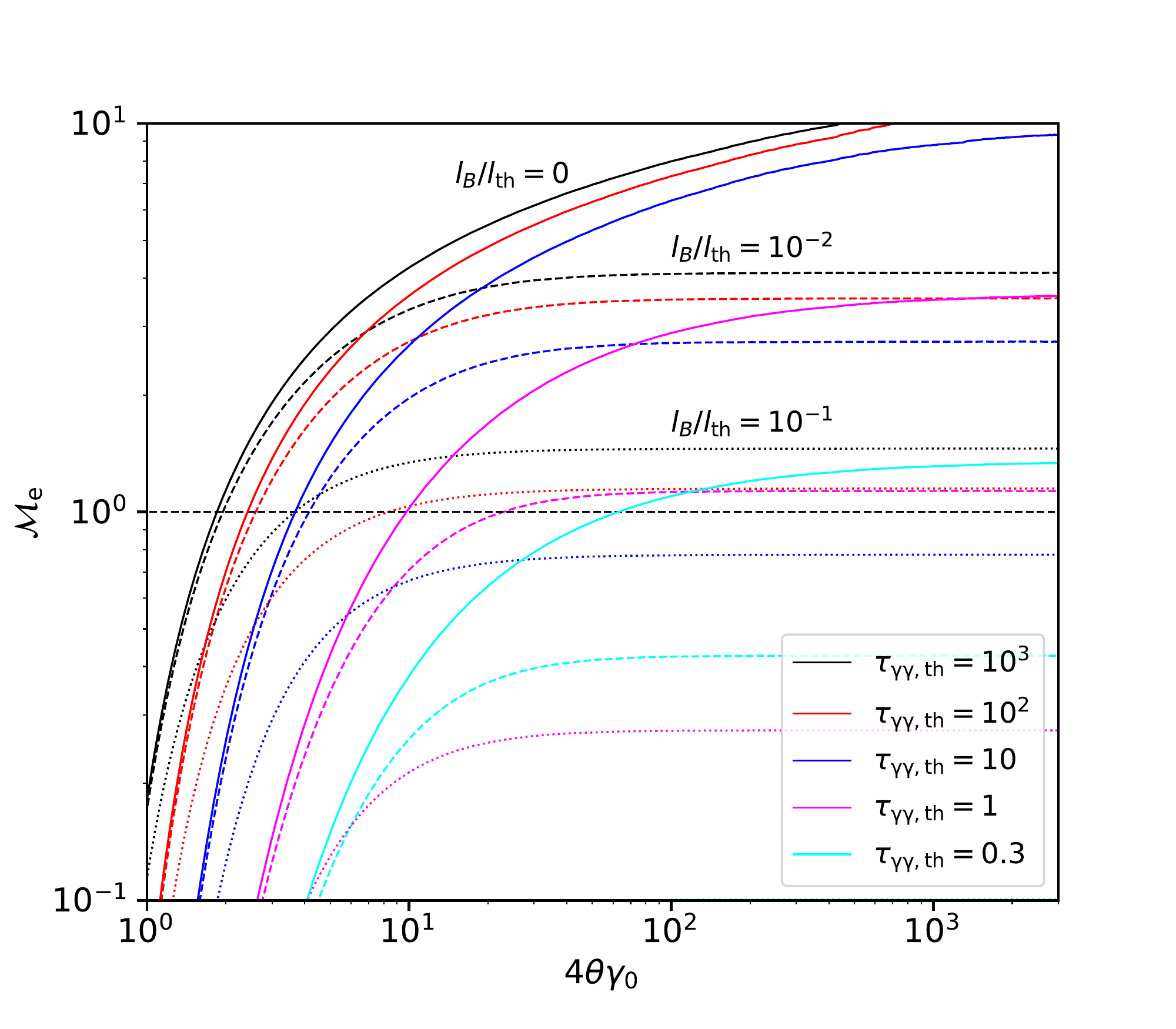}
%\vspace{-0.4cm}
\caption{Pair multiplicity $\mathcal{M}_{\rm e}$ generated in the pulsar wind by $\gamma\gamma$ interactions as a function of $4\gamma_0 \theta$, where $\gamma_0$ is the average Lorentz factor of heated leptons entering the nebula and $\theta \equiv kT_{\rm eff}/m_e c^{2}$ is the dimensionless temperature of the background target radiation field (assumed here to be thermal radiation with $T_{\rm eff} \sim 10^{4}-10^{5}$ K, corresponding to $\theta \sim 10^{-5}-10^{-6}$).  Separate lines show the results for different values of $\tau_{\gamma\gamma,\rm th}$ (the $\gamma\gamma$ pair creation optical depth through the wind region; eq.~\ref{eq:taugg}) and the ratio of magnetic to thermal compactness $\ell_{\rm B}/\ell_{\rm th}$ in the nebula (eq.~\ref{eq:lB_lth}).  As described in the text, $\gamma\gamma$ pair loading of the pulsar wind regulates $\mathcal{M}_{\rm e} \sim 1$ and hence $\gamma_0 \sim 1/\theta$ so long as $\tau_{\gamma\gamma,\rm th} \gtrsim 1$ and $\ell_{\rm B}/\ell_{\rm th} \lesssim 0.1$, conditions typically satisfied for months to years following the supernova explosion.}
\label{fig:pairmult}
\end{figure}

%We note that, although we have proposed an equilibrium feed-back process, the stability of this process is not obvious.  In Appendix \ref{sec:app:stability} we explore this question in greater detail, finding that...{\bf Insert}. 

%{\bf [[In progress: purpose is to illustrate that injection $\gamma$ stays near the KN transition ($4\theta\gamma \sim$ a few) as long as $\tauggth > 1$ at $4\theta x \approx 1$, and increases rapidly once $\tauggth < 1$.]]}

%{\bf BDM: we also need to include an estimate of the thickness of the pair loading zone (optical depth of high energy photons into the wind region) and estimate the synchrotron or IC cooling timescale of the wind particles relative to their expansion time through this region.  We can uses the "loaded" gamma-factor in performing the calculation. }

%% Further issues to consider: the effect of magnetization, pair cascade.

\section{Monte Carlo Simulations}
\label{sec:model}

\begin{table}
  \begin{center}
    \caption{Magnetar model parameters used in Monte Carlo simulations}
    \label{tab:models}
    \begin{tabular}{c|c|c|c|c|c|c} 
      SN & $B_{\rm d}$ & $P_0$ & $M_{\rm ej}$ & $v_{\rm ej}$ & $\kappa_{\rm opt}$ & $t_{\rm e}$ \\
      \hline
      - & ($10^{14}$ G) & (ms) & ($M_{\odot}$) & (km s$^{-1}$) & (cm$^{2}$ g$^{-1}$) & (d) \\
      \hline
\hline
%    2015bn & 0.3 & 2.2 & 11.7 & 5500 & 0.19 & 75 \\
    2015bn & 0.43 & 2.6 & 11.7 & 5400 & 0.1 & 54 \\
    2017egm$^{\dagger}$ & 1.0 & 5 & 3 & 7000 & 0.15 & 38 \\
    \end{tabular}
    $\dagger$Broadly motivated by Bayesian fits of the \citet{Kasen&Bildsten10} model to supernova optical light curve data
    \citep{Nicholl+17b}.
%    \citep{Nicholl+17b,Nicholl+18}.
  \end{center}
\end{table}

\subsection{Details of the Numerical Model}
\label{sec:setup}

We perform three-dimensional Monte Carlo simulations which track the energy evolution of electron/positron pairs and photons in the magnetar nebula and supernova ejecta, as well as the outwards radial diffusion of both high-energy and optical photons through the ejecta.  We employ a two-zone approach, which considers separately a homogeneous inner ``wind/nebula'' zone and an outer ``ejecta'' zone (Fig.~\ref{fig:cartoon}).  Photons and electron/positron pairs can occupy both zones, but ions and their associated electrons are restricted to the ejecta shell. Conversely, the magnetic field is restricted to the wind/nebula zone and is assumed to be negligible in the ejecta zone.  

The radial density profile of the ejecta is taken to be constant from $R_{\rm n} = R_{\rm ej}/2$ to $R_{\rm ej}$ and zero outside this range.  The time-dependent Thomson optical depth of the ejecta shell $\tau_{\rm T} \propto t^{-2}$ (eq.~\ref{eq:tauT}) is determined from the assumed fixed ejecta velocity $v_{\rm ej}$ and mass $M_{\rm ej}$.  We do not account for acceleration of the ejecta by the nebula, because throughout most of the evolution of interest the nebula/ejecta is radiative (such that most of the engine's luminosity is lost to radiation rather than converted into bulk kinetic energy).  The true ejecta distribution will undoubtedly be more complex than the constant density shell we have assumed, for instance being compressed into a thin radial shell by the nebula (e.g.~\citealt{Kasen&Bildsten10}) or characterized by inhomogeneities and mixing resulting from Rayleigh-Taylor instabilities at the nebula-ejecta interface (e.g.~\citealt{Chen+16,Suzuki&Maeda20}).  However, we do not expect our qualitative conclusions to depend on such details.

Primary pairs are injected into the nebula at the radius $R_{\rm inj} = R_{\rm n}$ corresponding to the termination shock at the rate $\dot{N}_{\pm}$ specified by Equation (\ref{eq:Ndot}) for an assumed Goldreich-Julian pair multiplicity $\mu_{\pm} = 10^{4}$.  The multiplicity is uncertain theoretically, but our key results turn out to be insensitive to its precise value because the mass loading of the wind over timescales relevant to optical and gamma-ray observations is dominated by secondary pairs generated by $\gamma\gamma$ interactions in the wind zone (Section \ref{sec:gamma}).  The latter are explicitly tracked after being generated as they are advected outwards in the wind.  

Upon reaching the termination shock radius, both the primary and secondary pairs are energized based on the current pulsar spindown luminosity, $L_{\rm e}$ (eq.~\ref{eq:Le}).  Depending on the total rate that particles enter the shock, $\dot{N}_{\pm}$, each particle experiences an energy gain of $\Delta \gamma \propto L_{\rm e}/\dot{N}_{\pm}$ (eq.~\ref{eq:deltagamma}).  All pairs entering the shock receive the same energy gain, regardless of their initial energy and we do not account for potential nonthermal particle acceleration at the shock.  At early times, when $\gamma\gamma$ pair-loading dominates the particle budget of the wind, the two-way feedback between the number of secondary pairs generated in the wind zone and the dissipated energy per particle $\Delta \gamma$ (Section \ref{sec:gamma}), introduces oscillations in $\Delta \gamma$ causing some ``jitter'' in our results on top of that present due to the Monte Carlo sampling.
%(see Appendix \ref{sec:app:stability} for a discussion of the stability of the feedback process).

We assume that pairs radiate their energy in the nebula isotropically with negligible bulk velocity.  In principle this assumption might be questioned because at early times of interest the $\gamma\gamma$ pair creation$-$and concomitant mass-loading and heating of the wind$-$occur over a region of finite radial thickness ahead of the wind termination shock roughly equal to the mean-free path of high-energy photons to $\gamma\gamma$ interactions.  If the hot pairs entrained in the wind were to cool radiatively over this radial scale faster than being advected into the nebula, then the engine's luminosity would instead be radially beamed due to the relativistic motion of the wind, violating the assumption of isotropic emission and altering the energy spectrum of the pairs' emission. 

In Appendix \ref{app:coolingwind} we show that IC cooling can be neglected within the pair loading region so long as the pulsar wind remains relativistic all the way to the termination shock.  The condition for negligible synchrotron cooling is somewhat more stringent and requires either the wind possess a bulk $\Gamma \gg f_{\rm th}^{-1/2} \sim 10-100$ or low magnetization $\sigma \ll f_{\rm th}\Gamma^{2}$ by the radius of pair-loading.  These conditions could be satisfied if the Poynting flux of the wind is largely converted into heat or bulk kinetic energy by the termination shock (e.g. as is inferred to occur in the Crab pulsar wind; \citealt{Kennel&Coroniti84}), or if the process of $\gamma\gamma$ pair-loading the wind itself acts to reduce its magnetization.  Though beyond the scope of this paper, we plan to explore the latter possibility in future work (see Section \ref{sec:latetime} for additional discussion).  

As long as the pairs do not cool in the relativistic wind zone, the precise location at which they are energized is not critical because the particles radiate most of their energy isotropically once reaching the nebula.  This allows us to forgo treating the wind zone as a separate region regarding synchrotron and IC processes.  

The magnetic field within the wind/nebula region is given by Equation (\ref{eq:Bn}), i.e. we assume the magnetic energy to be a fixed fraction $\varepsilon_{\rm B}$ of the total energy deposited by the engine over the ejecta expansion time.  We generally assume modest values of the nebula magnetization $\varepsilon_{\rm B} \lesssim 1$, motivated by the likelihood of magnetic reconnection within the sub-relativistically expanding nebula (e.g.~\citealt{begelman98,Porth+13}) and the potential reduction in wind magnetization even prior to the termination shock due to $\gamma\gamma$  pair-loading as mentioned above.  
%As we shall discuss, a very low magnetization in the nebula appears necessary to explain the late-time optical light curves of SLSNe in the magnetar model (Section \ref{sec:latetime}).
%This oversimplification can be improved upon in future work.
%separate treatment of synchrotron and IC processes in the wind zone.
%However, in Appendix \ref{app:coolingwind} we show that on timescales of interest, cooling with the pair-loading region of the wind can be ignored as long as...

All of the photon destruction and creation processes discussed in Section \ref{sec:absorption} are included in both the nebula and ejecta.  In addition to the high-energy (nonthermal) radiation, we also explicitly follow the creation and transport of optical (thermal) photons. The optical photons are created by thermalization processes described in Section \ref{sec:thermalization}, namely Compton downscattering on cold electrons, Coulomb losses by nonthermal electrons/positrons and photoionization. While the detailed spectrum of optical radiation is not considered, the code follows the frequency-integrated energy of the optical radiation field represented by MC particles. Their interactions with the ejecta material as they diffuse outward are assumed to take place with a constant opacity $\kappa_{\rm opt}$.  The radial expansion of the ejecta is taken into account in the interaction events (modeled as coherent scattering in the local instantaneous rest frame), thereby automatically accounting for adiabatic losses and hence suppression of the early optical radiation when $\tau_{\rm opt} > c/v_{\rm ej}$.

As summarized in Table \ref{tab:models}, we adopt parameters for the ejecta ($M_{\rm ej}$; $v_{\rm ej}$; $\kappa_{\rm opt}$) and magnetar engine ($B_{\rm d}$; $P_0$) which are broadly motivated by model fits to the optical light curve data on two well-studied SLSNe, 2015bn and 2017egm \citep{Nicholl+17b,Nicholl+18}.  Once the basic ejecta and engine parameters have been specified, the main free parameter is the nebula magnetization, whose value we vary from $\varepsilon_{B} = 0$ to $\varepsilon_{B} = 10^{-2}$ (similar to the Crab Nebula).  

%In Section \ref{sec:other} we expand briefly on the expected diversity of behavior for other potential engine-powered transients, such as FBOTs and neutron star mergers, which are characterized by lower ejecta masses and higher ejecta velocities.  

\begin{figure*}
\vspace{-0.2cm}
\includegraphics[width=0.5\textwidth]{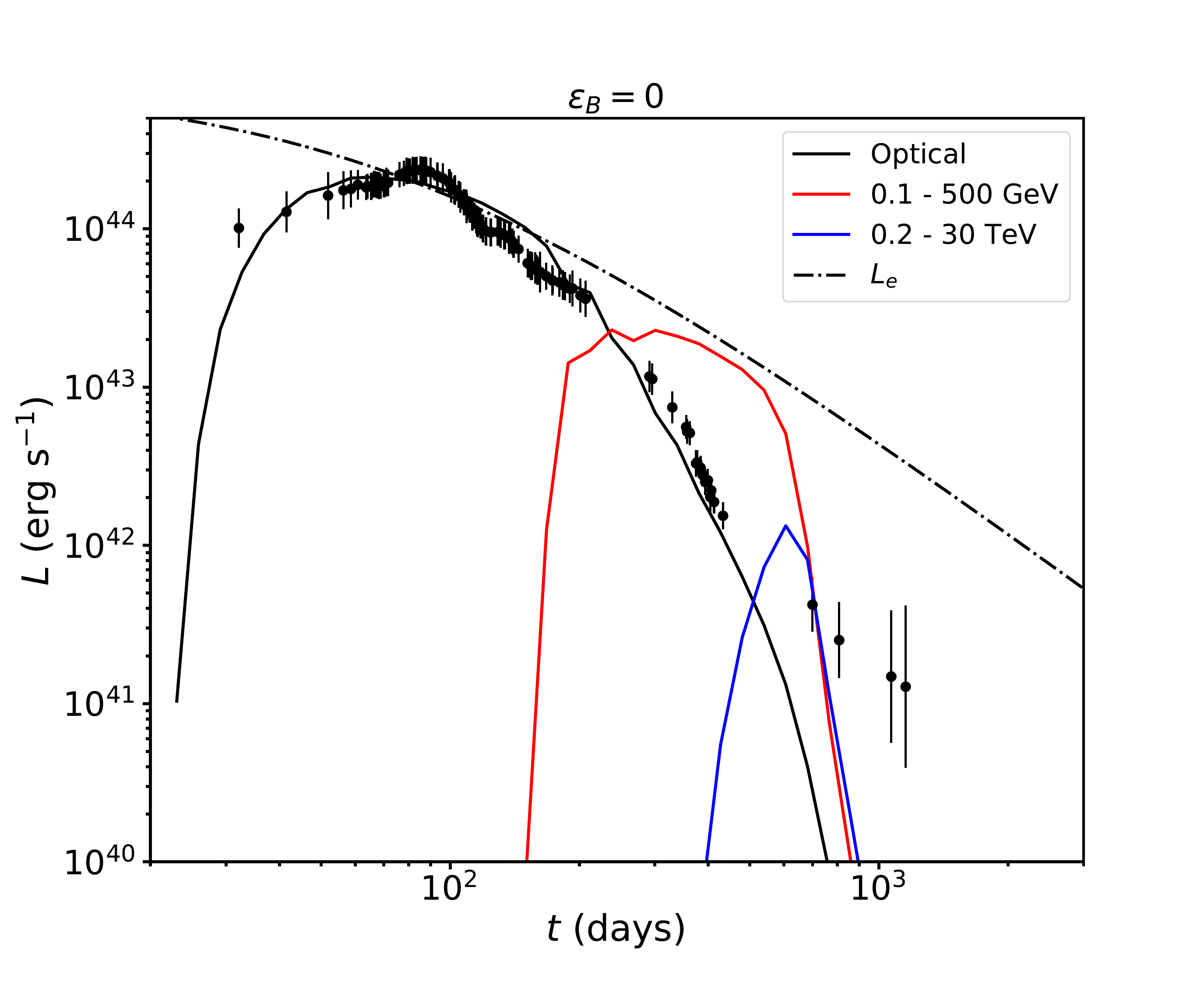}
\vspace{-0.15cm}
\includegraphics[width=0.5\textwidth]{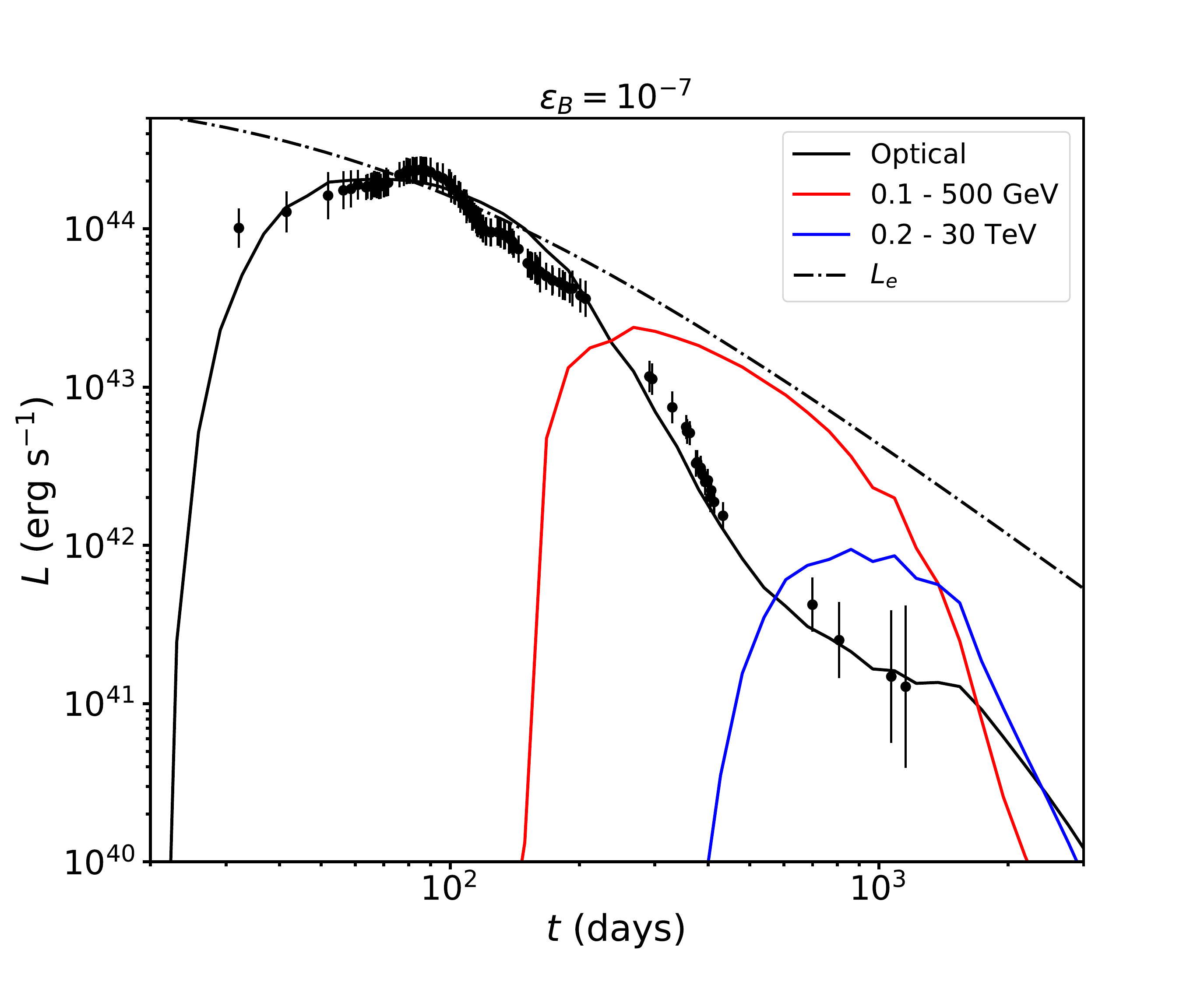}
\vspace{-0.15cm}
\includegraphics[width=0.5\textwidth]{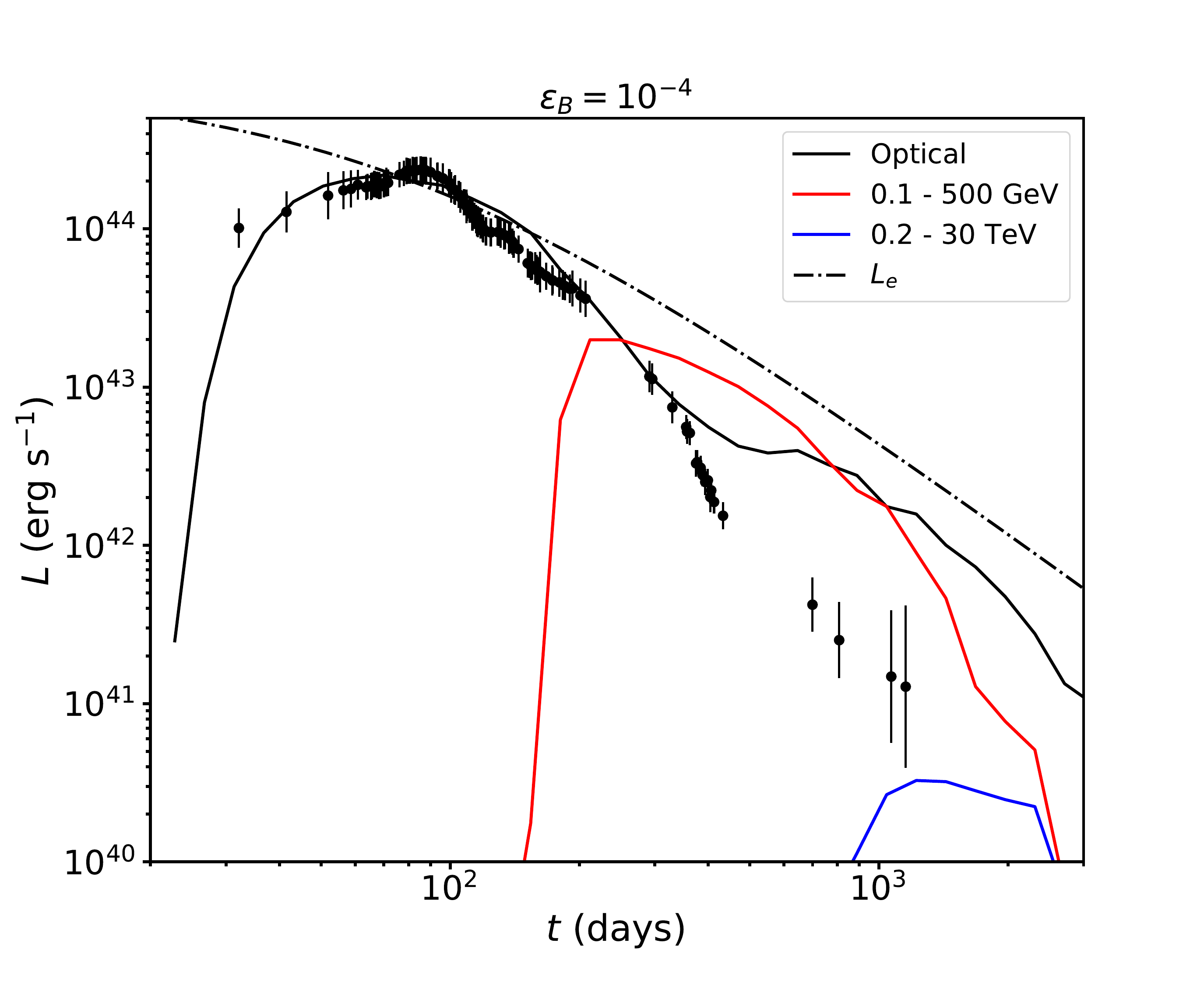}
\includegraphics[width=0.5\textwidth]{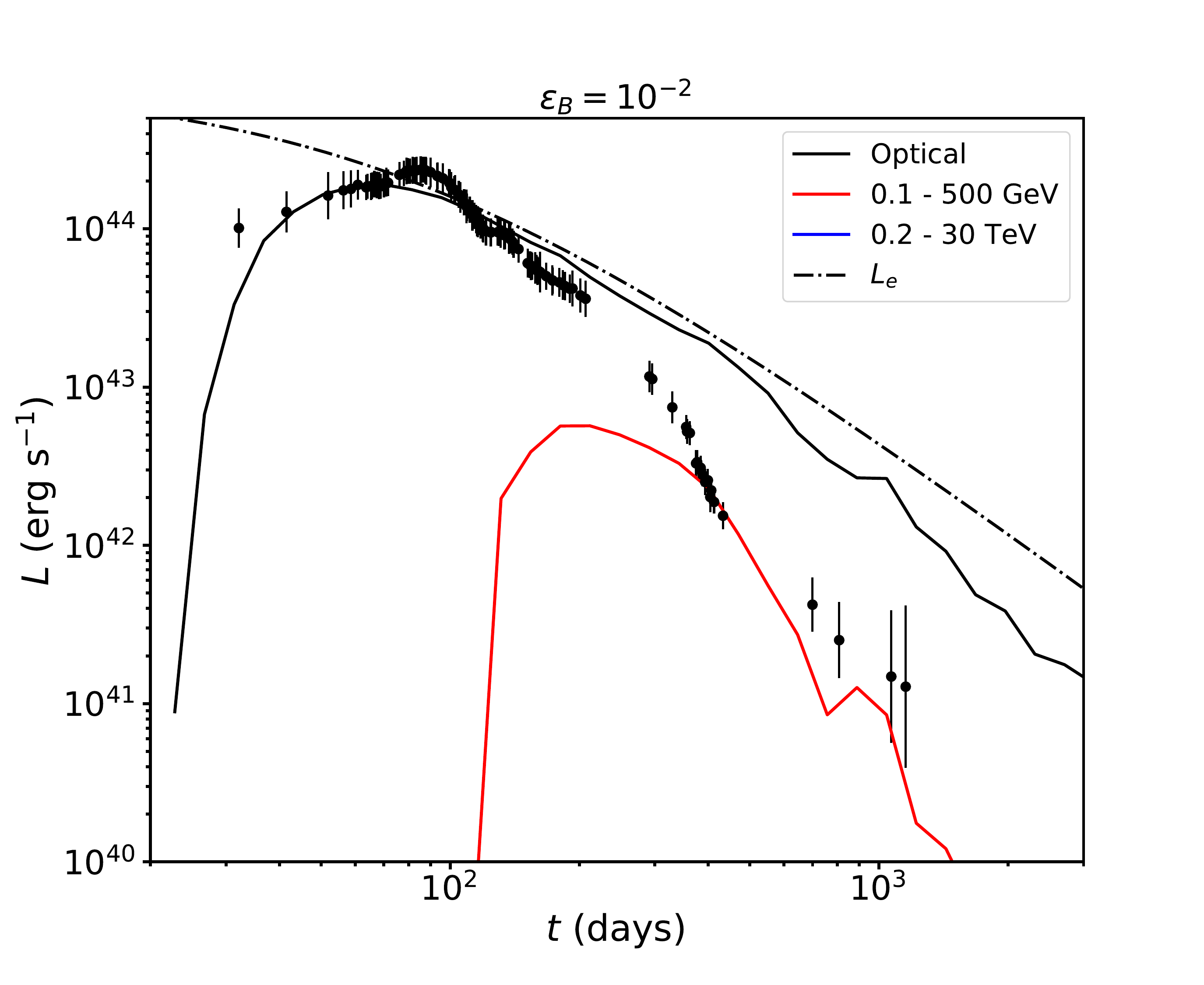}
%\vspace{-0.2cm}
\caption{Optical and gamma-ray light curves of magnetar-powered supernovae, calculated for different values of the nebula magnetization $\varepsilon_{B} = 0, 10^{-7}, 10^{-4}, 10^{-2}$ as marked above each plot.  The magnetar and ejecta parameters (Table \ref{tab:models}) follow model fits to the optical light curve data for SN2015bn (shown as black dots; \citealt{Nicholl+16a}).  A dot-dashed line shows the engine luminosity $L_{\rm e}$ (eq.~\ref{eq:Le2}).}
\label{fig:15bn}
\end{figure*}

\begin{figure*}
%\vspace{-0.2cm}
%\includegraphics[width=0.45\textwidth]{LC_opt_GeV_TeV_2017egm_sim_31053_nosynch_sm.pdf}
\includegraphics[width=0.5\textwidth]{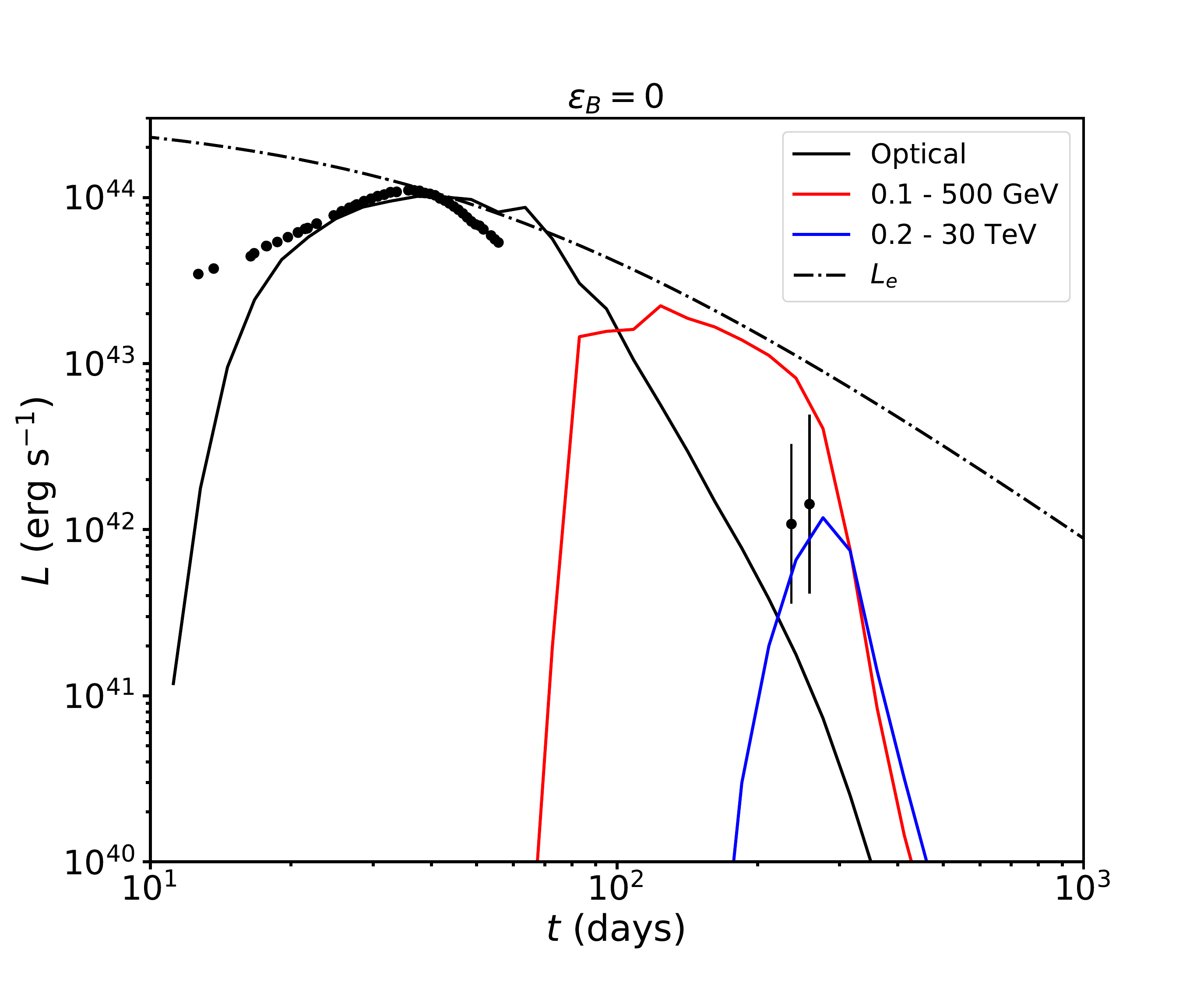}
%\vspace{-0.15cm}
\includegraphics[width=0.5\textwidth]{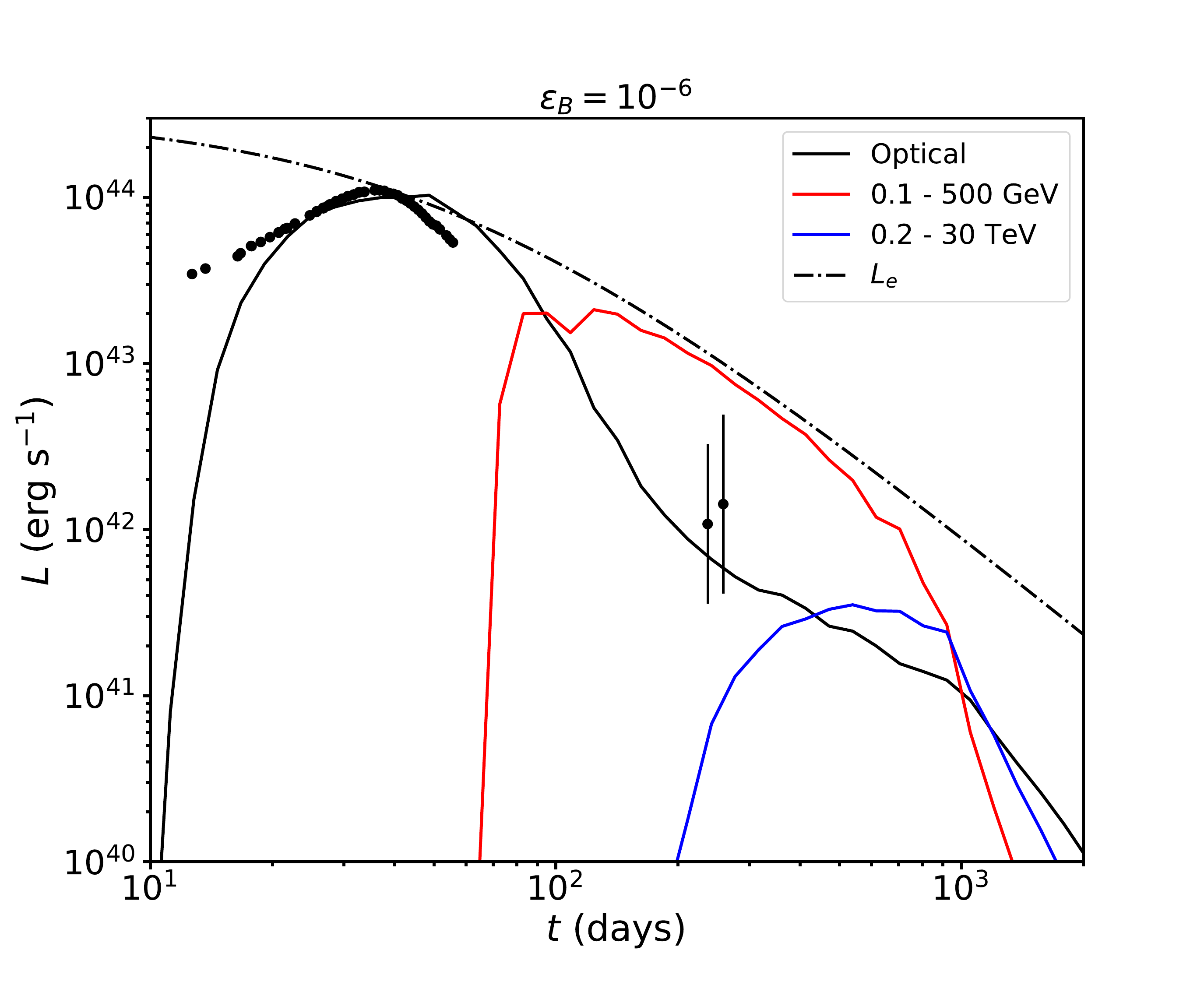}
%\includegraphics[width=0.45\textwidth]{LC_opt_GeV_TeV_2017egm_Titan_sim_31951d_epsB_1emin6_sm.pdf}
%\vspace{-0.15cm}
%\includegraphics[width=0.45\textwidth]{LC_opt_GeV_TeV_2017egm_Titan_sim_31952_epsB_1emin4_sm.pdf}
\includegraphics[width=0.5\textwidth]{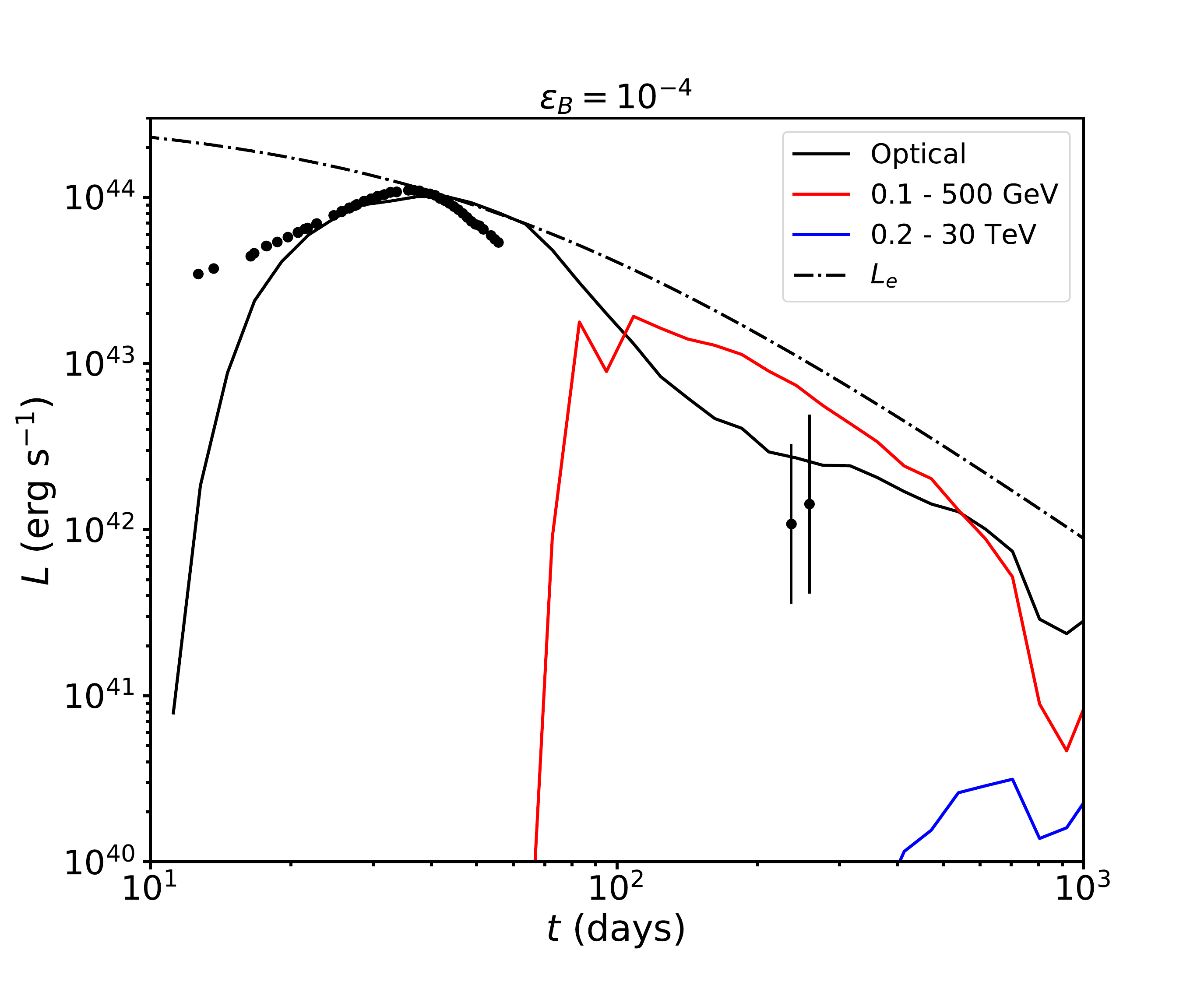}
\includegraphics[width=0.5\textwidth]{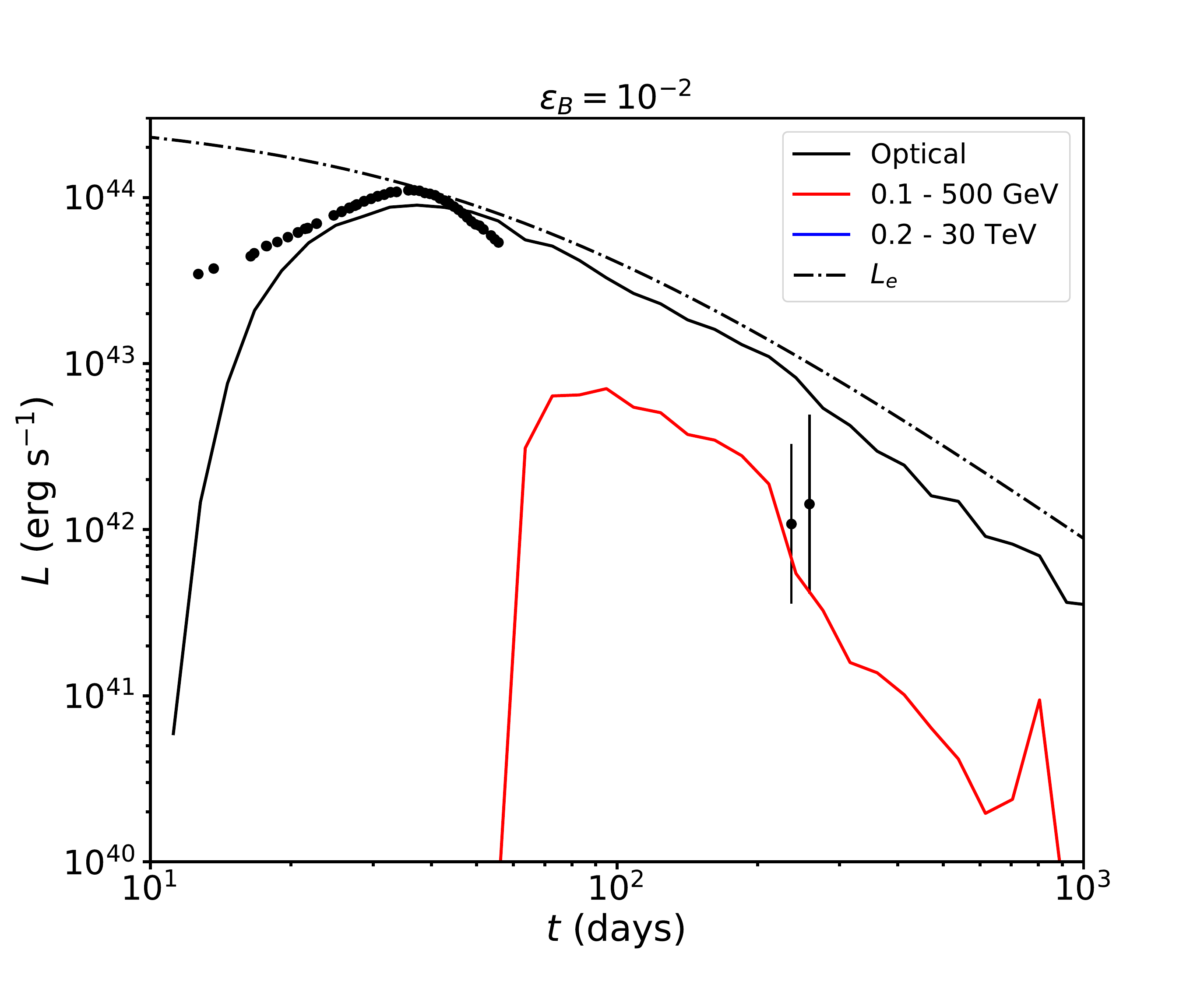}
%\vspace{-0.2cm}
\caption{Same as Figure \ref{fig:17egm} but for magnetar model parameters fit to SN2017egm (optical light curve data from \citealt{Nicholl+17b,Bose+18}).}
\label{fig:17egm}
\end{figure*}

\begin{figure*}
\includegraphics[width=1.0\textwidth]{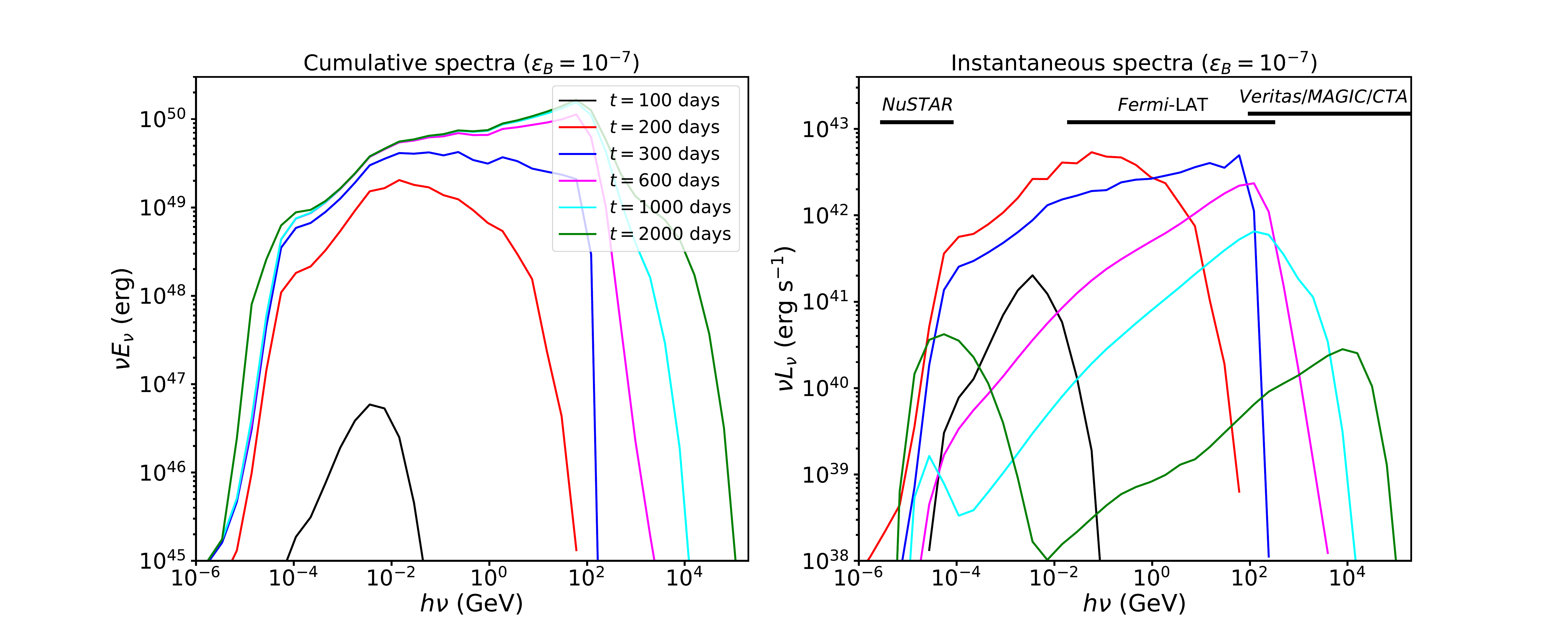}
\includegraphics[width=1.0\textwidth]{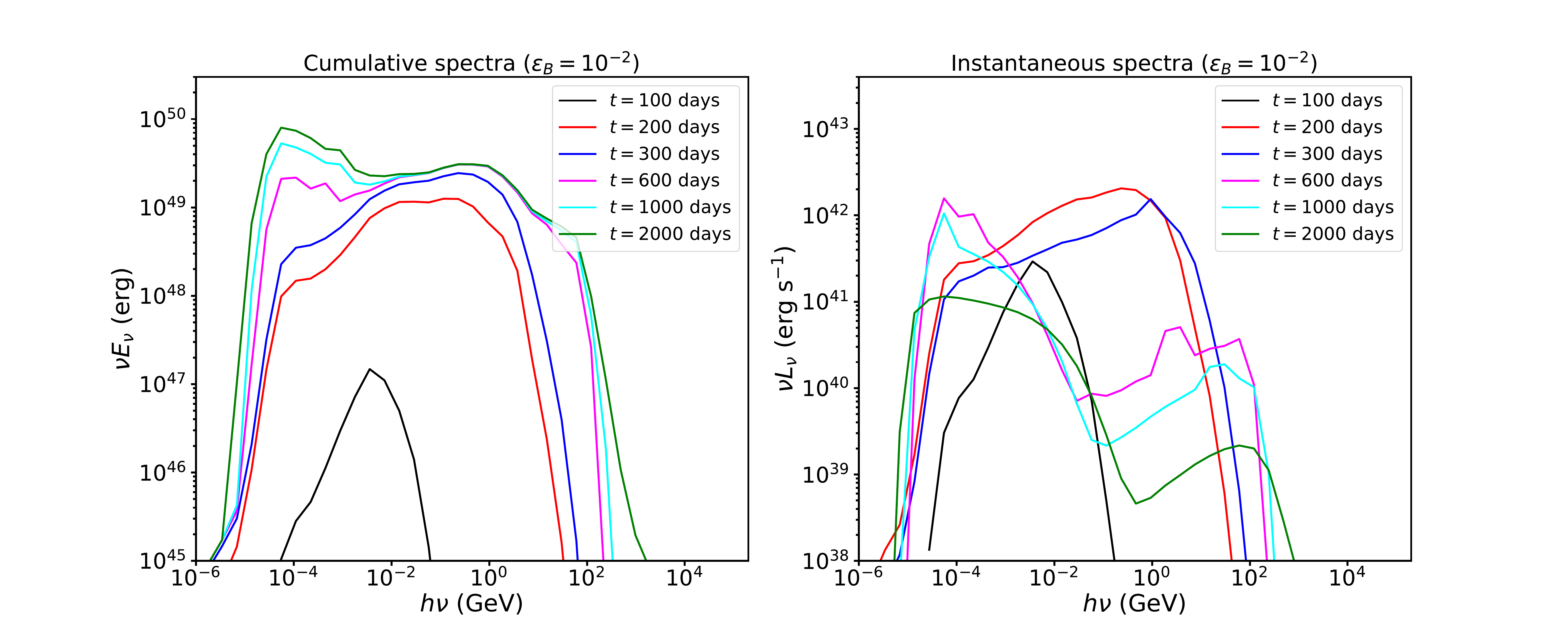}
%\vspace{-0.4cm}
\caption{Cumulative and instantaneous energy spectra of the escaping high-energy radiation from the $\varepsilon_{\rm B} = 10^{-7}$ and $10^{-2}$ models for SN2015bn, corresponding to the second and last panels of Fig.~\ref{fig:15bn}.}
\label{fig:15bn_spec}
\end{figure*}

\begin{figure}
\includegraphics[width=0.5\textwidth]{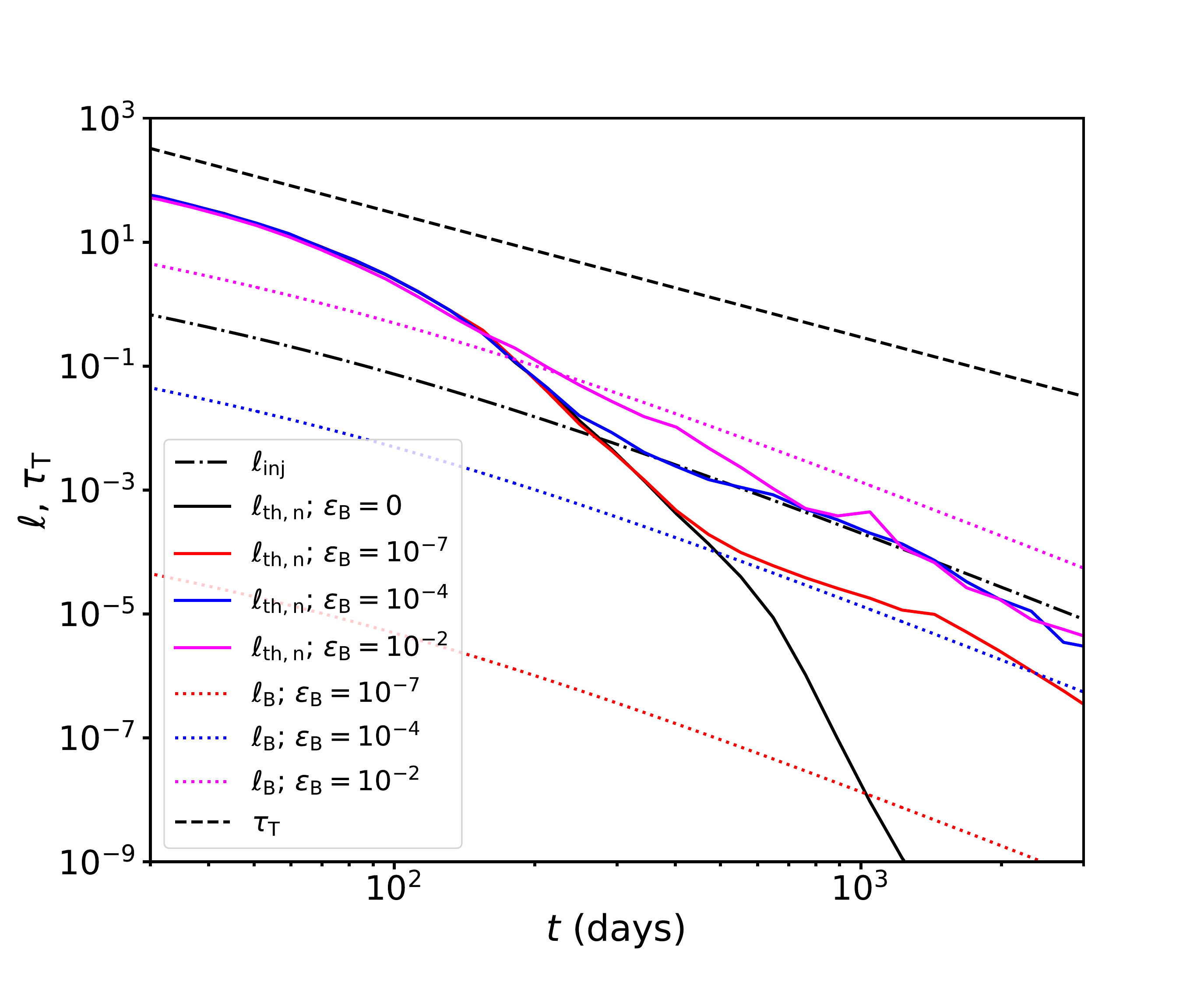}
\caption{Time evolution of injection, thermal and magnetic compactnesses and Thomson optical depth (eqs.~\ref{eq:ellinj},\ref{eq:ellopt},\ref{eq:lBn},\ref{eq:tauT}) for the SN2015bn models shown in Figure \ref{fig:15bn}.
}
\label{fig:15bn_ell}
\end{figure}

\begin{figure*}
\includegraphics[width=0.45\textwidth]{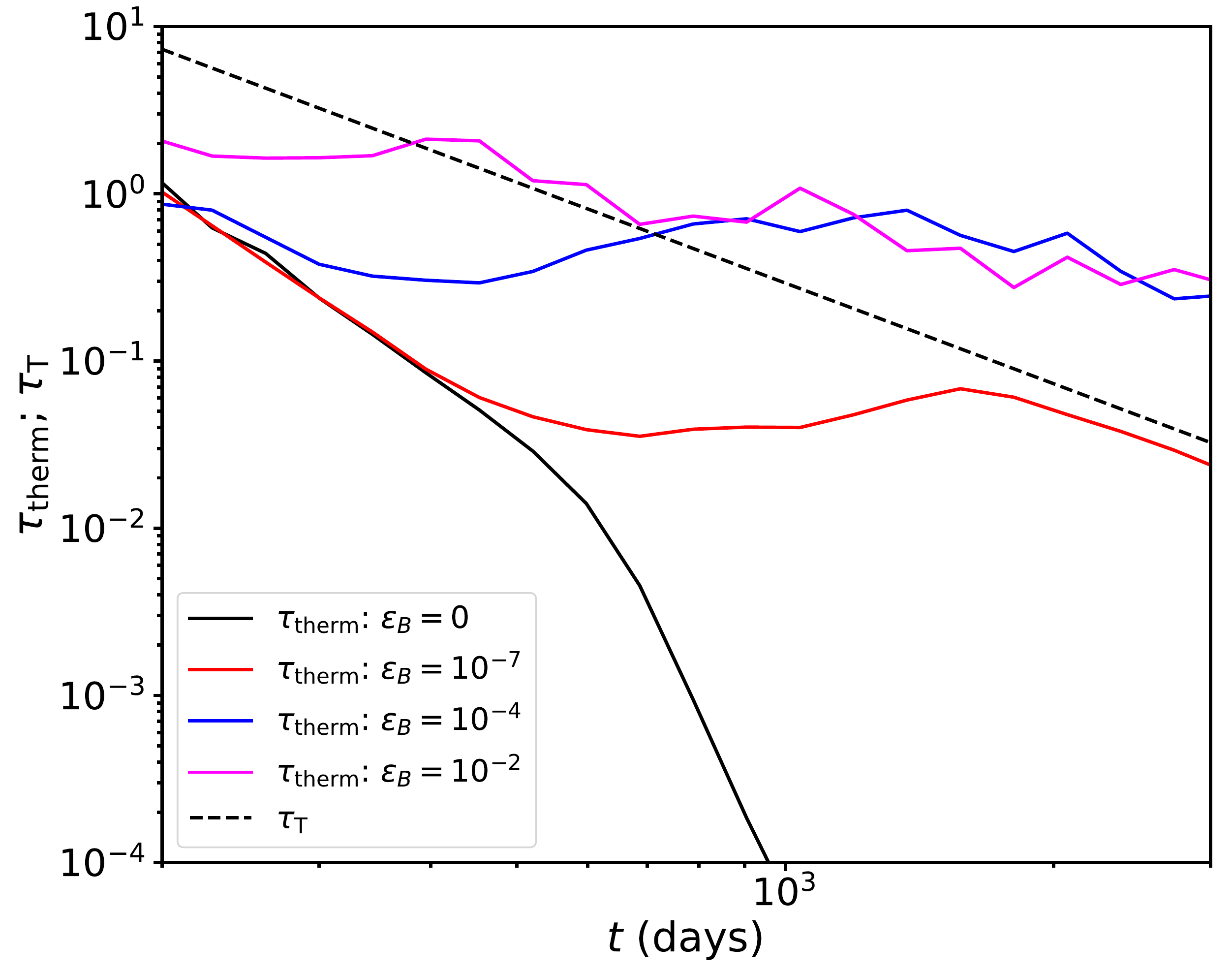}
\hspace{0.4cm}
\includegraphics[width=0.5\textwidth]{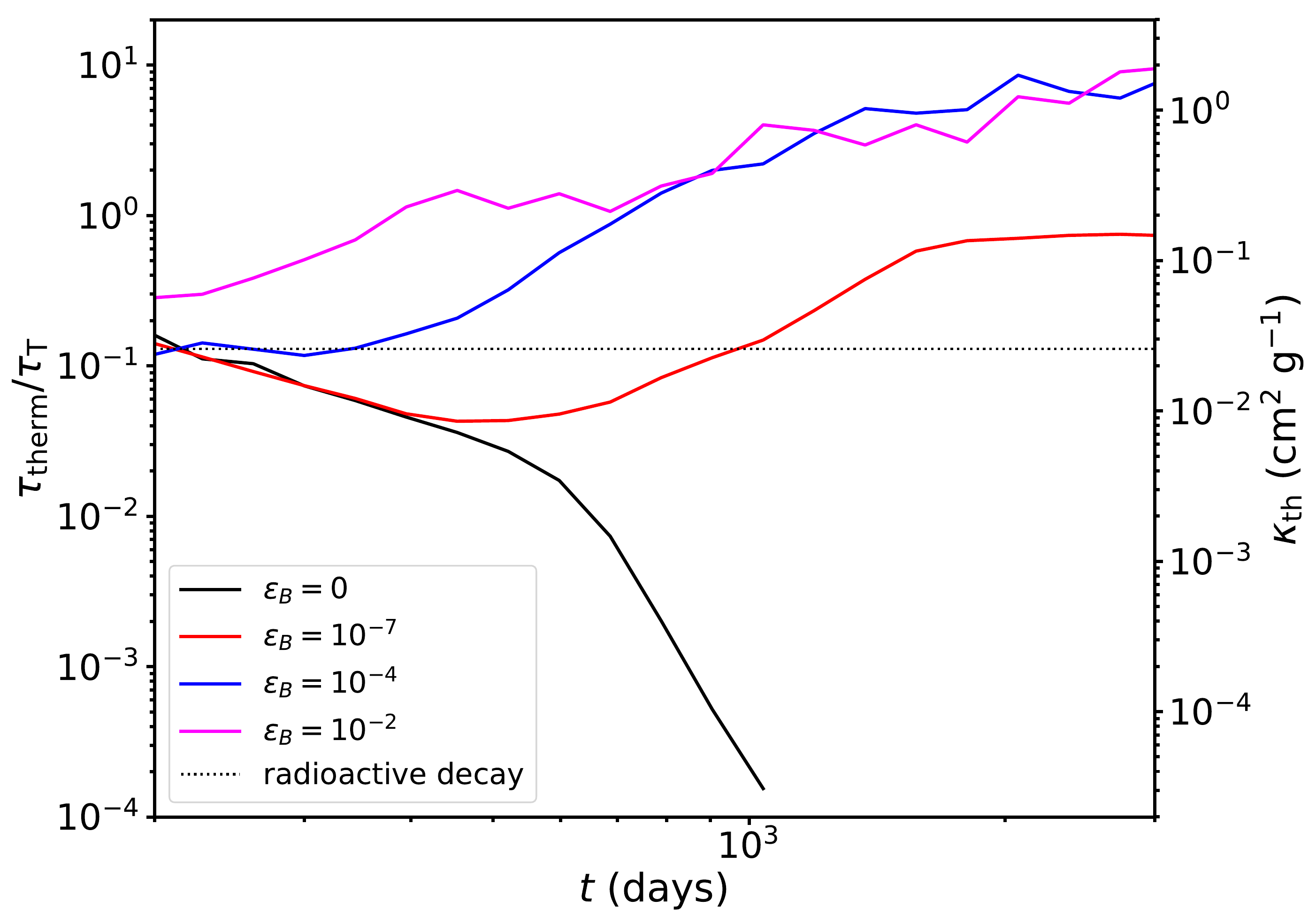}
%\vspace{-0.4cm}
\caption{Evolution of the thermalization optical depth $\tau_{\rm therm}$ (eq.~\ref{eq:tautherm})
for the SN2015bn models in Figure \ref{fig:15bn}.  On the right panel we have normalized $\tau_{\rm therm}$ to the Thomson optical depth $\tau_{\rm T} \propto t^{-2}$ (eq.~\ref{eq:tauT}), which is shown separately with a dashed line on the left panel.
The ratio $\tau_{\rm therm}/\tau_{\rm T} = \kappa_{\rm th}/\kappa_{\rm es}$ can also be interpreted in terms of the gamma-ray thermalization opacity $\kappa_{\rm th}$ introduced in previous SLSNe modeling (e.g.~\citealt{Wang+15,Nicholl+17d}; which however assume a constant value of $\kappa_{\rm th}$).  A dotted horizontal line shows the constant $\kappa_{\rm th} = 0.13\kappa_{\rm es}$ which approximates thermalization of $^{56}$Ni and $^{56}$Co radioactive decay in ordinary hydrogen-poor supernovae (\citealt{Swartz+95}). }
\label{fig:15bn_ther}
\end{figure*}

\begin{figure}
\includegraphics[width=0.5\textwidth]{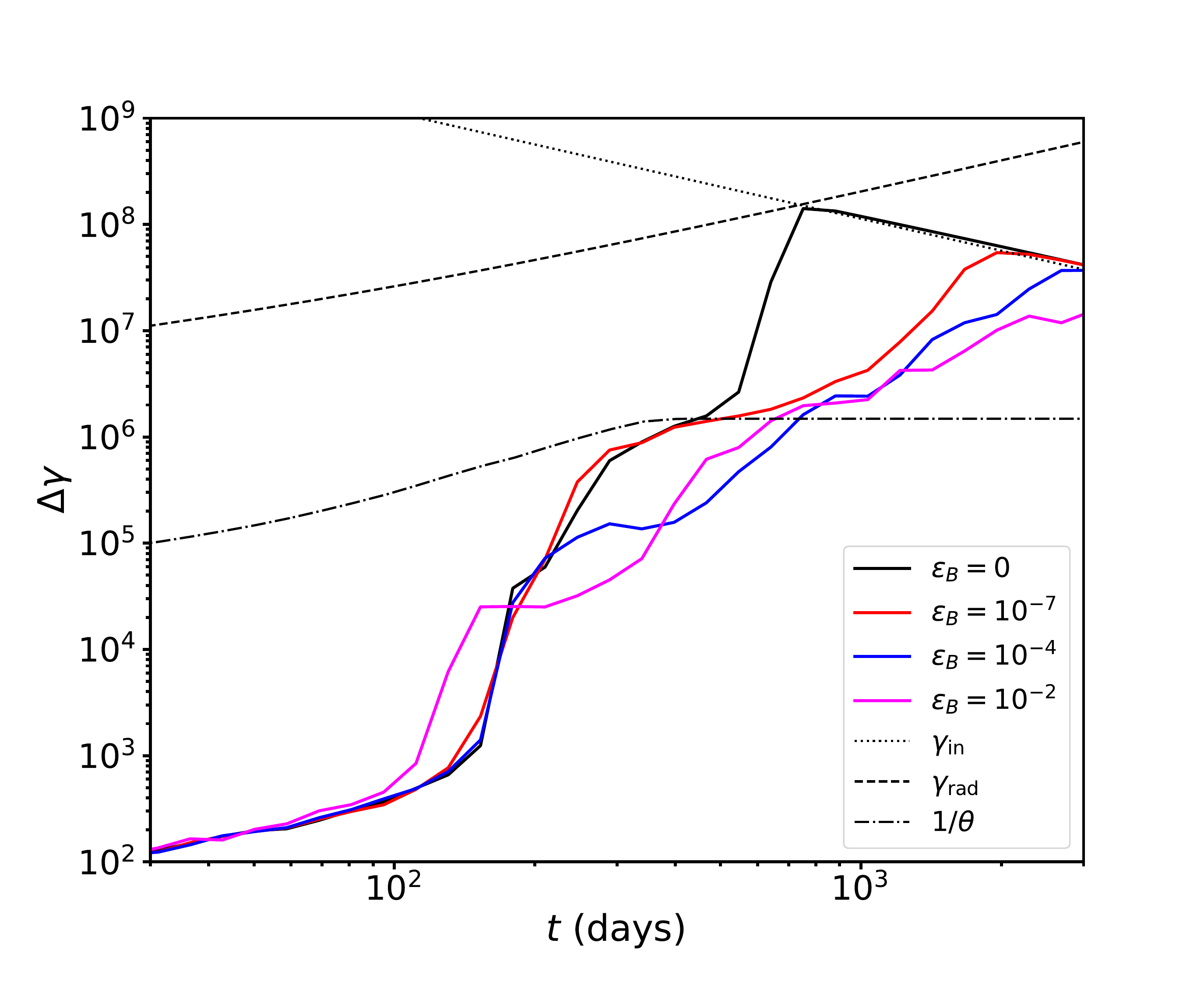}
\caption{Energy received per lepton at the wind termination shock for the SN2015bn models (Fig.~\ref{fig:15bn}).  Shown for comparison are: $\gamma_{\rm in}$ (eq.~\ref{eq:gammapm}), the mean energy pairs would acquire entering the nebula absent $\gamma\gamma$ loading; $1/\theta$, an analytic estimate of the self-regulated particle energy when $\gamma\gamma$ pair production on thermal radiation regulates the wind pair loading (see Section \ref{sec:gamma}, Fig.~\ref{fig:pairmult}); and $\gamma_{\rm rad}$ (eq.~\ref{eq:gammarad}), the maximum particle energy corresponding to the synchrotron burn-off limit.  }
\label{fig:15bn_param}
\end{figure}

\subsection{Results}
\label{sec:results}

Figures \ref{fig:15bn} -- \ref{fig:15bn_param} summarize our results for for SN2015bn and 2017egm for different values of the nebula magnetization.  For the illustration of key observables, such as the optical and gamma-ray light curves, we present results for both supernovae.  However, given that the two models exhibit a qualitatively similar evolution, for the more interpretation-focused figures we present just the SN2015bn cases.  Insofar as these two supernovae are also fairly representative of the larger SLSNe population (Fig.~\ref{fig:SLSNe}), we expect qualitatively similar results to those presented to hold more broadly.  
%Figure \ref{fig:15bn_ell} shows the time evolution of the Thomson optical depth, and the thermal and magnetic compactnesses, for the SN2015bn models.

The optical light curves (Figs.~\ref{fig:15bn}, \ref{fig:17egm}) around and just after peak are largely insensitive to magnetization and, in fact, more generally to the details of energy deposition and reprocessing of high-energy radiation within the nebula/ejecta. This is because reprocessing at this early stage is nearly 100\% efficient; thus, most of the high-energy radiation gets downgraded into the optical band where opacity is lower compared to higher frequencies.  Defining a ``thermalization'' optical depth,
\be \tau_{\rm therm} = -\ln(1 - L_{\rm opt}/L_{\rm e}),
\label{eq:tautherm}
\ee
we find $\tau_{\rm therm} > 1$ at times $t\lesssim 100$~days (Figure \ref{fig:15bn_ther}).  The opacity $\kappa_{\rm th} \equiv (\tau_{\rm therm}/\tau_{\rm T})\kappa_{\rm es}$ which corresponds to $\tau_{\rm thermal}$ is equivalent to the gamma-ray thermalization opacity introduced in previous SLSNe modeling (e.g., \citealt{Wang+15,Chen+15}), with the important difference that its value is calculated here self-consistently and is not constant in time.

Our theoretically predicted optical light curves rise somewhat more abruptly than SLSNe observations (shown as black points in Figs.~\ref{fig:15bn}, \ref{fig:17egm}) prior to optical maximum.  This is partly a consequence of our assumption of initially cold ejecta and the sole energy source being the centrally concentrated nebula.  In the physical case, other sources of volumetrically-distributed heating of the ejecta not accounted for in our model may contribute to the thermal emission at early times and flatten the rise.  Such additional heating sources include the shock driven into the ejecta by the nebula at early times (e.g.~\citealt{Kasen+16,Chen+16,Suzuki&Maeda19}) or a relativistic jet from the central engine (e.g., \citealt{Margalit+18b}).

Starting a few months after the explosion, $\tau_{\rm therm}$ falls below unity and high-energy radiation begins leaking out of the ejecta.  Most of the gamma-rays which escape at early times are generated by IC scattering in the nebula.  The ejecta first become transparent in the energy range $0.1 - 10$~GeV, within the bandpass accessible to {\it Fermi}-LAT (Section \ref{sec:absorption}).  The corresponding $\sim$GeV light curves (red lines in Figs.~\ref{fig:15bn}, \ref{fig:17egm}) rise to a maximum at $\sim 200$~days and $t\sim 100$ days, respectively, at which stage they carry a sizable fraction of the engine luminosity; any higher energy gamma-ray emission at this epoch is also reprocessed down into the $\sim$ GeV band.

The emission properties at yet higher energies $\gtrsim 100$\,GeV rely on both the pulsar wind's ability to inject sufficiently energetic pairs to create such photons, as well as the ability of those photons to escape the nebula and ejecta in face of $\gamma\gamma$ absorption.  Regarding the first condition, Figure \ref{fig:15bn_param} shows that the average energy of the injected pairs has already increased to $\Delta\gamma \gtrsim 10^5$ ($\gtrsim 100$~GeV) by $\sim 200$~days (as set by the regulated $\gamma\gamma$ pair-loading in the wind; Section \ref{sec:gamma}).  These energetic pairs mainly cool by IC radiation on optical targets in the Klein-Nishina regime (at the higher magnetizations, synchrotron also competes), depositing most of their energy into photons with energies comparable to the cooling lepton (see eq.~\ref{eq:EIC}).  

The delayed onset of $\gtrsim 100$~GeV until $t \sim 100-200$~days is thus mainly controlled by opacity, which at these energies is dominated by $\gamma\gamma$ interactions on the optical radiation field (Fig.~\ref{fig:taueff1}).  Because opacity controls the ability of the highest-energy gamma-rays to escape, the peak timescale and luminosity of the $\gtrsim 100$ GeV emission is strongly dependent on the thermalization efficiency (eq.~\ref{eq:tautherm}) insofar as the latter determines the fraction the engine's luminosity which is reprocessed into optical target radiation.  A steeper(shallower) decay of the optical light curve hastens(lengthens) the $\gamma\gamma$ optically-thin transition and generally increases(decreases) the peak luminosity in the $\gtrsim 100$ GeV bands.

On the other hand, a very steeply decaying optical light curve (e.g. the $\varepsilon_{\rm B} = 0$ cases in Figs.~\ref{fig:15bn}, \ref{fig:17egm}) also acts to suppress the late high energy emission.  This is because, absent synchrotron cooling, a weak target radiation field for IC cooling hastens the transition of the nebula into an adiabatic regime, after which the injected pairs are no longer able to cool efficiently over a dynamical time.  This results in a turnover in the light curves at all energy bands for $\varepsilon_{\rm B} \approx 0$.

Figure \ref{fig:15bn_spec} shows the cumulative and instantaneous spectra at different times for SN2015bn in the $\varepsilon_{\rm B} = 10^{-7}$ and $\varepsilon_{\rm B} = 10^{-2}$ models.  Gamma-ray emission initially appears in the $0.1-10$~MeV range around peak optical light.  Before $t \approx 100$~days the emission is significantly attenuated by Compton recoil losses.  At the low-energy end the spectrum is truncated by bound-free absorption, whereas the high-energy turnover represents the maximum energy of IC upscattered optical photons.  The latter is limited by the fact that the energies of the pairs in the nebula responsible for the upscattering at this stage are regulated to a low value (Figure \ref{fig:15bn_param}) due to efficient pair production on nonthermal radiation (Table \ref{tab:regulation}).

As the transparency window expands, the spectrum broadens and becomes limited by photo-electric absorption and thermal pair production at the low and high-energy ends, respectively. At $t\sim 300$~days the spectrum is relatively flat in $\nu F_{\nu}$ space, softened relative to the classical cooling spectrum $\nu F_{\nu} \propto \nu^{1/2}$ by pair-photon cascades. As the overall compactness decreases, the cascade ceases and the spectrum correspondingly hardens. The bulk of the gamma-ray energy at $\sim 1$~year emerges just below the frequency above which the source is optically thick to pair production on the exponential tail of the thermal photon spectrum.

Starting approximately $2-3$~years after the explosion the $\gamma\gamma$ optical depth drops below unity and all the radiated TeV photons can escape from the source. This also marks the time when the $\gamma\gamma$ pair-loading of the wind begins to wane, causing the energy per lepton to increase (Fig.~\ref{fig:15bn_param}; see discussion in Section \ref{sec:gamma}). As a result, the IC scattering evolves even deeper into the Klein-Nishina regime, which has two significant consequences: 1. synchrotron emission can become the dominant cooling mechanism even if $\ell_{\rm B}/\ell_{\rm th} \ll 1$, and 2. in case the electron cooling is still dominated by IC radiation (e.g. in the $\varepsilon_{\rm B} = 10^{-7}$ model) then the transition to the adiabatic regime takes place earlier than it would for lower $\Delta\gamma$; both have a suppressing effect on high-energy gamma ray emission. The emergence of an energetically dominant synchrotron component is seen at 
$t \sim 600$ and $2000$~days in the $\varepsilon_{\rm B} = 10^{-2}$ and $10^{-7}$ cases, respectively, at $10 - 100$~keV (magenta and green lines, right panels of Fig.~\ref{fig:15bn_spec}).
% $t \sim 1000$~days, at $10 - 100$~keV (cyan and green lines, right panel of Fig.~\ref{fig:15bn_spec}).
Prior to this time most of the escaping gamma-rays were generated by IC scattering. The time-integrated energy radiated in the synchrotron component is seen to be greater for larger values of $\varepsilon_{B}$. 

Figure \ref{fig:15bn_ther} shows the time evolution of the thermalization optical depth of the ejecta (eq.~\ref{eq:tautherm}), shown separately on its own (left panel) and normalized to the Thomson optical depth $\tau_{\rm T}$ (right panel).  The $\tau_{\rm therm}/\tau_{\rm T}$ ratio is proportional to the thermalization opacity $\kappa_{\rm th}$ employed in previous SLSNe model and shown on the right vertical axis.  The effective $\kappa_{\rm th}$ is larger for higher values of $\varepsilon_{\rm B}$ due to the greater role of synchrotron emission cooling in the nebula than IC scattering for stronger magnetic fields: the lower energies of the synchrotron photons result in them being thermalized with higher efficiency.  The rise of $\tau_{\rm therm}/\tau_{\rm T}$ at hundreds of days occurs once synchrotron emission begins to dominate the thermalized luminosity (even if synchrotron does not necessarily yet dominate the total nebula cooling because $\ell_{\rm th} \gg \ell_{\rm B}$; Fig.~\ref{fig:15bn_ell}).  The abrupt plunge in $\tau_{\rm therm}$ at $t \sim 1$ year in the unmagnetized case ($\varepsilon_{B} = 0$) case is again the result of the nebula becoming adiabatic to IC cooling as a result of the reduced seed field due to energy escaping the nebula.  All models depart at late times from the approximately constant value of $\tau_{\rm therm}/\tau_{\rm T} \approx 0.1$ expected for thermalization of radioactive decay products (e.g., \citealt{Swartz+95}) and employed in previous SLSNe modeling (e.g., \citealt{Chen+15}).

\begin{figure*}
\includegraphics[width=0.5\textwidth]{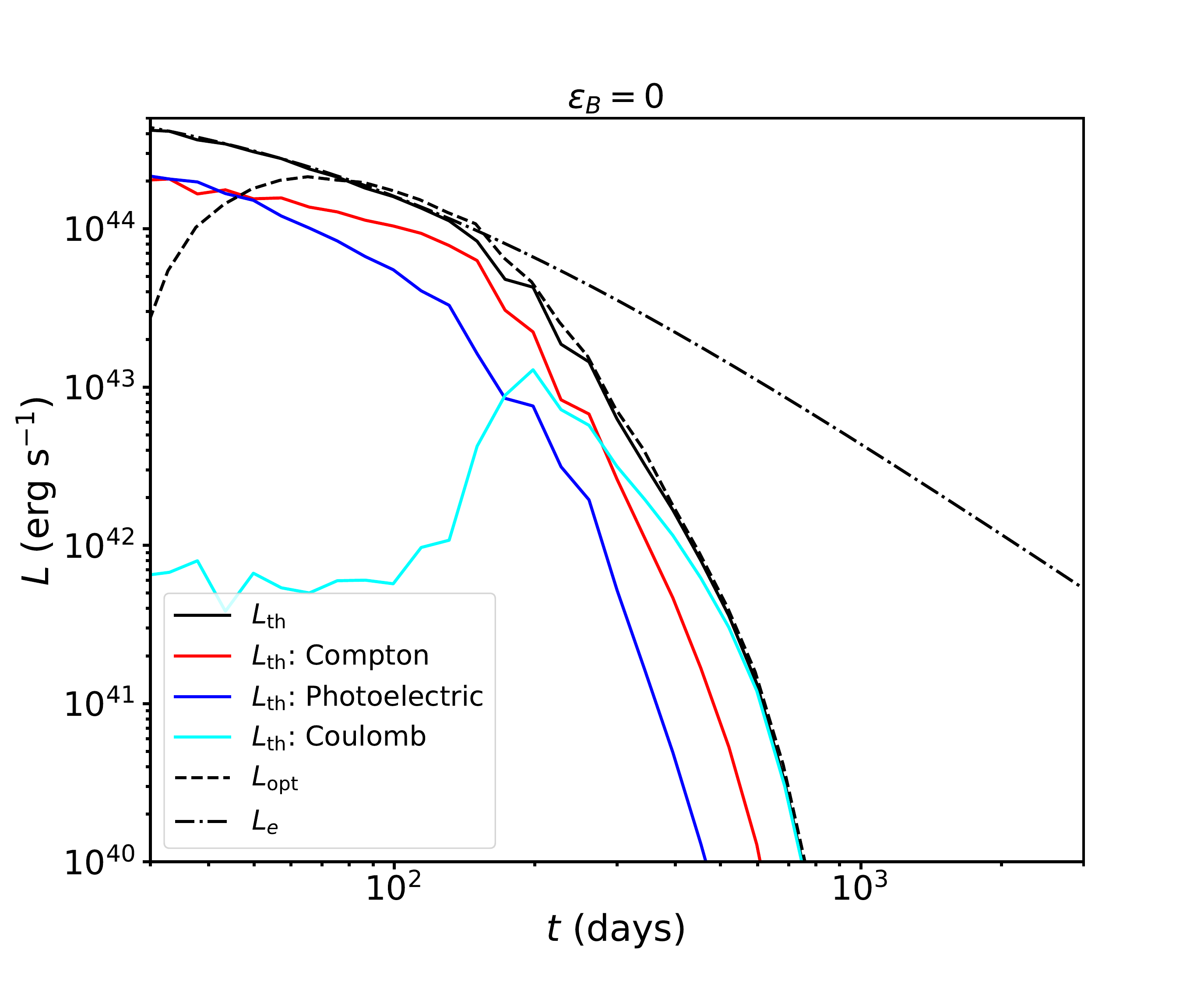}
%\vspace{0.2cm}
%\includegraphics[width=0.45\textwidth]{L_therm_2017egm_Titan_sim_34951d_epsB_1emin6_sm.pdf}
\includegraphics[width=0.5\textwidth]{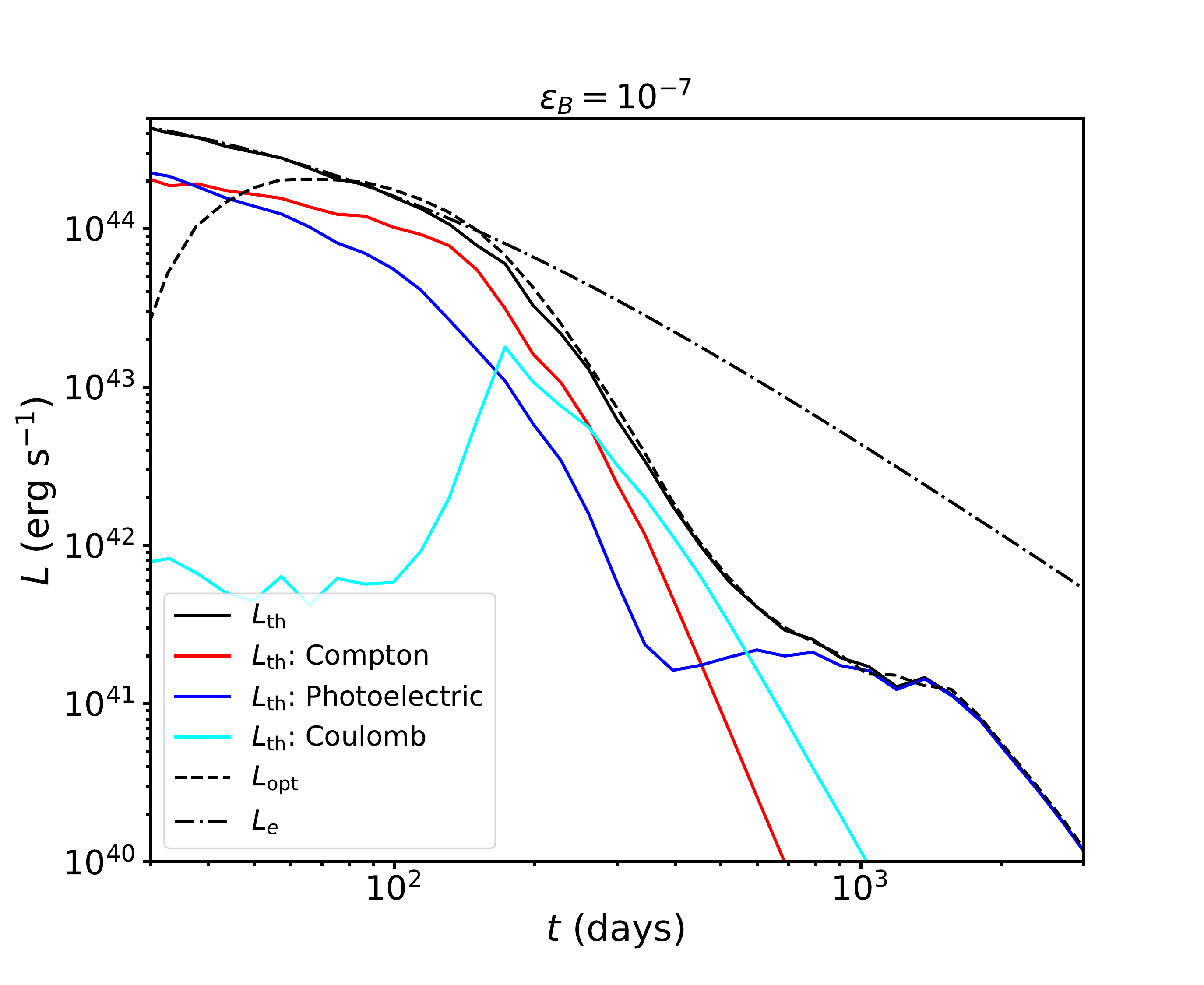}
%\vspace{0.2cm}
%\includegraphics[width=0.45\textwidth]{L_therm_2017egm_Titan_sim_34952a_epsB_1emin4_sm.pdf}
%\includegraphics[width=0.5\textwidth]{L_therm_2015bn_Dione_sim_56952_epsB_1emin4.pdf}
\includegraphics[width=0.5\textwidth]{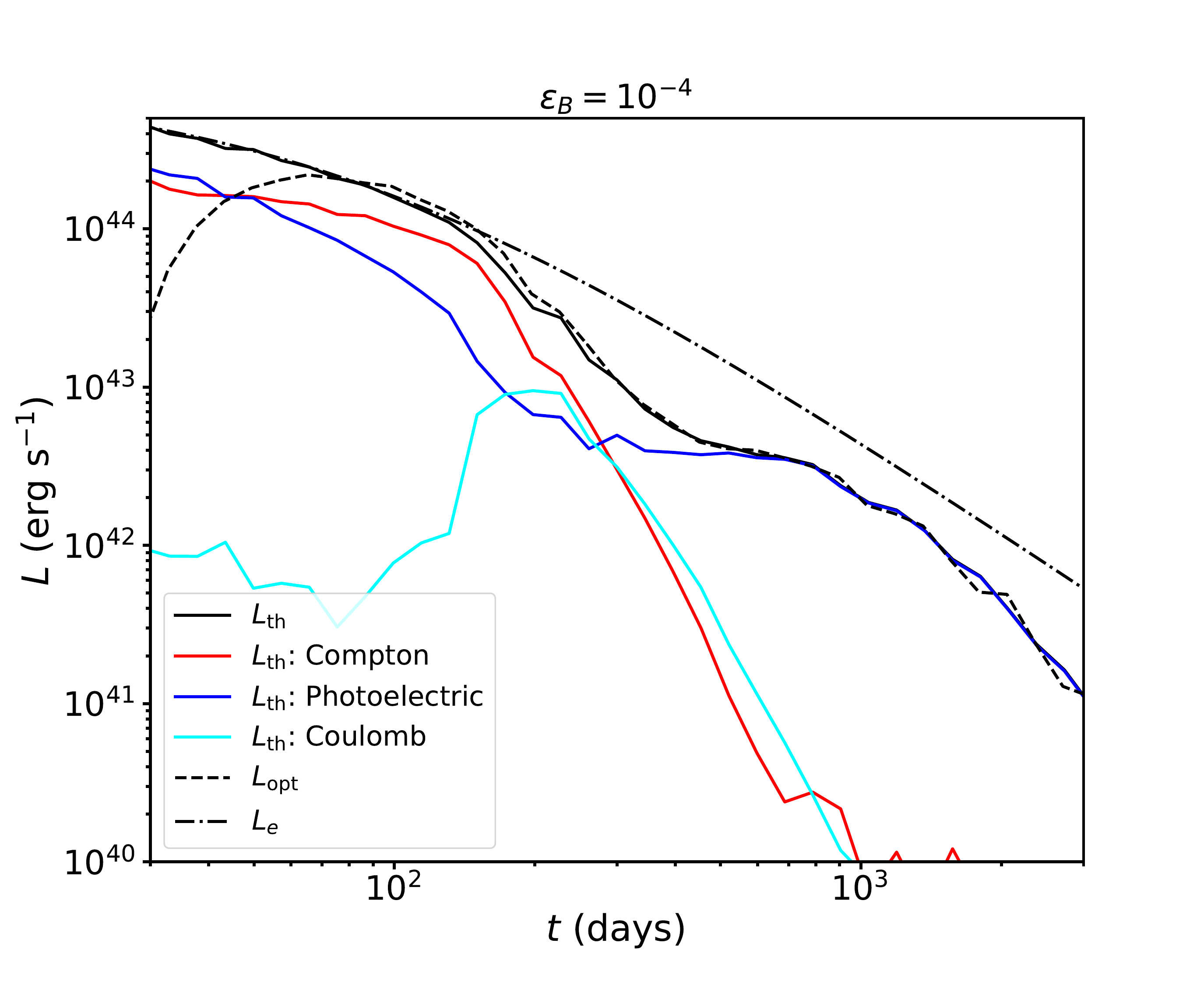}
%\vspace{0.2cm}
%\includegraphics[width=0.5\textwidth]{L_therm_2015bn_Dione_sim_56955_epsB_1emin2.pdf}
\includegraphics[width=0.5\textwidth]{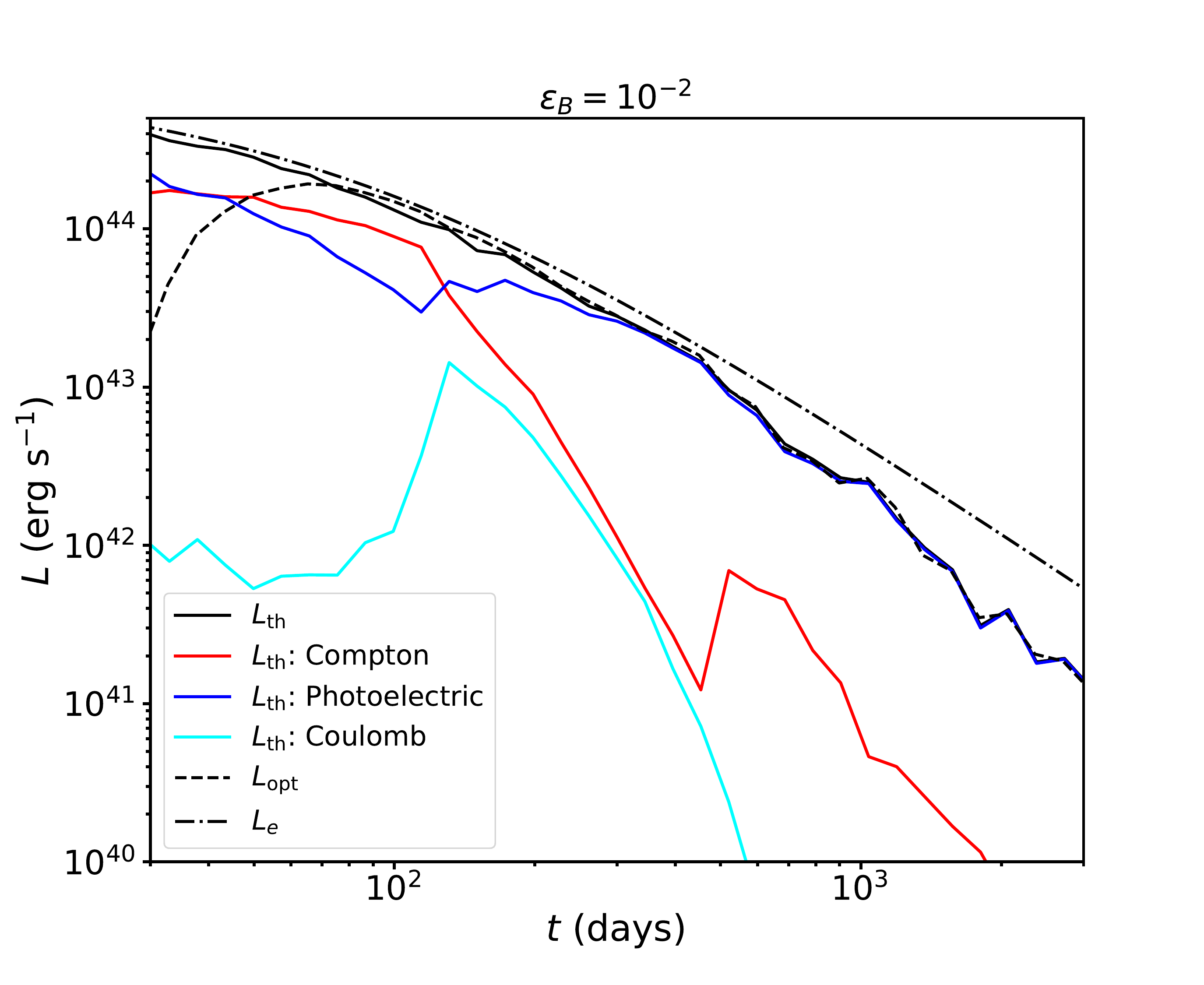}
%\vspace{0.2cm}
\caption{Thermal luminosity $L_{\rm th}$ responsible for powering the optical light curves of SLSNe, and its partition into different channels of thermalization (Compton, Coulomb, Photoelectric) described in Section \ref{sec:thermalization}, for the models of SN2015bn shown in Figure \ref{fig:15bn}. Photoelectric absorption of soft synchrotron X-ray photons dominate the thermalization of the nebula radiation at late times, unless the nebula magnetization is very low (see eq.~\ref{eq:tautherm_analytic} and surrounding discussion).}
\label{fig:Lth}
\end{figure*}

Figure \ref{fig:Lth} separates the ejecta heating into the individual contributing processes: photoelectric absorption, Compton downscattering, and Coulomb scattering (Section \ref{sec:thermalization}).  For the first few months after the explosion the thermalized power is shared roughly equally by Compton downscattering of MeV photons and photoelectric absorption. At these early times, $\gamma\gamma$ pair production in the pulsar wind on IC-generated X-rays regulates the injected pair energy to a relatively low value ($\Delta\gamma \sim 1/\sqrt{\theta} \sim 300$, see Figure \ref{fig:15bn_param}, Table \ref{tab:regulation}). The leptons energized in the wind/nebula cool primarily by upscattering optical photons into the MeV domain.  These photons subsequently experience Compton recoil losses as they diffuse through the ejecta (Figure \ref{fig:taueff1}, top panel). The portion of the IC spectrum below a few tens of keV is absorbed by the photoelectric process, which also contribute appreciably to the total heating at early times.

The channels for loading secondary pairs into the pulsar wind on nonthermal IC radiation becomes less efficient a few hundred days after the explosion. The available energy per lepton then increases until thermal pair production takes control of the pair loading, enforcing $\Delta\gamma \sim 1/\theta$ (Table \ref{tab:regulation}). The overall thermalization efficiency decreases, as the bulk of the dissipated luminosity now escapes as MeV-GeV photons. Albeit with decreasing efficiency, the low-energy part of the IC spectrum continues to contribute to thermalization via the photoelectric and Compton recoil processes (Figure~\ref{fig:Lth}), roughly according to the effective opacity shown in Figure~\ref{fig:taueff1}. In addition, the lower-energy ($\gamma \ll \Delta\gamma$) electron-positron pairs generated within the ejecta by photon-matter and photon-photon pair production contribute to heating/thermalization via Coulomb losses.

However, after $\sim 100 - 200$~days synchrotron emission becomes the dominant contributor to the thermalized luminosity, unless the nebula magnetization is extremely low.  At this stage, the synchrotron spectrum typically peaks in the X-ray domain (eq.~\ref{eq:num}) and a large fraction of its luminosity is thermalized by photo-electric absorption. In this regime one can derive a simple approximate relation between the injected and thermalized luminosities (i.e. thermalization efficiency). Assuming that the nebula pairs cool predominantly by IC upscattering thermal photons and the cooling is still fast compared with the dynamical time ($t < t_{\rm c}$; eq.~\ref{eq:tcool}), one can write
\begin{align}
L_{\rm opt} \approx L_{\rm e} \frac{\dot{\gamma}_{\rm syn}}{\dot{\gamma}_{\rm IC}} \, (1 - e^{-\overline{\tau}_{\rm bf}})
= L_{\rm e} \frac{u_{\rm B, n}}{F_{\rm KN} u_{\rm th, n}} \, (1 - e^{-\overline{\tau}_{\rm bf}}), 
\label{eq:Losyn}
\end{align}
where $u_{\rm B, n}$ and $u_{\rm th, n}$ are magnetic and thermal radiation energy densities within the nebula, respectively, and $\overline{\tau}_{\rm bf}$ is an appropriately defined ``average'' photo-electric opacity of the synchrotron photons. Using equation (\ref{eq:Bn}), the magnetic energy density can be expressed as
\begin{align}
u_{\rm B} = \frac{B_{\rm n}^2}{8\pi} = \frac{L_{\rm e}}{4\pi c R_{\rm n}^2} \frac{3\varepsilon_{\rm B} c}{v_{\rm n}},
\end{align}
whereas the thermal energy density is $u_{\rm th} = L_{\rm opt}/(4\pi c R_{\rm n}^2)$. Substituting $u_{\rm B}$ and $u_{\rm th}$ into equation (\ref{eq:Losyn}), we obtain
\begin{align}
\tau_{\rm therm} \approx \frac{L_{\rm opt}}{L_{\rm e}} \approx \sqrt{\frac{3\varepsilon_{\rm B} c}{F_{\rm KN} v_{\rm n}} \, (1 - e^{-\overline{\tau}_{\rm bf}})}. \label{eq:tautherm_analytic}
\end{align}

For example, in our SN2015bn model with $\varepsilon_{\rm B} = 10^{-7}$, 
synchrotron emission is efficiently absorbed the by bound-free process for the first $\sim 3$~years, i.e $\overline{\tau}_{\rm bf} \gg 1$. Between $t\approx 1$ and $3$~years, the lepton injection energy is regulated to $\Delta\gamma \sim 1/\theta$, such that $F_{\rm KN} \approx 0.05$ (Section \ref{sec:gamma}). This, along with $v_{\rm n} \approx v_{\rm ej}/2 \approx 2700$~km~s$^{-1}$, yields $\tau_{\rm therm} \approx L_{\rm opt}/L_{\rm e}\approx 0.024$. This is a slight underestimate compared with the numerically computed $\tau_{\rm therm}$ (Figure~\ref{fig:15bn_ther}, left panel), owing to contributions from Compton and Coulomb thermalization at earlier times as well as somewhat higher $\Delta\gamma$ (and thus lower $F_{\rm KN}$) towards the end of the considered time interval. These are also the reasons for the slight upward curvature of $\tau_{\rm therm}$, which would otherwise remain constant as long as the above assumptions are valid.

%roughly in agreement with the numerical result (Fig.~\ref{fig:17egm}). As long as the above assumptions hold, the thermalization efficiency remains approximately constant, thus accounting for the upturn of the $\tau_{\rm therm}/\tau_{\rm T} \propto t^{2}$ curves at $t \approx 200$~days in Fig.~\ref{fig:15bn_ther}.  {\bf Should we change example to 15bn?}

\section{Discussion}
\label{sec:discussion}

\subsection{Late-time optical light curves as probes of the nebula magnetization}
\label{sec:latetime}

If SLSNe are powered from within by reprocessing of a central gamma-ray source, this should manifest indirectly in the late-time behavior of their optical light curves.  Our self-consistent models of the nebula gamma-ray emission and its thermalization provide an opportunity to explore this question on solid theoretical footing and explore its implications for the magnetar model.

\citet{Nicholl+18} fit SLSNe light curves to the magnetar model, including a suppression factor $f_{\rm th} = (1-e^{-\tau_{\rm therm}})$ on the engine luminosity to account for incomplete thermalization of the engine gamma-rays at late times, where $\tau_{\rm therm} \propto \kappa_{\rm th}\tau_{\rm T} \propto \kappa_{\rm th}t^{-2}$ and $\kappa_{\rm th}$ is taken to be a constant gamma-ray thermalization opacity \citep{Wang+15,Chen+15}.  At late times, when $\tau_{\rm therm} \ll 1$, this model predicts a decay of the supernova luminosity $L_{\rm opt} \propto L_{\rm e}f_{\rm th} \propto L_{\rm e}\tau_{\rm therm} \propto t^{-4}$ for a magnetar engine with $L_{\rm e} \propto t^{-2}$ (eq.~\ref{eq:Le}), in reasonable accord with the best-fit power-law decay $L_{\rm opt} \propto t^{-3.8}$ measured for SN2015bn over the time interval $t = 200-1100$ days (\citealt{Nicholl+18}; Fig.~\ref{fig:15bn}).  
A similar late-time steepening below the putative magnetar spin-down luminosity is seen in the optical light curve of SN2017egm (Fig.~\ref{fig:17egm}).

The best-fit values of $\kappa_{\rm th} \sim 0.01-0.1$ cm$^{2}$ g$^{-1}$ inferred from the SLSNe light curve fits (e.g.~\citealt{Nicholl+17d}) are similar to the value $\kappa_{\rm th} \sim 0.02$ cm$^{2}$ g$^{-1}$ expected from the thermalization of $\sim$MeV gamma-rays generated by the radioactive decay of $^{56}$Ni $\rightarrow ^{56}$Co $\rightarrow ^{56}$Fe in ordinary stripped envelope supernovae.  In the case of radioactive decay, $\kappa_{\rm th}$ is indeed expected to be roughly constant over the first $\sim 1000$ days after the explosion (e.g.~\citealt{Swartz+95}).
As our results illustrate, however, for engine-powered supernovae the spectrum of the central gamma-ray emission spans a much wider energy range than the predominantly $\sim$MeV spectrum of radioactive decay products (e.g., Fig.~\ref{fig:15bn_spec}) and there is no reason {\it a priori} to expect a similar value of $\kappa_{\rm th}$ to ordinary supernovae, much less one which is constant in time.  

On timescales of months after the explosion our Monte Carlo simulations do in fact reveal an effective value $\tau_{\rm therm}/\tau_{\rm T} \sim 0.1-1$ corresponding to $\kappa_{\rm th} \sim 0.01-0.1$ cm$^{2}$ g$^{-1}$  (Fig.~\ref{fig:15bn_ther}).  However, the predicted time evolution strongly deviates from $\kappa_{\rm th} = const$, with $\kappa_{\rm th}$ initially decreasing before reaching a minimum and then beginning to rise.  At late times the reprocessed optical luminosity approaches a fixed fraction of the engine luminosity ($\tau_{\rm therm}/\tau_{\rm T} \propto \kappa_{\rm th} \propto t^{2}$; eq.~\ref{eq:tautherm_analytic}).  The transition to a rising $\kappa_{\rm th}$ is driven by synchrotron emission becoming the dominant mode of thermalization and hence occurs earlier for larger values of the nebula magnetization, $\varepsilon_{\rm B}$.  In particular, to match the steep optical light curve decay $L_{\rm opt} \propto t^{-3.8}$ of SN2015bn out to $t \gtrsim 1000$ days, we require $\varepsilon_{\rm B} \lesssim 10^{-6}$ (Fig.~\ref{fig:15bn}).  For SN2017egm, matching the optical luminosity at $t \sim 250$ d requires $\varepsilon_{\rm B} \lesssim 10^{-4}$ (Fig.~\ref{fig:17egm}).

Stated another way, our models with $\varepsilon_{\rm B} \gtrsim 10^{-6}-10^{-4}$ significantly overpredict the late-time optical light curves of SN2015bn and SN2017egm.  If the magnetar scenario for these SLSNe (as commonly understood) is correct, then our results show that highly efficient dissipation of the wind's magnetic field must occur in the wind or nebula (near the light cylinder, the energy of the pulsar wind is carried almost entirely in Poynting flux; \citealt{Goldreich&Julian70}).  Magnetic dissipation can take place in the nebula due to non-axisymmetric instabilities arising from the dominant toroidal magnetic field geometry \citep{begelman98}; however, the relatively high residual magnetization observed in ordinary pulsar wind nebulae such as the Crab Nebula $\varepsilon_{\rm B} \gtrsim 10^{-2}$ (e.g.~\citealt{Kennel&Coroniti84,Begelman&Li92}) suggest that such a dissipation process alone may not be sufficient to reach the low values of $\varepsilon_{B}$ required to explain SLSNe.  

One mechanism for converting Poynting flux into kinetic or thermal energy is magnetic reconnection in the pulsar wind ahead of the termination shock, as may occur due to compression of the pulsar's ``striped wind'' of radially-alternating magnetic polarity (e.g.~\citealt{Lyubarsky_03}).  Supporting this possibility, \citet{Lander&Jones20} show that soon after birth, magnetars are likely to evolve to a configuration in which their rotational and magnetic dipole axes are orthogonal.  This geometry maximizes the efficiency of magnetic dissipation in the striped wind (\citealt{Komissarov13,Margalit+18b}).  

Another potentially important source of magnetic dissipation, which would not be present in older pulsar wind nebulae due to their low compactness, is the result of pair-loading due to $\gamma\gamma$ interactions (Section \ref{sec:gamma}).  Loading of the wind with relativistically hot pairs may not only act to decelerate the outflow, but also reduce the wind magnetization as a result of the electric current induced by the deposited pairs prior to their isotropization in the co-moving frame of the wind (D.~Giannios, private communication).  Though beyond the scope of the present study, we plan to explore this possibility in future work.  

Absent a viable mechanism for dissipating the nebula's magnetic field with extraordinary efficiency, we conclude that the magnetar model for SLSNe as presently envisioned may be theoretically challenged.  We encourage additional late-time optical observations of SLSNe to ascertain whether the steepening seen in SN2015bn and SN2017egm are generic and to search for the predicted late-time flattening arising from the turnover in $\tau_{\rm therm}/\tau_{\rm T}$.

\subsection{Prospects for gamma-ray detection of SLSNe}

Our models predict that high-energy gamma-ray emission will accompany SLSNe months to years after the explosion, starting at $\sim$MeV energies ($t \gtrsim 50$ days) and then moving up to $\sim$ GeV ($t \gtrsim 100$ days) and eventually $\sim$ TeV energies ($t \gtrsim 1$ yr).  

Unfortunately, the relatively low sensitivity of existing MeV gamma-ray telescopes render the predicted fluxes challenging to detect given the typically large distances of SLSNe.  Upper limits on the $0.6-600$ GeV luminosities of SLSNe by {\it Fermi} LAT were compiled by \citet{Renault-Tinacci+18}. On an individual basis these limits on Type I SLSNe, $L_{\rm LAT} \lesssim 10^{43}-10^{46}$ erg s$^{-1}$ over a 6-month window following the explosion, are in general not constraining on the luminosities predicted by our models of $\lesssim$ few $10^{42}$ erg s$^{-1}$ (Figs.~\ref{fig:15bn}, \ref{fig:17egm}, \ref{fig:15bn_spec}).
%\footnote{Tighter upper limits were placed on the gamma-ray luminosity of the particularly nearby supernova CSS140222; however, this event was classified as Type IIn and hence is more likely to be powered by circumstellar interaction than a central engine.}  

Higher energy gamma-rays $\gtrsim 100$ GeV can be observed by atmospheric Cherenkov telescopes, such as the future planned Cherenkov Telescope Array \citep{CTA19}, which can perform pointed observations that achieve deeper $\nu F_{\nu}$ sensitivity than obtained by {\it Fermi} LAT in survey mode.  However, these advantages are mitigated by the fact that the $\sim$ TeV gamma-ray emission predicted by our models only rises on a timescale of years, due primarily to the high $\gamma\gamma$ opacity of the supernova optical light (Fig.~\ref{fig:SLSNe}).  Our low-$\varepsilon_{B}$ models, which best reproduce the optical light curves of SLSNe, fortunately also possess the greatest TeV luminosities (Figs.~\ref{fig:15bn},\ref{fig:17egm}).  However, even in these best-case scenarios the predicted $\sim$TeV peak luminosities are an order of magnitude smaller than that in the $\sim$GeV band and a factor of several below the instantaneous engine luminosity at this late epoch.  

Despite these challenges, observational efforts to observe nearby and optically bright SLSNe in the gamma-ray bands on timescales of months to years are strongly encouraged.  A gamma-ray detection would represent a convincing confirmation of engine-powered models \citep{Kotera+13,Murase+15}.

\subsection{Implications for late-time spectral features}

\citet{Nicholl+16b} propose that the atypical oxygen emission line features observed in the nebular spectra of SN2015bn may be formed at the dense inner edge of the ejecta shell and arise due to excitation by radiation or energetic particles from the central engine (see also \citealt{Jerkstrand+17,Nicholl+19}). \citet{Milisavljevic+18} argue that similar features in the nebular spectra of the (atypical but non super-luminous) Type Ib SN 2012au result from photoionization of the ejecta shell by a central pulsar wind nebula.  

Our models explicitly identify what physical processes are responsible for heating the ejecta in engine-powered models.  As shown in Figure \ref{fig:15bn_ther}, at early times $t \lesssim 200$~d a sizable fraction of the optical light curve is powered by Compton thermalization, in which the upscattered electron shares a portion of its energy directly with the plasma via Coulomb scattering and another part via photoionization from secondary photons created by IC scattering off the optical radiation.  By contrast, at late times $t \gtrsim 200$~d, almost all of the thermalization is due to photoionization by synchrotron photons.  Thus, insofar as ``cosmic ray'' versus photo-ionization energy deposition will affect the ionization state evolution of the ejecta in distinct ways, our models' predictions for the temporal and radial distribution of the heating through these different channels, could serve as important input to future models of SLSNe nebular spectra.

\subsection{Late-time synchrotron radio emission}

As electrons cool in the nebula via synchrotron radiation, they will produce emission extending down into the radio bands $\lesssim 100$ GHz \citep{Murase+16,Metzger+17,Omand+18}.  The supernova ejecta typically become optically thin at GHz frequencies on a timescale of several decades after the explosion (e.g.~\citealt{Margalit+18}).  \citet{Eftekhari+19} discovered radio emission from the location of SLSN PTF10hgi at $t \approx 7.5$ yr following the explosion, which could originate from an engine-powered nebula (\citealt{Eftekhari+20} obtain upper limits on the late-time radio emission from 43 other SLSNe and long GRBs).  

Our models predict that on timescales of decades after the explosion when radio emission from the nebula would become visible, the regulation process due to $\gamma\gamma$ pair deposition in the wind has typically ended ($t \gtrsim t_{\pm}$).  Therefore, the mean energy per pair injected into the nebula at these lates time can be very large $\Delta \gamma \sim {\rm min}[\gamma_{\rm in},\gamma_{\rm rad}] \sim 10^{7}-10^{9}$ (eqs.~\ref{eq:gammapm},\ref{eq:gammarad}) for physical values of the primary Goldreich-Julian pair multiplicity of the wind $\mu_{\pm} \lesssim 10^{5}$ \citep{Timokhin&Harding19}.  Given that these high average energies result in synchrotron emission peaking in the X-ray or gamma-ray band (eq.~\ref{eq:num}), we expect that pulsar/magnetar wind nebulae may be challenged to produce the radio luminosity seen in PTF10hgi, which is a significant fraction of the total magnetar spin-down power on this timescale.
%(however, see \citealt{Omand+18}).  

Other sources of electron injection and heating, such as those which accompany mildly relativistic ejections of baryon-rich matter from magnetically powered flares and which can shock heat electrons entering the nebula to Lorentz factors $\sim 100$ in the appropriate range for radio emission, are potentially more promising for producing luminous late-time radio emission from magnetar engines \citep{Margalit&Metzger18}.

\subsection{Application to other engine powered transients}
\label{sec:other}

Although we have focused on engine and ejecta parameters appropriate to SLSNe, the general physical set-up presented may apply to other optical transients which have been considered to be powered by a millisecond pulsar/magnetar or accreting black hole central engine.  The primary requirement to preserve the qualitative picture is only that the engine generate a relativistic pair outflow behind an expanding ejecta shell, with a sufficiently high per-particle energy to activate the $\gamma\gamma$ regulation process described in Section \ref{sec:gamma}.  

As one prominent example, the optical and X-ray/$\gamma$-ray light curves of FBOTs such as AT2018cow may be powered by a magnetar or black-hole formed in a core collapse explosion characterized by ejecta significantly less massive ($M_{\rm ej} \lesssim 1M_{\odot}$) and faster ($v_{\rm ej} \gtrsim 0.1$ c) than in SLSNe (e.g., \citealt{Prentice+18,Perley+19,Margutti+19,Quataert+19}).  Likewise, the gravitational wave driven coalescence of some neutron star binaries may give rise to a long-lived millisecond magnetar remnant (e.g., \citealt{Metzger+08b,Bucciantini+11,Giacomazzo&Perna13}) which powers extended X-ray and optical emission for minutes to days following the merger.  The physical picture of the latter is qualitatively similar to the SLSNe case, arising from the interaction of the magnetar wind with the merger ejecta  (e.g.~\citealt{Yu+13,Metzger&Piro14,Siegel&Ciolfi16a}), the latter possessing an even lower mass and higher velocity ($M_{\rm ej} \lesssim 0.1M_{\odot}$; $v_{\rm ej} \gtrsim 0.5 c$) than in the FBOT case.  Yet another potential application of our model is to tidal disruption events (TDE) of stars by massive black holes (e.g., \citealt{Rees88}), in which gravitational or rotational energy released as the stellar debris is accreted by the supermassive black hole may be ``reprocessed'' by the unbound debris from the disruption, playing an important role in powering the optical light curves of TDEs (e.g., \citealt{Guillochon+14,Metzger&Stone16}) 

Throughout Section \ref{sec:preliminaries} we have attempted couch key physical quantities in terms of dimensionless parameters such as compactness, which can be readily scaled to other systems.  The peak time of optical transients is usually given by the photon diffusion time (eq.~\ref{eq:tpk}), which for fixed optical opacity scales as $t_{\rm pk} \propto M_{\rm ej}^{1/2}v_{\rm ej}^{-1/2}$.  For a magnetar engine with a spin-down time $t_{\rm e} \lesssim t_{\rm pk}$, the peak optical luminosity from Arnett's Law is $L_{\rm pk} \sim L_{\rm e}(t_{\rm pk}) \propto B_{\rm d}^{-2}t_{\rm pk}^{-2} \propto B_{\rm d}^{-2}M_{\rm ej}^{-1}v_{\rm ej}$ (eq.~\ref{eq:Le}) and hence $B_{\rm d} \propto M_{\rm ej}^{-1/2}v_{\rm ej}^{1/2}L_{\rm pk}^{-1/2}$.

With this information in hand, we can now scale the other key timescales in the problem (Table \ref{tab:timescales}) to the peak time $t_{\rm pk}$ and luminosity $L_{\rm pk}$.  This gives:
\be
\frac{t_{\gamma\gamma}}{t_{\rm pk}} \propto M_{\rm ej}^{-1/10}v_{\rm ej}^{-3/10}L_{\rm pk}^{1/5};\,\,\,\,\,\,\,\,\,\frac{t_{\rm T}}{t_{\rm pk}} \propto v_{\rm ej}^{-1/2}
\ee
\be
\frac{t_{\rm B}}{t_{\rm pk}} \propto constant;\,\,\,\,\,\,\,\,\,\,\,\,\,\,
\frac{t_{\rm c}}{t_{\rm pk}} \propto L_{\rm pk}^{1/3}M_{\rm ej}^{-1/6}v_{\rm ej}^{-5/6}
\ee
These ratios reveal that, even scaling to their faster optical rise times, FBOTs and neutron star merger transients with lower $M_{\rm ej}$ and higher $v_{\rm ej}$ will generally pass through the critical stages of evolution faster than SLSNe.  In particular, their shorter time spent at high gamma-ray compactness (smaller $t_{\gamma\gamma}$) could lead to the injected energy of the nebular pairs $\Delta \gamma$ (and their resulting gamma-ray emission) rising more rapidly in time.  Given also that the ejecta becomes Thomson thin faster (smaller $t_{\rm T}$), the engine's luminosity may begin to leak out as gamma-rays faster, causing the reprocessed optical emission after peak light to decrease faster.  Furthermore, our implicit assumption throughout this work has been that the ejecta is largely neutral and will efficiently absorb soft X-rays may also break down if the nebular radiation re-ionizes the ejecta and reduces its bound-free opacity (\citealt{Metzger+14,Metzger&Piro14,Margalit+18}), a task made easier for lower ejecta densities $\propto M_{\rm ej}/v_{\rm ej}^{3}$.  

The X-ray/gamma-ray emission observed over the first several weeks to months of AT2018cow could be attributed to synchrotron emission from a magnetar or black hole nebula escaping from a highly-ionized aspherical ejecta shell (e.g.~\citealt{Margutti+19}).  The engine-powered X-ray light curve furthermore showed a break from $L_{\rm X} \propto t^{-2}$ to $\propto t^{-4}$ at approximately 30 days after the explosion, potentially consistent with the spectral energy distribution of the nebula shifting to higher photon energies once $\gamma\gamma$ regulation of the engine's wind mass-loading ceases.  In SLSNe this transition occurs roughly on the timescale $t_{\gamma\gamma} \sim 3$ yr  (Fig.~\ref{fig:15bn_param}), so scaling to the earlier $t_{\rm pk} \sim 3$ d of AT2018cow would indeed predict a transition time of $\sim 1$ month.

\subsection{Caveats and uncertainties}

The model we have presented is subject to several uncertainties and simplifying assumptions which require clarification in future work.  Some of these assumptions, such as of isotropic emission by the energized pairs after entering the nebula, were already discussed in Section \ref{sec:setup} (see also Appendix \ref{app:coolingwind}).  

Another earlier-mentioned uncertainty is the intrinsic Goldreich-Julian pair multiplicity of young pulsar winds, $\mu_{\pm}$. However, unless the multiplicity is orders of magnitude larger than generally assumed in pulsar wind nebulae, our results are not dependent on $\mu_{\pm}$ because the wind mass-loading is dominated by secondary pairs generated by $\gamma\gamma$ interactions in the wind just ahead of the termination shock (Fig.~\ref{fig:15bn_param}).  For the same reason, the conclusions we have drawn for relativistic pulsar winds would be similar if the engine were instead the ultra-relativistic jet of an accreting black hole carrying the same power and a similarly high initial per-particle energy (though not necessarily if the engine were a mildly relativistic accretion disk wind with a comparatively low per-particle energy). 

Our calculations neglect any dynamical effects of cooling in the nebula.  Except in the case of an unmagnetized nebula, the pairs remain radiative for decades to centuries after the explosion (eq.~\ref{eq:tcool}).  Absent additional pressure support (e.g. turbulence or an ion component of the magnetar ejecta), the ram pressure of pulsar wind could compress the nebula into a thin layer, of radial thickness $\Delta R_{\rm n}/R_{\rm n} \sim v_{\rm ej}/c \sim 0.05$, potentially leading to dynamical effects or instabilities not captured by our model.  

More broadly, our calculations assume the nebula and ejecta shell are spherically symmetric.  \citet{Suzuki&Maeda20} present 2D radiative hydrodynamical simulations which show that multi-dimensional mixing of the supernova ejecta with the nebular material may aid the escape of high-energy radiation.  This could result in the gamma-ray escaping earlier than predicted by our calculations, which assume a homogeneous ejecta shell.  Polarization studies of hydrogen-poor SLSNe at early times following the explosion show little evidence for significant asphericity of the outer ejecta layers \citep{Leloudas+15,Saito+20}; however, higher polarizations have been measured during the later nebular phase revealing more aspherical inner ejecta \citep{Inserra+16,Saito+20}.  Nevertheless, the implied deviations from spherical symmetry are not likely to alter our results at the qualitative level.  

By contrast, several observations indicate that in AT2018cow the ejecta was highly aspherical, warranting more attention to this issue when expanding the techniques in this paper to the FBOT case.  The optical and X-ray light curves and spectra of AT2018cow can be understood by at least two distinct ejecta components: a fast polar outflow of velocity $\gtrsim 0.1$ c bounded by a slower equatorially concentrated shell of velocity $\lesssim 5000$ km s$^{-1}$ (e.g.~\citealt{Margutti+19}).  In particular, the lower densities and high ionization state of the fast polar outflow may have enabled soft X-rays to escape from the polar channel even at early times around optical peak, while absorption of the same central X-ray source by the dense slower equatorial ejecta was responsible for the optical emission (see also \citealt{Piro&Lu20}).

%{\bf Add more caveats here.}

%If significant dissipation were instead to occur on much smaller radial scales in the wind (e.g. \citealt{Cerutti+20}), then the high compactness in this region could convert a large fraction of the pulsar luminosity into a pair fireball.  Given the low baryon loading of the wind one would expect most of the pulsar energy to emerge as $\sim 0.1-1$ MeV photons (Pacyznski 1986, Goodman 1986, Shemi \& Piran 1990).  In this case the value of $\kappa_{\gamma}$ assumed by Nicholl et al. might be OK as the thermalization would indeed resemble that from radioactive decay products in ordinary (e.g. Type Ia) supernovae.  

\section{Conclusions}
\label{sec:conclusions}

We have presented the first radiative transfer simulations of supernova light curves powered by the rotational wind of a pulsar/magnetar that self-consistently calculate the escape of high-energy radiation from the nebula and its thermalization by the ejecta, taking into account a wide range of physical processes responsible for exchanging energy between the matter and radiation fields.  

Our results can be summarized as follows:
\begin{itemize}
    \item While some energy from the central engine may be transferred directly to the ejecta at early times after the explosion (e.g., via mechanical work of a shock driven into the ejecta by the nebula; \citealt{Kasen+16}), the bulk of the optical radiation from SLSNe around after peak light must be powered by reprocessing of nonthermal radiation from the nebula by the supernova ejecta.
    
    Several processes in the ejecta are capable of absorbing high-energy photons (Fig.~\ref{fig:taueff}), but an initial absorption of energy is no guarantee of its ultimate thermalization.  Thermalization requires reprocessing the engine's luminosity into low-energy photons (which readily share their energy with the thermal pool by photoelectric absorption or Compton down-scattering) or low-energy particles (which readily share their energy with the ejecta by Coulomb scattering).  This behavior cannot be captured by a single constant thermalization opacity for the nonthermal photons.  
    
    \item Gamma-rays are produced in the nebula by a combination of IC scattering and synchrotron emission, which dominate the nebula luminosity at early and late-times, respectively (the cross-over time $t \sim t_{\rm B}$; eq.~\ref{eq:tB}). Synchrotron emission tends to produce lower energy photons, which are more readily absorbed and thermalized by the ejecta, while IC emission produces higher energy gamma-rays which more readily escape without thermalizing.  The relative importance of these processes is sensitive to the nebula magnetization $\varepsilon_{B}$, which turns out to be the most important free variable in the problem (once the engine and ejecta properties have been chosen to match the optical light curve near maximum).

    \item Except in the case of very low magnetization $\varepsilon_{B} \lesssim 10^{-6}$, the nebula remains radiative for years to decades or longer following the explosion, i.e. the bolometric output tracks the spin-down power of the pulsar.  However, the spectral energy distribution of the nebular radiation depends sensitively on the mean per-particle energy $\Delta \gamma \simeq L_{\rm e}/(\dot{N}_{\pm} m_e c^{2})$ pairs gain upon entering the nebula from the pulsar wind.  The latter depends on ratio of the wind luminosity to the pair-loading rate, $\dot{N}_{\pm}$.  
    
    Although the contribution to $\dot{N}_{\pm}$ that arises directly from the inner magnetosphere is uncertain, this contribution is subdominant.  Over the first several years after the explosion, $\dot{N}_{\pm}$ is dominated by pairs generated by $\gamma\gamma$ interactions in the upstream wind region.  We identify a new feedback mechanism between the nebula radiation and $\gamma\gamma$ pair creation in the wind that regulates the mean energy of the pairs entering the nebula.  At very early and late times, nonthermal photons generally serve as the targets for pair production, while the targets are thermal at intermediate times when the ejecta is becoming transparent to the highest energy gamma-rays (Table \ref{tab:regulation}, Appendix \ref{app:regulation}). This self-regulation process has the benefit of rendering our results insensitive to the uncertain intrinsic pair-multiplicity of the wind.
    
    On the other hand, the reduced per-particle energy in the pulsar wind due to $\gamma\gamma$ mass-loading (occurring over timescales during which most of the spin-down power is released) may also negatively impact the efficacy of baryon cosmic ray acceleration and associated high-energy neutrino emission in very young magnetar winds (e.g., \citealt{Arons03,Murase+15,Kotera+15,Fang&Metzger17,Fang+19}).  However, a quantitative exploration of the implication of our results for ion acceleration will require a more detailed model for the dissipation in the wind.   
    
    \item For the first several months after the explosion, including and following optical maximum, the ejecta is opaque across all photon energies.  During this phase the engine's luminosity is efficiently reprocessed into thermal optical radiation, independent of any assumptions about the nebula.  This provides a rigorous justification for the assumption of many previous works which calculate detailed optical/UV light curves and spectra of magnetar-powered SLSNe around and following peak light by depositing 100\% of the magnetar's spin-down luminosity as thermal energy at the inner edge of the ejecta (e.g.~\citealt{Kasen&Bildsten10,Dessart+12,Inserra+13,Metzger+15,Nicholl+17}).  Furthermore, these models can indeed be used to obtain accurate constraints on the engine properties, such as the dipole magnetic field strength and birth spin period, provided that the engine lifetime $t_{\rm e}$ (i.e. magnetar dipole spin-down time) is shorter than a few months.
    
    \item Gamma-rays leak out of the ejecta starting months after the explosion, beginning in the $\sim 0.1-100 $ GeV band accessible to {\it Fermi LAT} and then moving up to the $\gtrsim 100$ GeV band accessible to atmospheric Cherenkov telescopes.  The shift in the gamma-ray spectra to higher energies is driven by a combination of the rising of mean particle energy $\Delta \gamma$ entering the nebula as the $\gamma\gamma$ pair-loading of the pulsar wind subsides and the opacity window through the nebula and ejecta expands to higher photon energies (both processes being driven by the decreasing $\gamma\gamma$ optical depth, in the pulsar wind and ejecta, respectively).
    
    The escaping gamma-ray luminosity is most sensitive to the nebula magnetization, with lower $\varepsilon_{B}$ producing higher peak gamma-ray luminosities and, conversely, lower late-time optical luminosities.  This is due to the tendency of more magnetized nebulae to radiate a greater fraction of the engine's energy in low-frequency synchrotron radiation, which is more readily absorbed by the ejecta than the higher frequency IC emission. 
    
    \item Reproducing the steep post-maximum decays in the optical light curves of SN2015bn and SN2017egm requires weakly magnetized nebulae, $\varepsilon_{B} \lesssim 10^{-6}-10^{-4}$.  These magnetizations are much lower than inferred from other young pulsar wind nebula like the Crab Nebula.  However, the physical conditions in the extremely young, rapidly spinning and highly magnetized pulsars/magnetars considered here, may substantially differ from those of older pulsar nebulae frequently observed in the Milky Way.  The particular impact of $\gamma\gamma$ pair loading on potentially reducing the magnetization of the ultra-relativistic pulsar/magnetar wind feeding the nebula, merits further study.    
    
    \item Alternatively, if such a low nebula magnetization is deemed to be unphysical, our results suggest that the magnetar model as currently envisioned may be an incorrect or at least incomplete explanation for SLSNe.  If the true luminosity of the central engine were to decrease faster in time (e.g., $L_{\rm sd} \propto t^{-3.8}$ in SN2015bn) than the canonical $\propto t^{-2}$ magnetic dipole spin-down rate (e.g., due to a growing magnetic dipole moment or the effects of fall-back accretion on the magnetar wind; \citealt{Metzger+18b}), then even a sustained high level of thermalization (as achieved in our high $\varepsilon_B$ models) would be consistent with the late-time optical light curve data.  In such a model there would also be no heretofore unobserved large escaping gamma-ray flux.  
    
    Absent such alternatives, other models such as CSM shock interaction or mildly-relativistic accretion disk outflows, could supplant ultra-relativistic magnetar winds or black hole jets as the favored engines of SLSNe.
    
    \item  Our models predict a late-time flattening in the optical light curve, once synchrotron overtakes IC emission in the nebula on a timescale $\sim t_{\rm B}$ (eq.~\ref{eq:tB}) and the effective thermalization opacity begins to rise (Fig.~\ref{fig:15bn_ther}).  Hints of such flattening behavior are evident in SN 2015bn (Fig.~\ref{fig:15bn}).
    
    \item A definitive confirmation of the central engine model for SLSNe would come from the detection of the leaking nebular gamma-rays.  However, this is an observational challenge due to the limited sensitivity of gamma-ray telescopes and the delayed onset of the gamma-ray flux, which reduce their luminosity relative to the optical peak.  Nevertheless, our models predict the expected gamma-ray emission given as the output of models which can self-consistently reproduce the observed optical light curve.  A detection may be possible for a particularly nearby future SLSNe with {\it Fermi} LAT at $\sim$ GeV energies, or present and future atmospheric Cherenkov telescopes such as the CTA at $\sim$ TeV energies. 
    
    \item Although our models are focused on SLSNe, the general scenario we have outlined could have broader applicability to other engine-powered transients such as FBOTs, tidal disruption events, and post-merger counterparts of neutron star mergers.  All else being equal, the prospects may be better for detecting the escaping high-energy emission from these events due to the shorter timescales over which the ejecta becomes transparent to gamma-rays (Section \ref{sec:other}).  Indeed, the X-ray/gamma-ray emission from a magnetar or black hole powered nebula may have already been detected in AT2018cow (e.g., \citealt{Margutti+19}).
     
\end{itemize}

%Figure \ref{fig:F_esc} shows the escaping nebular flux as a function of time, calculated for the sample of SLSNe introduced in Section \ref{sec:SLSNe}.  Given large uncertainties in the nebular spectral energy distribution, each panel corresponds to a different assumed energy range, under the assumption that all of the engine luminosity is emitted in that band.  Shown for comparison are the approximate sensitivities of representative current or future X-ray/gamma-ray telescopes, such as NuSTAR, {\it Fermi} LAT, AMEGO, and CTA.  

%\begin{figure*}
%\includegraphics[trim = 2cm 2cm 2cm 2cm, width=1.0\textwidth]{F_esc_atten_4panel_100_GeV.eps}
%\includegraphics[width=1.0\textwidth]{F_esc_atten_4panel_100_GeV.eps}}
%\vspace{-0.4cm}
%\caption{In the final panel we show two cases, corresponding to different models for the assumed temperature of the target optical background radiation.  }
%\label{fig:F_esc}
%\end{figure*}

\bigskip

We thank Ke Fang, Keiichi Maeda, Matt Nicholl, and Akihiro Suzuki for providing helpful information and comments on an early draft of the paper.  IV acknowledges support by the ETAg grant PRG1006 and by EU through the ERDF CoE grant TK133.  BDM acknowledges support from the NASA Astrophysics Theory Program (grant number NNX17AK43G), Fermi Guest Investigator Program (grant number GG016287) and the NSF through the AAG Program (grant number GG016244).

\appendix

\section{Radiative diffusion with incoherent scattering}
\label{sec:app:diffusion}

In this Appendix we provide an approximate analytical prescription for computing the fraction of the energy injected into the ejecta by the central engine that emerges without significant attenuation. Neglecting emission from secondary pairs generated by the high-energy photons, the main remaining nontrivial issue in the problem is the characterization of incoherent (Compton) scattering which degrades the photon energy without destroying it.

To keep the problem simple while retaining the salient features, we treat Compton recoil as an energy sink rather than a frequency redistribution mechanism. All spectral information is lost in this approach, which however is acceptable for the present purpose. This simplification allows us to consider radiative transfer independently at each frequency (without coupling between frequencies).  The equation of radiative diffusion is given by
\begin{align}
\frac{dI}{ds} = -\alpha I + j + (1-\kappa_{\rm C}) \sigma J, 
\label{eq:RTE}
\end{align}
where $I$ is the specific intensity, $\alpha$ is the absorption coefficient, $j$ is the emissivity, $\sigma$ is the scattering coefficient, $J = (4\pi)^{-1}\int_{4\pi} I d\Omega$ is the angle-averaged intensity,
$\kappa_{\rm C}$ is the fraction of the photon energy lost to recoil in a scattering event. 
Since we are considering high-energy radiation in an essentially cold medium, the emissivity $j$ can be neglected compared with the last term in Equation (\ref{eq:RTE}).

Rewriting the transfer equation in spherical coordinates and taking the first two angular moments (where $\mu \equiv \cos \theta$) yields
\begin{align}
&\frac{1}{r^2} \frac{\partial}{\partial\tau}\left( r^2  H \right) = -\varepsilon J 
\label{eq:mom1} \\
&\frac{\partial K}{\partial \tau} - \frac{1}{\tau} \left( J - 3K \right) = -H
\label{eq:mom2}
\end{align}
where $H = (4\pi)^{-1}\int_{4\pi} I \mu d\Omega$ and $K = (4\pi)^{-1}\int_{4\pi} I \mu^2 d\Omega$ are the first and second moments of the intensity, respectively, $\tau = (\alpha + \sigma) r $, and
\begin{align}
\varepsilon = \frac{\alpha+\kappa_{\rm C}\sigma}{\alpha+\sigma}    
\end{align}
is defined such that $1-\varepsilon$ is the effective single-scattering albedo. In this simplified treatment $\epsilon$ is the average fraction of the photon energy lost (either by absorption or recoil) in a single interaction.

Following standard treatments (e.g. \citealt{Rybicki&Lightman}) we use the Eddington approximation $K \approx J/3$ to close the system of equations (\ref{eq:mom1}) - (\ref{eq:mom2}) and obtain a diffusion equation for the angle-averaged intensity
\begin{align}
\frac{\partial^2 J}{\partial \tau_{\star}^2} + \frac{2}{\tau_{\star}}\frac{\partial J}{\partial \tau_{\star}} - J = 0,
\label{eq:diff}
\end{align}
where $\tau_{\star} = \sqrt{3\varepsilon} \, \tau$. Equation (\ref{eq:diff}) can be solved by standard methods, yielding
\begin{align}
J = C_1 \frac{e^{\tau_{\star}}}{\tau_{\star}} + C_2 \frac{e^{-\tau_{\star}}}{\tau_{\star}}.
\end{align}
The integration constants $C_1$ and $C_2$ can be found by prescribing the flux at the inner boundary $r_{\rm in}$ and requiring that no radiation impinges on the outer boundary from above. The escaping flux can then be obtained from Equation (\ref{eq:mom2}) (again using using $J = K/3$), $H = - (1/3) \partial J/\partial\tau$.
The ratio of escaping energy to that input at the center of the spherical cloud is given by
\begin{align}
    \frac{r_0^2 H(r_0)}{\left. r^2_{\rm in} H(r_{\rm in}) \right|_{r_{\rm in} \rightarrow 0}} = \frac{2\tau^2_{\star, 0}}{[ (\sqrt{\epsilon} - 1) \tau_{\star, 0} + \sqrt{\epsilon}] \, e^{-\tau_{\star, 0}} + [(\sqrt{\epsilon} + 1) \tau_{\star, 0} - \sqrt{\epsilon}] \, e^{\tau_{\star, 0}}},
\label{eq:escfrac}    
\end{align}
where $\tau_{\star, 0}$ is the effective optical thickness of the cloud. Using the appropriate absorption and scattering opacities (see Section \ref{sec:absorption}) in Equation (\ref{eq:escfrac}) yields the approximate fraction of energy that escapes without significant attenuation/reprocessing, for any given input photon energy.  Figure \ref{fig:taueff} compares the effective opacity versus photon energy derived from this result (dashed lines) compared to the result of a full Monte Carlo calculation (solid lines), demonstrating reasonable agreement at high values of $\tau_{\rm T} \gg 1$.

\section{Pair cooling timescales}
\label{sec:app:cooling}

The photon absorbing processes described in Section \ref{sec:absorption} result in the generation of energetic electrons or positrons, which cool and emit secondary radiation over a range of frequencies. The efficiency of this energy reprocessing depends on the ability of pairs to cool efficiently, i.e. faster than the expansion time; the relative importance of the different cooling mechanisms
depends on the radiation and matter field densities and has an impact on both the high-energy spectral evolution as well as the fraction of energy that ultimately gets thermalized.  This Appendix covers the cooling rates of relativistic leptons of Lorentz factor $\gamma$ due to different processes and compares the corresponding timescales with each other as well as the dynamical time $t_{\rm dyn} = R_{\rm ej}/v_{\rm ej}$ over which adiabatic losses occur due to the expanding ejecta/nebula.

\begin{figure}
\includegraphics[width=1.0\textwidth]{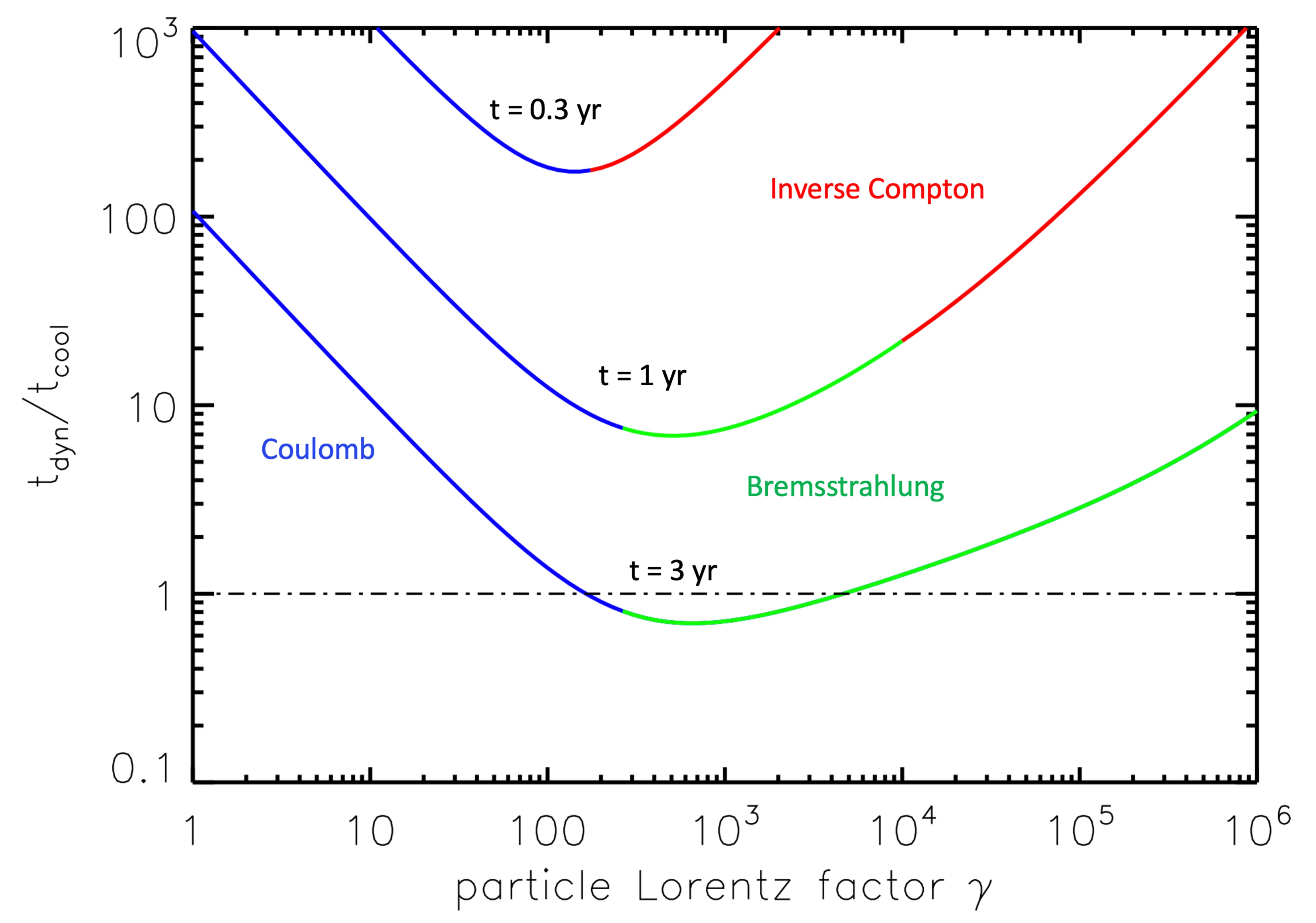}
%\includegraphics[width=1.0\textwidth]{F_esc_atten_4panel_100_GeV.eps}}
%\vspace{-0.4cm}
\caption{Ratio of the dynamical timescale to cooling timescale in the supernova ejecta as a function of electron/positron Lorentz factor $\gamma$, shown at three epochs after explosion ($t =$ 0.3 yr, 1 yr, 3 yr as marked).  The dominant cooling process for each range of $\gamma$ is denoted by the line color as marked.  We have assumed magnetar and ejecta parameters corresponding to SN2017egm (Table \ref{tab:models}).  The key point of this figure is to illustrate that relativistic electron/positrons are fast-cooling in the ejecta across all relevant particle energies for several years after the explosion. }
\label{fig:tcool}
\end{figure}

For relativistic bremsstrahlung emission, the particle cooling rate is $\dot{\gamma}_{\rm br} \approx 5c\alpha_{\rm fs} \tauT \gamma^{1.2} (\overline{Z^2}/\overline{Z} + 1)/(6R_{\rm ej})$ (\citealt{Vurm&Metzger18}). The ratio of the expansion to cooling timescales is
\begin{eqnarray}
\frac{t_{\rm dyn}}{t_{\rm br}} &=& \frac{\dot{\gamma}_{\rm br}}{\gamma} \frac{R_{\rm ej}}{v_{\rm ej}} =
\frac{5c \alpha_{\rm fs} \tauT \, \gamma^{0.2}}{6 v_{\rm ej}} \left( \frac{\overline{Z^2}}{\overline{Z}} + 1 \right) %\nonumber \\ 
%&\approx& 
\approx
6.2 \gamma_3^{0.2} M_{10}v_{9}^{-3} t_{\rm yr}^{-2}
\label{eq:t_brem}
\end{eqnarray}
where in the final step $\gamma_3 \equiv \gamma/10^3$ and we have assumed oxygen-rich ejecta, $\overline{Z} = (\overline{Z^{2}})^{1/2} = 8$.

For IC cooling in the Thomson regime ($\gamma \lesssim 10^{6}$), for which the particle cooling rate is
$\dot{\gamma}_{\rm IC} = 4c \ell_{\rm th} \gamma^2/(3R_{\rm ej})$, we obtain
\be
\frac{t_{\rm dyn}}{t_{\rm IC}} = \frac{\dot{\gamma}_{\rm IC}}{\gamma} \frac{R_{\rm ej}}{v_{\rm ej}} =
\frac{4 \, c \, \ell_{\rm th} \gamma}{3 v_{\rm ej}}
= 0.88 \, \gamma_3 M_{10}v_{9}^{-4}B_{14}^{-2}t_{\rm yr}^{-5}
\label{eq:t_IC}
\ee
where $\ell_{\rm th}$ is the thermal compactness (eq.~\ref{eq:ellopt}) and in the final step we have assumed $\kappa_{\gamma}/\kappa_{\rm es} = 0.1$ and $\tau_{\rm opt} \le 1$.\footnote{The target radiation for IC cooling should in fact include all photons in the Thomson regime, i.e. with $x \lesssim 1/\gamma$ including UV/X-rays.  However, as those photons are efficiently photoelectrically absorbed by the ejecta, the error made by using just the thermal radiation density is modest.}

Finally, cooling due to Coulomb scattering $\dot{\gamma}_{\rm Coul} = 3c\ln\Lambda \tauT/(2R_{\rm ej})$, results in
\be
\frac{t_{\rm dyn}}{t_{\rm Coul}} = \frac{\dot{\gamma}_{\rm Coul}}{\gamma} \frac{R_{\rm ej}}{v_{\rm ej}} =
\frac{3\ln\Lambda \, c \, \tauT}{2 \, v_{\rm ej} \, \gamma} = 
1.1\gamma_{3}^{-1} M_{10} v_{9}^{-3}t_{\rm yr}^{-2},
%3.6\gamma_{3}^{-1} M_{10} v_{9}^{-3}t_{\rm yr}^{-2},
\label{eq:t_Coul}
\ee
where in the final step we have taken $\ln\Lambda \approx 25$ for the Coulomb logarithm. As long as the sum of (\ref{eq:t_brem}) - (\ref{eq:t_Coul}) exceeds unity for a given set of parameters,
the high-energy electron/positron is in the fast-cooling regime and its adiabatic losses can be neglected in the lowest order.
%{\bf [[[Plot cooling timescales for some characteristic values of $\tauT$ and $\ell_{\rm th}$]]]}

%One can also express the critical Lorentz factors at which processes have (pair-wise) identical cooling rates, one or the other becoming dominant at higher/lower $\gamma$.
The relative rates of bremsstrahlung and Coulomb cooling can be expressed as
\be
\frac{\dot{\gamma}_{\rm br}}{\dot{\gamma}_{\rm Coul}} =
\left( \frac{\gamma}{\gamma_{\star}} \right)^{1.2}, \label{eq:coolingratesratio}
\ee
where
\be
\gamma_{\star} =
\left[ \frac{9\ln\Lambda}{5\,\alpha_{\rm fs} (\overline{Z^2}/\overline{Z} + 1)} \right]^{0.83}
%\left( \frac{9\ln\Lambda}{10\,\alpha_{\rm fs}} \right)^{0.83}
\approx 810 \left( \frac{\overline{Z^2} + \overline{Z}}{2\overline{Z}}\right)^{-0.83}
\approx 230
\label{eq:gstar}
\ee
and in the last step we have again assumed oxygen-dominated composition.

Similarly, for bremsstrahlung and IC one obtains
\be
\frac{\dot{\gamma}_{\rm IC}}{\dot{\gamma}_{\rm br}} = \left( \frac{\gamma}{\gamma_{\star\star}} \right)^{0.8},
\ee
where
\be
\gamma_{\star\star} = \left[ \frac{5 \tauT \, \alpha_{\rm fs} (\overline{Z^2}/\overline{Z} + 1)}{8 \ell_{\rm th}} \right]^{1.25}
%\gamma_{\star\star} = \left( \frac{5 \tauT \, \alpha_{\rm fs}}{4 l_{\rm th}} \right)^{1.25}
\approx 8.9\times 10^4 \, \ell_{\rm th, -6}^{-1.25} \, \tauT^{1.25}  \left( \frac{\overline{Z^2} + \overline{Z}}{2\overline{Z}}\right)^{1.25}
\approx 5.8\times 10^5 \, \ell_{\rm th, -6}^{-1.25} \, \tauT^{1.25}
\label{eq:gammastarstar}
\ee
and $\ell_{\rm th, -6} = \ell_{\rm th}/10^{-6}$.

Finally, the ratio of IC and Coulomb cooling rates is
\be
\frac{\dot{\gamma}_{\rm IC}}{\dot{\gamma}_{\rm Coul}} = \left( \frac{\gamma}{\gamma_{\dagger}} \right)^{2}, \label{eq:coolingratesratio2}
\ee
where
\be
\gamma_{\dagger} = \left( \frac{9 \ln\Lambda \,\tauT}{8\, \ell_{\rm th}} \right)^{1/2}
\approx 5.3\times 10^3 \, \tauT^{1/2} \, l_{\rm th, -6}^{-1/2}.
\label{eq:gdagger}
\ee
Figure \ref{fig:tcool} shows the ratio of the dynamical timescale to the total cooling timescale in the supernova ejecta as a function of $\gamma$ at three epochs after explosion ($t =$ 0.3 yr, 1 yr, 3 yr) as calculated from the above estimates for engine/ejecta parameters corresponding to SN2017egm.  At all epochs Coulomb losses dominate the energy loss of low-$\gamma$ leptons, while the high-$\gamma$ electrons cool through primarily IC emission at early times and via bremsstrahlung at later times.  For all electron energies $\gamma$ adiabatic losses can be neglected for the first several years after the explosion (i.e. $t_{\rm dyn}/t_{\rm cool} > 1$).  Furthermore, at even later times most of the thermalization occurs due to photoelectric absorption of the nebular synchrotron radiation (Fig.~\ref{fig:15bn_ther}) which does not involve the production of relativistic electrons.

\section{Radiative cooling in the pulsar wind}
\label{app:coolingwind}

This Appendix addresses whether the relativistically hot pairs deposited by $\gamma\gamma$ interactions in the pulsar/magnetar wind (Section \ref{sec:gamma}) are able to cool radiatively before reaching the termination shock, or whether they will enter the nebula with their thermal energy intact and hence emit the engine luminosity isotropically as assumed in our Monte Carlo simulations (Section \ref{sec:setup}).

Consider an electron or positron at the termination shock with an energy $\gamma_0$ that emits IC photons on thermal target radiation of temperature $\theta$. Well after the optical peak the self-regulation mechanism of pair loading generally ensures that $4\theta\gamma_0 > 1$ (Section \ref{sec:gamma}), so that most of the IC power is radiated near $x \approx \gamma_0$. A fraction of the upscattered photons propagates back into the unshocked wind and produces pairs via $\gamma\gamma$ interactions on the thermal radiation. The average energy of the generated electron-positron pair is $\gamma_{\pm, 0} \approx x/2 \approx \gamma_0/2$. If the pulsar wind is relativistic (bulk radial Lorentz factor $\Gamma\gg 1$) and the generated electron/positron has time to isotropize in the wind frame, its energy in the lab frame after being picked up by the wind is %$\gamma_{\pm} \approx \Gamma^2 \gamma_{\pm, 0} \approx \Gamma^2\gamma_0/2.
\begin{align}
\gamma_{\pm} \approx \Gamma\gamma_{\pm}^{\prime} \approx \Gamma^2 \gamma_{\pm, 0} \approx \frac{\Gamma^2\gamma_0}{2} \approx \frac{\Gamma^2 x}{2},
\label{eq:gpm}
\end{align}
where a prime denotes the wind rest frame.

First consider IC cooling.  The IC cooling time within the ambient thermal radiation field is (in the lab frame)
\begin{align}
    t_{\rm IC} = \frac{3R_{\rm t}}{4c \ell_{\rm th, w} \gamma_{\pm} F_{\rm KN}}, % \left( 1 + \frac{\ell_{B}}{\ell_{\rm th} F_{\rm KN}} \right)^{-1}
    \label{eq:tIC}
\end{align}
where $\ell_{\rm th, w}$ is the thermal compactness in the pulsar wind (see below) and 
\begin{align}
F_{\rm KN} \approx \frac{9}{8(4\theta\gamma_{\pm})^2} \left[\ln(4\theta\gamma_{\pm}) - \frac{11}{6}\right].
\label{FKN}
\end{align}
is the Klein-Nishina correction factor for $4\theta\gamma > 1$.

Using equations (\ref{eq:gpm}) and (\ref{FKN}), one can write 
\begin{align}
    t_{\rm IC} = \frac{R_{\rm t}}{3c\ell_{\rm th, w} x} \Gamma^2 \frac{(4\theta\gamma_0)^2}{\ln(4\theta\gamma_{\pm}) - 11/6}.
    \label{eq:tICb}
\end{align}

The cooling time (\ref{eq:tICb}) should be compared with the time it takes for the generated pair to be advected back to the shock. On average, a high energy photon will propagate a distance $\Delta = \min(R_{\rm t},R_{\rm t}/\tau_{\gamma\gamma, {\rm th}})$ back into the wind zone before producing a pair, so the average advection time is $\Delta/c$. The pair production opacity is
\begin{align}
\tau_{\gamma\gamma, {\rm th}}
\approx \ell_{\rm th, w} x \, \frac{\ln(1 + x\theta)}{5.4 \, (x\theta)^2} \exp\left(-\frac{0.9}{\theta x}\right).
\label{eq:taugg2}
\end{align}
Substituting $\ell_{\rm th, w} x$ from Equation (\ref{eq:taugg2}) back into Equation (\ref{eq:tIC}), using Equation (\ref{eq:gpm}) and rearranging terms, we obtain
\begin{align}
    \frac{t_{\rm IC}}{t_{\rm adv}} \approx \Gamma^2 \frac{\ln(1+\theta\gamma_0) \, e^{-0.9/\theta\gamma_0}}{\ln(4\theta\gamma_0) + \ln(\Gamma^2/2) - 11/6},
    \label{eq:tIC2}
\end{align}
where
\begin{align}
t_{\rm adv} = \frac{\Delta}{c} = \frac{R_{\rm t}}{c\tau_{\gamma\gamma, {\rm th}}}
\end{align}
and we have assumed that $\tau_{\gamma\gamma, {\rm th}} > 1$. 

As long as $4\theta\gamma_0 > 1$, the last fraction in Equation (\ref{eq:tIC2}) is comparable to unity, so we conclude that in a relativistic wind the generated pairs do not have time to cool by inverse Compton emission before being advected back to the shock.

Now consider synchrotron cooling.  The lab-frame synchrotron cooling time is
\begin{align}
    t_{\rm syn} = \Gamma t_{\rm syn}^{\prime} = \Gamma \frac{3m_{\rm e} c }{4\sigma_{\rm T} U_B^{\prime} \gamma_{\pm}^{\prime}} =
    \frac{3m_{\rm e} c \Gamma^2}{4\sigma_{\rm T} U_B \gamma_{\pm, 0}} = \frac{3 \Gamma^2}{4 \ell_{B, {\rm w}} \gamma_{\pm, 0}} \frac{R_{\rm t}}{c},
    \label{eq:tsyn}
\end{align}
where again $\gamma_{\pm, 0}$ is the energy of the created pair before being picked up by the wind. The ratio of cooling and advection times is
\begin{align}
\frac{t_{\rm syn}}{t_{\rm adv}}
= \frac{3\Gamma^2}{4 \ell_{B, {\rm w}} \gamma_{\pm, 0}}
\ell_{\rm th, w} x \, \frac{\ln(1 + x\theta)}{5.4 \, (x\theta)^2} \exp\left(-\frac{0.9}{\theta x}\right)
\approx \frac{9}{2} \frac{\ell_{\rm th, w}}{\ell_{B, {\rm w}}} \Gamma^2 \frac{\ln(1+\theta\gamma_0) \, e^{-0.9/\theta\gamma_0}}{(4\theta\gamma_0)^2},
\end{align}
where again $\tau_{\gamma\gamma, {\rm th}} > 1$ has been assumed.

In analogy with Equation (\ref{eq:lBn}), the magnetic compactness in the pulsar wind is defined as
\begin{align}
\ell_{B, {\rm w}} = \frac{\sigma_{\rm T} U_B R_{\rm t}}{m_{\rm e} c^2}
= \ell_{\rm inj} \frac{\sigma}{\sigma + 1} \frac{R_{\rm ej}}{R_{\rm t}},
\label{eq:ellBw}
\end{align}
where $U_B = B_{\rm w}^2/8\pi$, and the magnetic field and injection compactness are (eq.~\ref{eq:Bw})
\begin{align}
B_{\rm w} = \left[\frac{2 L_{\rm e} \sigma}{(\sigma + 1)c R_{\rm t}^2}\right]^{1/2}
\end{align}
and
\begin{align}
\ell_{\rm inj} = \frac{\sigma_{\rm T}L_{\rm e}}{4\pi m_{\rm e} c^3 R_{\rm ej}},
\end{align}
respectively. The thermal compactness in the wind zone is
\begin{align}
\ell_{\rm th, w}
= \frac{\sigma_{\rm T} R_{\rm t} u_{\rm th, w}}{m_{\rm e} c^2}
\approx \frac{\sigma_{\rm T}  R_{\rm t}}{m_{\rm e} c^2} \frac{L_{\rm e} f_{\rm th} (1+\tau_{\rm opt})}{4\pi c R_{\rm t}^2} = \ell_{\rm inj} \, f_{\rm th} \, (1+\tau_{\rm opt}) \frac{R_{\rm ej}}{R_{\rm t}}.
\end{align}

Using the above definitions, the ratio of the synchrotron cooling and advection times becomes (again for $\tau_{\gamma\gamma, {\rm th}} > 1$)
\begin{align}
\frac{t_{\rm syn}}{t_{\rm adv}}
\approx \frac{9}{2}
\frac{\sigma+1}{\sigma}
f_{\rm th} \, (1+\tau_{\rm opt}) \Gamma^2 \,
\frac{\ln(1+\theta\gamma_0) \, e^{-0.9/\theta\gamma_0}}{(4\theta\gamma_0)^2}.
\label{eq:tsyn2}
\end{align}
Given that pair creation in the wind regulates $\theta\gamma_0 \sim$~few (Section \ref{sec:gamma}), we see that even for a strongly magnetized wind $\sigma \gg 1$, synchrotron cooling is negligible ($t_{\rm syn} \gg t_{\rm adv}$) provided that the wind is relativistic $\Gamma \gtrsim f_{\rm th}^{-1/2} \sim 10-100$ by the radii at which substantial pair loading takes place.

At late times when $\tau_{\gamma\gamma, {\rm th}} < 1$ and $t_{\rm adv} = R_{\rm t}/c$, the critical ratio instead becomes
\begin{align}
\frac{t_{\rm syn}}{t_{\rm adv}} = \frac{3 \Gamma^2}{4 \ell_{B, {\rm w}} \gamma_{\pm, 0}}
= \frac{3 \Gamma^2}{2 \ell_{\rm inj} \gamma_{0}} \frac{\sigma+1}{\sigma} \frac{R_{\rm t}}{R_{\rm ej}},
\label{eq:tsyn3}
\end{align}
which, by definition, must be smaller than the value given by Equation (\ref{eq:tsyn2}).  In this case, synchrotron cooling is only negligible if the wind remains highly relativistic and/or the wind magnetization has dropped to low values by the time it reaches the pair loading zone.  However, once $\tau_{\gamma\gamma, {\rm th}} < 1$ the wind will no longer be heavily pair-loaded anyways and hence a much smaller fraction of the spin-down power will go into heating pairs over an extended region.

Future work is needed to study the impact of $\gamma\gamma$ pair loading on the radial evolution of the Lorentz factor and magnetization of the pulsar wind (see \S\ref{sec:discussion}) and to assess the conditions required for particle cooling as described in this section within such a self-consistent framework.  However, the results derived here make it plausible that radiative losses in the wind may be neglected to first order, supporting the assumptions of our simulations (Section \ref{sec:setup}).

\section{Regulation of the particle number and mean particle energy in the wind dissipation region}
\label{app:regulation}

Upon entering the nebula, a heated lepton cools rapidly by both IC scattering on thermal photons and synchrotron emission (Section \ref{sec:nebula}), with a fraction $\xi \approx 1/2$ of the IC upscattered photons propagating back into the wind region.  The probability that the IC upscattered photon generates a secondary pair is
$1-e^{-\tau_{\gamma\gamma}}$.
%$1-e^{-\tauggth}$.
Putting this together with Equation (\ref{eq:regulation}) yields the following condition:
\begin{align}
    {\cal M}_{\rm e} = 2\xi \int \left.\frac{dN_{\rm ph}}{d\ln{x}}\right|_{\rm 1 el.}
%    (1-e^{-\tauggth})
    (1-e^{-\tau_{\gamma\gamma}})
    \, d\ln{x} = 1,
    \label{eq:Me}
\end{align}
where the factor $2$ accounts for the two leptons created in a single pair-production event, and $dN_{\rm ph}/d\ln{x}$ is the spectrum emitted by a single electron over its cooling history.

To make further progress on must specify the spectrum emitted by the electron as it cools down from its initial $\gamma_0 \approx \Delta\gamma$, as well as the source of $\gamma\gamma$ opacity. It is useful to discuss separately the cases when the opacity is provided by nonthermal or thermal target photons, as well as to separate the cases based to the physical mechanisms giving rise to the high-energy and target photons.  Results for these different cases are summarized in Table \ref{tab:regulation}.

\subsection{Nonthermal pair loading}

\subsubsection{IC photons + IC targets}

Close to the optical peak the injection compactness $\ell_{\rm inj} \sim 0.1 - 1$ for typical SLSN parameters  (eq.~\ref{eq:ellinj}), hence the radiation compactness generated by the cooling leptons is also close to unity. The opacity for pair production on the nonthermal radiation field (eq.~\ref{eq:tauggnth}) is therefore non-negligible and is responsible for pair loading at the early stages of the event. In this high compactness phase efficient pair loading within the wind keeps the average energy per lepton sufficiently low so that their IC cooling on the (energetically dominant) thermal target radiation field takes place in the Thomson regime. The pair loading is dominated by interactions between thermal photons upscattered into the MeV domain and slightly lower-energy ($\lesssim 1$~MeV) IC photons.

For the present purpose it is sufficient to employ a simple delta-function approximation for the IC spectrum of a single electron of energy $\gamma$ \citep{Vurm&Metzger18}
\begin{align}
\left.\frac{dN_{\rm ph}}{d\ln{x} \, dt}\right|_{\rm IC, 1 el.} = \dot{\gamma}_{\rm IC} \, \delta\left( x - 4\theta\gamma^2 \right).
\label{eq:elIC1}
\end{align}
The spectrum emitted by the electron as it cools down from its initial $\gamma_0 \approx \Delta\gamma$
is obtained by integrating Equation (\ref{eq:elIC1}) over time $dt = d\gamma/\dot{\gamma}$, yielding
\begin{align}
\left.\frac{dN_{\rm ph}}{d\ln{x}}\right|_{\rm IC, 1 el.} = \frac{1}{2(4\theta x)^{1/2}}, \hspace{5mm} x \le 4\theta\gamma_0^2,
\label{eq:elICcool1}
\end{align}
where we have assumed $\dot{\gamma} \approx \dot{\gamma}_{\rm IC}$.
The deposited (but not necessarily escaping) IC luminosity per logarithmic photon interval is
\begin{align}
\frac{dL}{d\ln x} = \dot{N}_{\pm} x \left.\frac{dN_{\rm ph}}{d\ln{x}}\right|_{\rm IC, 1 el.}
= L_{\rm e} \frac{x}{\gamma_0} \left.\frac{dN_{\rm ph}}{d\ln{x}}\right|_{\rm IC, 1 el.},
\end{align}
where $\dot{N}_{\pm}$ denotes the number of pairs, including secondaries, that share the dissipated power.
The frequency-dependent compactness (eq.~\ref{eq:ellnu}) that enters opacity (eq.~\ref{eq:tauggnth}) can now be written as
\begin{align}
\ell_{\nu} &= \frac{\sigma_{\rm T}}{m_{\rm e} c^2} R_{\rm n} \frac{du_{\gamma}}{d\ln x}
= \frac{\sigma_{\rm T}}{m_{\rm e} c^2} R_{\rm n} \frac{dL}{d\ln x} \frac{t_{\rm res}}{V_{\rm n}} \nonumber \\
&= 3 \frac{\sigma_{\rm T}}{m_{\rm e} c^3} \frac{L_{\rm e}}{4\pi R_{\rm n}} \frac{x}{\gamma_0}  \left.\frac{dN_{\rm ph}}{d\ln{x}}\right|_{\rm IC, 1 el.} \frac{t_{\rm res}}{t_{\rm LC}}
= \frac{3}{2} \, \ell_{\rm inj} \, \frac{R_{\rm ej}}{R_{\rm n}} \left( \frac{x}{x_{\rm IC}} \right)^{1/2} \frac{t_{\rm res}}{t_{\rm LC}},
\label{eq:ellnuIC}
\end{align}
where $t_{\rm LC} = R_{\rm n}/c$ is the light crossing time of the wind/nebula, $x_{\rm IC} = 4\theta\gamma_0^2$ is the characteristic IC photon energy, and the injection compactness $\ell_{\rm inj}$ is defined by equation (\ref{eq:ellinj}). The target photon residence time $t_{\rm res}$ in the wind/nebula is typically of the same order as $t_{\rm LC}$, since the nebula is baryon starved and both photoionization and Compton recoil losses are negligible.

Using equations (\ref{eq:tauggnth}), (\ref{eq:elICcool1}) and (\ref{eq:ellnuIC}) in equation (\ref{eq:Me}) and assuming $\tau_{\gamma\gamma, {\rm nth}} < 1$ (i.e. that nonthermal photons can penetrate the entire wind region), one obtains the condition
\begin{align}
    {\cal M}_{\rm e} \approx \frac{3}{2} \, \xi \, \eta(\alpha) \, \ell_{\rm inj} \, \frac{R_{\rm ej}}{R_{\rm n}} \, \frac{\ln x_{\rm IC}}{4\theta\gamma_0} = 1.
    \label{eq:MeIC}
\end{align}

A few remarks should be made about equation (\ref{eq:MeIC}). First, equation (\ref{eq:MeIC}) can only be satisfied if $x_{\rm IC} = 4\theta\gamma_0^2 > 1$, i.e. at least some IC photons must be above the pair production threshold. Secondly, near the threshold $x_{\rm IC} \gtrsim 1$, the term $(4\theta\gamma_0)^{-1} = (4\theta x_{\rm IC})^{-1/2} = 500 \theta_{-6}^{-1/2} x_{\rm IC}^{-1/2} \gg 1$, so that ${\cal M}_{\rm e} > 1$ for a range of values $\gamma_0 > (4\theta)^{-1/2}$ as long as
$\ell_{\rm inj}$
%$\eta(\alpha) \ell_{\rm inj} \approx 0.1 \ell_{\rm inj}$
remains sufficiently high. Therefore, to satisfy ${\cal M}_{\rm e} = 1$, the pair loading maintains $\gamma_0$ sufficiently low so that only a small fraction of IC photons exceed the pair production threshold. In other words, $x_{\rm IC} \approx 1$ and $\gamma_0 \approx (4\theta)^{-1/2}$, as long as
\begin{align}
\ell_{\rm inj} \gtrsim \frac{(4\theta)^{1/2}}{\eta(\alpha)} \frac{R_{\rm n}}{R_{\rm ej}} \approx 0.02 \,  \theta_{-6}^{1/2} \frac{R_{\rm n}}{R_{\rm ej}},
\label{eq:ell_thr}
\end{align}
where we have used $\eta(\alpha) \approx 0.1$.  Using equation (\ref{eq:ellinj}) for $\ell_{\rm inj}$, this occurs on a timescale
\begin{equation}
t_{\pm, \rm IC} \approx 0.28 \, B_{14}^{-2/3}v_{9}^{-1/3}\theta_{-6}^{-1/6}\,{\rm yr}.
\label{eq:tpmnth}
\end{equation}

Finally, once (\ref{eq:ell_thr}) is no longer satisfied ($t > t_{\rm \gamma\gamma,nth})$, the nonthermal pair regulation rapidly breaks down: since ${\cal M}_{\rm e} \propto \gamma_0^{-1}$, the feedback process is unstable away from the threshold (at $x_{\rm IC} \gg 1$); an infinitesimal decrease in $\dot{N}_{\pm}$ produces a small increase in $\gamma_0$, which leads to a decrease in ${\cal M}_{\rm e}$ and hence a further decrease in $\dot{N}_{\pm}$. The feedback then switches to the thermal mechanism described in the next subsection.

\subsubsection{Synchrotron photons + synchrotron targets}

\label{app:reg:syn}

Another mode of nonthermal pair loading can occur if at some stage synchrotron photons become sufficiently energetic for pair creation, i.e. if the dimensionless synchrotron frequency $x_{\rm syn} = h\nu_{\rm syn, 0}/(\me c^2) = \gamma_0^2 x_{\rm B} \ge 1$, where $x_{\rm B}$ is the dimensionless Larmor frequency. The argument is analogous to the one presented above, except one has to replace $\dot{\gamma}_{\rm IC}$, $4\theta$ and $x_{\rm IC}$ with $\dot{\gamma}_{\rm syn}$, $x_{\rm B}$ and $x_{\rm syn}$, respectively, in equations (\ref{eq:elIC1})-(\ref{eq:MeIC}). This yields a condition analogous to (\ref{eq:MeIC}), under the assumption that pair cooling is dominated by sychrotron radiation. One can account for additional IC losses on thermal radiation by writing $\dot{\gamma} = \dot{\gamma}_{\rm syn} + \dot{\gamma}_{\rm IC} = \dot{\gamma}_{\rm syn} (1 + \ell_{\rm th} F_{\rm KN}/\ell_B)$, where
\begin{align}
F_{\rm KN} \approx 
\left\{\begin{array}{ll}
         1, 		                                            & 4\theta\gamma \ll 1 \vspace{1mm}\\
         \displaystyle\frac{9}{8(4\theta\gamma)^2} \left[\ln(4\theta\gamma) - \displaystyle\frac{11}{6}\right],  & 4\theta\gamma \gg 1.
                                       \end{array}
                               \right.
\end{align}
is the Klein-Nishina correction to the Thomson IC cooling rate. Using this in $dt = d\gamma/\dot{\gamma}$ when integrating equation (\ref{eq:elIC1}) (with the above substitutions) over the electron cooling history,
%one eventually arrives at the condition
the synchrotron spectrum emitted by a single electron becomes
\begin{align}
\left.\frac{dN_{\rm ph}}{d\ln{x}}\right|_{\rm syn, 1 el.} = \frac{1}{2(x_{\rm B} x)^{1/2} (1 + \ell_{\rm th} F_{\rm KN}/\ell_B)}, \hspace{5mm} x \le x_{\rm B}\gamma_0^2.
\label{eq:elsyncool1}
\end{align}
Instead of condition (\ref{eq:MeIC}) one now obtains
\begin{align}
    {\cal M}_{\rm e} \approx \frac{3}{2} \, \xi \, \eta(\alpha) \, \frac{\ell_{\rm inj}}{1 + \ell_{\rm th} F_{\rm KN}/\ell_B} \, \frac{R_{\rm ej}}{R_{\rm n}} \, \frac{\ln x_{\rm syn}}{x_{\rm B}\gamma_0} = 1.
    \label{eq:MeSyn}
\end{align}

%Based on equation (\ref{eq:num}),
The threshold condition $h\nu_{\rm syn,0} > \me c^2$ requires (eq. \ref{eq:num})
%a high $\gamma_0 > 10^6$ for typical parameters;
\begin{align}
    \gamma_0 > \gdag = 6.0\times 10^6 \, B_{14}^{1/2} \, \varepsilon_{\rm B, -2}^{-1/4} \, v_9^{3/4} \, t_{\rm yr}. \label{eq:gpmsyn}
\end{align}
Below we will show that thermal pair loading regulates the injected lepton energy to $\gamma_0 \approx 1/\theta \approx 10^6$, hence the synchrotron mechanism is unlikely to operate 
%hence the synchrotron mechanism is unlikely to be the dominant regulation mechanism
as long as thermal pair loading is effective. Once the thermal regulation fails, the synchrotron loading can operate as long as the ``unloaded'' $\gamma_{\pm} = \gamma_{\rm in}$ (eq. \ref{eq:gammapm}) satisfies the above threshold condition.

Neglecting IC cooling,
%and assuming $x_{\rm syn} \approx \gamma_0^2 x_{\rm B} \approx 1$,
the pair multiplicity (\ref{eq:MeSyn}) can be written approximately as
\begin{align}
{\cal M}_{\rm e} \sim 0.1 \, \ell_{\rm inj} \frac{\gamma_0 \ln x_{\rm syn}}{x_{\rm syn}} = 0.2 \, \ell_{\rm inj} \gdag \, \frac{\gdag \ln (\gamma_0/\gdag)}{\gamma_0}, \hspace{5mm} \gamma_0 \ge \gdag,
\label{eq:MeSyn2}
\end{align}
where we have used $\gdag^2 x_{\rm B} = 1$.
Intuitively, the multiplicity ${\cal M}_{\rm e}$ is a product of $\gamma_0/x_{\rm syn} \sim \gamma_0$ photons produced per lepton and a fraction $\tau_{\gamma\gamma} \sim 0.1 \ell_{\rm inj}$ of them producing secondary pairs. The second equality in (\ref{eq:MeSyn2}) shows that as long as $0.1 \, \ell_{\rm inj} \gdag > 1$ (approximately), then the condition ${\cal M}_{\rm e}(\gamma_0) = 1$ has two roots, $\gamma_0 \sim \gdag$ and another at $\gamma_0 > \gdag$. The latter solution is in the unstable region, where ${\cal M}_{\rm e} \propto \gamma_{0}^{-1}$, i.e. any initial perturbation of $\gamma_0$ is amplified until another mechanism takes over the pair regulation. Typically,  $\gamma_0$ is a growing function of time, so the lower-energy (stable) solution $\gamma_0 \sim \gdag$ is attained first; the system remains in that state as long as $0.1 \, \ell_{\rm inj} \gdag > 1$.

The pair loading due to synchrotron radiation, if it occurs at all, will take place at times $t_{\pm,\rm syn}$ obeying
\be
t_{\pm,\rm syn} < \min(t_{\pm,\rm syn}^{(2)}, t_{\pm,\rm syn}^{(1)}).
%t_{\pm,\rm syn}^{(1)}< t_{\pm,\rm syn} < t_{\pm,\rm syn}^{(2)}.
\label{eq:tppsyn}
\ee
The first timescale is the requirement that $0.1 \, \ell_{\rm inj} \gdag > 1$, which is satisfied before a time
\be
t_{\pm,\rm syn}^{(1)} \approx 11.6\,B_{14}^{-3/4} v_{9}^{-1/8}\varepsilon_{B,-2}^{-1/8}\,{\rm yr}.
\ee
The second timescale arises from the condition that the regulated $\gamma_0 \sim \gamma_{\dagger}$ be lower than the ``unregulated" value $\gamma_{\rm in}$ (eq. \ref{eq:gammapm}), which yields
\be
t_{\pm,\rm syn}^{(2)}= 7.2 \, B_{14}^{-3/4} v_{9}^{-3/8} \varepsilon_{B,-2}^{1/8} \left(\frac{\mu_{\pm}}{10^{4}}\right)^{-1/2} \, {\rm yr}.
\ee

%The first timescale is the requirement for synchrotron photons to be sufficiently energetic to pair produce ($h\nu_{\rm syn} > m_e c^{2}$), which using the regulated $\gamma_0$ (eq.~\ref{eq:gpmsyn}) in equation (\ref{eq:num}) is satisfied after a time
%\be
%t_{\pm,\rm syn}^{(1)} \approx 11.6\,B_{14}^{-3/4}v_{9}^{-1/8}\varepsilon_{B,-2}^{-1/8}\,{\rm yr}.
%\ee
%The second timescale corresponds to the maintaining the compactness condition $0.1\, \ell_{\rm inj} \gamma_{\rm in} %\gtrsim 1$, which using equations (\ref{eq:ellinj}, \ref{eq:gammapm}), gives
%\be
%t_{\pm,\rm syn}^{(2)} \approx 8.8 B_{14}^{-3/4}v_{9}^{-1/4}\left(\frac{\mathcal{M}_{\pm}}{10^{4}}\right)^{-1/4}\,{\rm yr}.
%\label{eq:tppsyn}
%\ee

\subsubsection{IC photons + synchroton targets}

The mechanisms outlined in the previous two sections operate in the early (close to the optical peak) and late (years) stages of the event, respectively. In addition to the thermal mechanism discussed below, a further mode of pair regulation is possible in the intermediate stage ($t \sim 1$~year) for sufficiently high nebula magnetizations, in which case IC upscattered thermal photons pair produce on UV and X-ray synchrotron radiation.

The threshold condition for pair production by this process is $x_{\rm IC} x_{\rm syn} = 4\theta x_{\rm B} \gamma_0^4 > 1$. The multiplicity ${\cal M}_{\rm e}$ is obtained from equation (\ref{eq:Me}) by using equation (\ref{eq:elICcool1}) for the IC spectrum emitted by a single electron (with an additional factor $(1 + \ell_{\rm B}/\ell_{\rm th} F_{\rm KN})^{-1}$ that accounts for synchrotron losses), and using the synchrotron spectrum (\ref{eq:elsyncool1}) for computing the pair production opacity. The result is 
\begin{align}
    {\cal M}_{\rm e} \approx \frac{3}{2} \, \xi \, \eta(\alpha) \, \ell_{\rm inj} \, \frac{\ell_B \ell_{\rm th} F_{\rm KN}}{(\ell_B + \ell_{\rm th} F_{\rm KN})^2} \, \frac{R_{\rm ej}}{R_{\rm n}} \, \frac{\gamma_0 \ln (x_{\rm IC} x_{\rm syn})}{(x_{\rm syn} x_{\rm B})^{1/2}} = 1.
    \label{eq:MeICSyn}
\end{align}
Following an argument analogous to the previous two sections, one concludes that
%as long as $0.1 \, \gamma_{0} \, \ell_{\rm inj} \ell_B \ell_{\rm th} F_{\rm KN}/(\ell_B + \ell_{\rm th} F_{\rm KN})^2 > 1$, then
pair loading maintains $\gamma_0 \sim (4\theta x_{\rm B})^{-1/4}$ (i.e. $x_{\rm IC} x_{\rm syn} \sim 1$),
\begin{align}
    \gamma_0 \approx 5.5\times 10^4 \, \theta_{-6}^{-1/4} B_{14}^{1/4} \, \varepsilon_{\rm B, -2}^{-1/8} \, v_9^{3/8} \, t_{\rm yr}^{1/2}, \label{eq:gpmICsyn}
\end{align}
as long as $0.1 \, \gamma_{0} \, \ell_{\rm inj} \ell_B \ell_{\rm th}/(\ell_B + \ell_{\rm th})^2 > 1$ (assuming $F_{\rm KN} \approx 1$). The latter condition is satisfied at $t < t_{\pm,\rm IC-syn}$, where
\begin{align}
t_{\pm,\rm IC-syn} \approx 1 \, \theta_{-6}^{-1/10} \, B_{14}^{-7/10} \, \varepsilon_{\rm B, -2}^{-1/20} \, v_9^{-1/4}
\left[
\frac{\ell_B \ell_{\rm th}}{(\ell_B + \ell_{\rm th})^{2}}
\right]^{2/5} \,{\rm yr}.
\label{eq:tppICsyn}
\end{align}

\subsection{Thermal pair loading}

In this regime pair loading is controlled by IC gamma-rays interacting with thermal target photons. To determine the pair multiplicity, we now need to use the IC spectrum of an electron cooling on the thermal radiation field in both Thomson and Klein-Nishina regimes. In the delta-function approximation, one can write
%
%To make further progress we must specify the spectrum emitted by the electron as it cools down from its initial $\gamma_0 \approx \Delta\gamma$.  Rather than use the exact IC spectrum, for the present purpose it is sufficient (and more instructive) to employ a simple delta-function approximation \citep{Vurm&Metzger18}
\begin{align}
    \left.\frac{dN_{\rm ph}}{d\ln{x} \, dt}\right|_{\rm 1 el.} = 
    \left\{\begin{array}{ll}
         \dot{\gamma}_{\rm IC} \, \delta\left( x - 4\theta\gamma^2 \right),  		    & 4\theta\gamma \le 1 \,\, \mbox{\rm (Thomson)} \vspace{1mm}\\
         \dot{\gamma}_{\rm IC} \, \delta\left( x - \gamma \right), 		                & 4\theta\gamma > 1 \,\, \mbox{\rm (Klein-Nishina)}.
                                       \end{array}
                               \right.
\label{eq:elIC}                               
\end{align}
The total cooling rate, also including synchrotron emission, can be written as\footnote{We have neglected Coulomb and bremsstrahlung cooling (Appendix \ref{sec:app:cooling}), as justified by the low Thomson optical depth and low expected baryon content of the wind and nebula.}
\begin{align}
\dot{\gamma} = \dot{\gamma}_{\rm IC} + \dot{\gamma}_{\rm syn}
= \dot{\gamma}_{\rm IC} \left( 1 + \frac{\ell_{B}}{\ell_{\rm th} F_{\rm KN}} \right) = \frac{4c}{3R_{\rm n}} \ell_{\rm th} \gamma^2 F_{\rm KN} \left( 1 + \frac{\ell_{B}}{\ell_{\rm th} F_{\rm KN}} \right).
\end{align}
%where 
%\begin{align}
%F_{\rm KN} \approx 
%\left\{\begin{array}{ll}
%         1, 		                                            & 4\theta\gamma \ll 1 \vspace{1mm}\\
%         \displaystyle\frac{9}{8(4\theta\gamma)^2} \left[\ln(4\theta\gamma) - \displaystyle\frac{11}{6}\right],  & 4\theta\gamma \gg 1.
%                                       \end{array}
%                               \right.
%\end{align}
%is the Klein-Nishina correction to the Thomson IC cooling rate.

The spectrum emitted by a cooling electron is obtained by integrating Equation (\ref{eq:elIC}) over time $dt = d\gamma/\dot{\gamma}$: 
\begin{align}
    \left.\frac{dN_{\rm ph}}{d\ln{x}}\right|_{\rm 1 el.} = 
    \left\{\begin{array}{ll}
         \displaystyle\frac{1}{2 (4\theta x)^{1/2} \, (1 + \ell_B/\ell_{\rm th})},    &  x \le \min(1/4\theta,4\theta\gamma_0^2) \vspace{1mm}\\
         \displaystyle\frac{1}{1 + \ell_B/(\ell_{\rm th} F_{\rm KN})}, 		          & 1/4\theta < x < \gamma_0 \vspace{1mm}\\
         0,                                                     & \mbox{else.}
                                       \end{array}
                               \right.
\label{eq:elICcool}                               
\end{align}

The pair production opacity $\tauggth$ that enters Equation~(\ref{eq:Me}) drops rapidly for $4\theta x < 1$ (eq.~\ref{eq:taugg}), such that Equation~(\ref{eq:Me}) can only be satisfied if $4\theta\gamma_0 \gtrsim 1$, i.e. the cooling electron initially scatters thermal photons in the Klein-Nishina regime.  On the other hand, the number of emitted IC photons {\it increases} with decreasing photon energy.  Therefore, the main contribution to the integral in eq.~(\ref{eq:Me}) comes from values of $4\theta x \approx$~few.

Using the spectrum (\ref{eq:elICcool}) we can rewrite condition (\ref{eq:Me}) as:
\begin{align}
    {\cal M}_{\rm e} \approx \xi \int^{1/4\theta} \frac{1}{(4\theta x)^{1/2} (1 + \ell_B/\ell_{\rm th})} \, (1-e^{-\tauggth}) \, d\ln{x}  
    + 2 \xi \int_{1/4\theta}^{\gamma_0} \frac{1}{1 + \ell_B/(\ell_{\rm th} F_{\rm KN})} \, (1-e^{-\tauggth}) \, d\ln{x}
    = 1.
    \label{eq:Me2}
\end{align}

Figure \ref{fig:pairmult} shows the numerically computed multiplicity ${\cal M}_{\rm e}$, using now exact IC/synchrotron cooling rates and spectra, as a function of $\gamma_0$ for different values of the gamma-ray optical depth $\tau_{\gamma\gamma,\rm th}$ through the wind region (evaluated at the peak value $x\theta \approx 2$) and different ratios of the magnetic to thermal compactness, $\ell_{\rm B}/\ell_{\rm th}$ (eq.~\ref{eq:lBn}).  As expected, the pair multiplicity increases with the  initial Lorentz factor of the radiating pair $\gamma_0$, the temperature of the background radiation field $\theta$, and the gamma-ray optical depth $\tau_{\gamma\gamma,\rm th}$.  However, to understand the detailed behavior of ${\cal M}_{\rm e}$ with increasing $\gamma_0 \theta$ we must rely on the approximate expression (\ref{eq:Me2}).  Through condition~(\ref{eq:regulation}) this determines the regulated Lorentz factor of the injected nebular pairs (eq.~\ref{eq:gammapm2}) and how it evolves as $\tau_{\gamma\gamma,\rm th}$ decreases and $\ell_{\rm B}/\ell_{\rm th}$ increases as time advances after the explosion.

First, consider the limit of low $\ell_B/\ell_{\rm th} \ll 1$ and high $\tauggth > 1$, which characterizes the nebula/wind region at early times of months to years (depending on the value of $\varepsilon_B$; eq.~\ref{eq:lB_lth}).  In this limit, the factor $1 - e^{-\tauggth} \approx 1$ near the peak, but drops rapidly at smaller $x\theta < 1$ due to the exponential dependence of $\tauggth$ on $x\theta$.  As a result, the first term on the right hand side of Equation~(\ref{eq:Me2}) is typically less than unity unless the thermal compactness is very high.  The second integral in Equation~(\ref{eq:Me2}) is trivial in the domain of interest as long as $\tauggth \gg 1$ and $\ell_B/\ell_{\rm th} \ll 1$; the condition ${\cal M}_{\rm e}=1$ yields $\ln(4\gamma_0\theta) \sim 1$, or $4\gamma_0\theta \sim$~a~few.  Consistent with this, Fig.~\ref{fig:pairmult} shows that the value $\gamma_0$ which satisfies ${\cal M}_{\rm e} = 1$ (eq.~\ref{eq:regulation}) depends only weakly on $\tauggth$ for $\tauggth > 1$.  

For larger values of $\ell_B/\ell_{\rm th}$ the integrands of both terms in Equation~(\ref{eq:Me2}) are smaller, such that a higher value of $\gamma_0$ is required to attain ${\cal M}_{\rm e} = 1$.  However, once $\ell_B/\ell_{\rm th}$ increases to a value $ \gtrsim 0.1$, the pair multiplicity drops significantly and eventually the thermal regulation condition ${\cal M}_{\rm e} = 1$ can no longer be satisfied for any $\gamma_0$.
Physically, this is because synchrotron emission (which dominates IC cooling for $\ell_B/\ell_{\rm th} \gtrsim 0.1$) cannot generate  energetic enough photons to pair produce on the thermal radiation background (eq.~\ref{eq:num}).  Similarly, once the wind/nebula becomes transparent to $\gamma\gamma$ interactions with optical photons (i.e. $\tauggth \ll 1$ at $x\theta \approx 2$), one has $(1-e^{-\tauggth}) \approx \tauggth \ll 1$ in Equation~(\ref{eq:Me2}), and again the condition ${\cal M}_{\rm e} = 1$ cannot be satisfied.

In case synchrotron photons become sufficiently energetic to pair produce between themselves (Section \ref{app:reg:syn}), the nonthermal mechanism can take over the pair regulation. For typical parameters this takes place only when the thermal mechanism has become inefficient.  
\end{document}